\documentclass[12pt]{article}

\usepackage{graphicx,psfrag,epsf,color}
\usepackage{amsmath,amssymb,amsfonts}
\usepackage{array}
\usepackage{cite}
\usepackage{multirow}
\usepackage{multirow}

\renewcommand{\arraystretch}{1.25}

\newcommand{\rescm}{\widehat}
\newcommand{\mmbar}{\widehat{m} \, \widehat{\overline{m}}}
\newcommand{\diffmmbar}{\Delta_m}
\newcommand{\mihat}{\widehat{m}_{i_1}}
\newcommand{\mjhat}{\widehat{m}_{i_2}}
\newcommand{\mAhat}{\widehat{m}_{A}}
\newcommand{\mBhat}{\widehat{m}_{B}}

\newcommand{\sumABij}{\Bigl\lbrace A \leftrightarrow B\, , i_1 \leftrightarrow i_2 \Bigr\rbrace}
\newcommand{\sumij}{\Bigl\lbrace  i_1 \leftrightarrow i_2 \Bigr\rbrace}

\begin{document}

\allowdisplaybreaks

\begin{titlepage}

\begin{flushright}
{\small
TUM-HEP-878/13\\
TTK-13-04\\
SFB/CPP-13-15\\
IFIC/13-05\\[0.2cm]

\today}
\end{flushright}

\vskip1cm
\begin{center}
\Large\bf\boldmath
Non-relativistic pair annihilation of  
nearly mass degenerate neutralinos and charginos
II. P-wave and next-to-next-to-leading order S-wave coefficients
\end{center}

\vspace{0.8cm}
\begin{center}
{\sc C.~Hellmann$^{a,b}$} and 
{\sc P. Ruiz-Femen\'\i a$^{c}$}\\[5mm]
{\it ${}^a$Physik Department T31,\\
James-Franck-Stra\ss e, 
Technische Universit\"at M\"unchen,\\
D--85748 Garching, Germany\\
\vspace{0.3cm}
${}^b$Institut f\"ur Theoretische Teilchenphysik und 
Kosmologie,\\
RWTH Aachen University, D--52056 Aachen, Germany\\
\vspace{0.3cm}
${}^c$Instituto de F\'\i sica Corpuscular (IFIC), 
CSIC-Universitat de Val\`encia \\
Apdo. Correos 22085, E-46071 Valencia, Spain}\\[0.3cm]
\end{center}

\vspace{1cm}
\begin{abstract}
\vskip0.2cm\noindent
This paper is a continuation of an earlier work (arXiv:1210.7928) which computed analytically the 
tree-level annihilation rates of a collection of non-relativistic neutralino and chargino two-particle
states in the general MSSM. Here we extend the results by providing the next-to-next-to-leading order 
corrections to the rates in the non-relativistic expansion in momenta and mass differences, which
include leading $P$-wave effects, in analytic form. The
results are a necessary input for the calculation of the Sommerfeld-enhanced dark matter
annihilation rates including short-distance corrections at next-to-next-to-leading order in the non-relativistic expansion in the general MSSM with neutralino LSP.

\end{abstract}
\end{titlepage}


\section{Introduction}
\label{sec:introduction}
The increasing precision on the experimental determination of the dark matter (DM) density,
which is expected to be further improved by the data from the PLANCK satellite, has 
brought a renewed interest in the impact of radiative corrections to 
the annihilation cross section of dark matter candidates of particle nature.
Particles with weak interaction strength and masses around the TeV scale that
dropped out of thermal equilibrium in the Early Universe yield the correct order
of magnitude for the relic density. An example of the latter is provided
by the lightest neutralino of the minimal supersymmetric
extension of the standard model (MSSM), perhaps the most promising
candidate for such weakly interacting dark matter. 
The potential to set stringent constraints on the parameter space of the MSSM
using the high precision measurements of
the dark matter density crucially depends on having an accurate calculation of the neutralino
relic abundance. 

A necessary input for this calculation is the
annihilation cross section of the lightest neutralino, and of all
possible co-annihilation processes. While programs exist that provide the tree-level
results numerically~\cite{Belanger:2010gh,Gondolo:2004sc}, at the one-loop
level such calculations are not available for a generic MSSM model, 
although they have been performed 
for some scenarios~\cite{Herrmann:2007ku,Herrmann:2009wk,Herrmann:2009mp,Baro:2007em,Baro:2009na}, 
or under certain approximations~\cite{Boudjema:2011ig,Chatterjee:2012db,Drees:2013er}. 

There is however a certain class of radiative corrections where higher-order loop diagrams contributing
to the DM annihilation amplitude
are not necessarily suppressed.
At the temperatures where freeze-out of the relic particle abundance takes place, the
dark matter particles are non-relativistic, with typical velocities of order
$v\sim 0.2\,c$.
Quantum loop corrections 
due to the exchange of light particles between the non-relativistic DM particles
before annihilation can become more and more important in situations where
the force coupling strength is larger
than the DM velocity and the mass of the force carrier is much lighter than the DM mass, eventually
requiring a resummation of the terms in the perturbative expansion to all loop orders.
This phenomenon has been termed as ``Sommerfeld effect'', and can lead to a significant enhancement 
of the DM annihilation rates, also of relevance for the calculation of primary decay
spectra in the present Universe. 
In the MSSM, Sommerfeld corrections may constitute the dominant radiative correction
when the lightest neutralino is much heavier than the electroweak 
gauge bosons and the Yukawa potential generated by their exchange becomes long-range.
For such heavy neutralinos mixing effects are suppressed by ${\cal O}(M_Z/m_{\rm LSP}$) 
and thus mass degeneracies arise in the neutralino-chargino sector, making necessary 
to account for co-annihilation processes in the relic density calculation.
Two prominent examples of this scenario are the MSSM wino- and Higgsino-limit, for which the
impact of the Sommerfeld effect has been extensively 
studied~\cite{Hisano:2004ds,Hisano:2006nn,Cirelli:2007xd,Hryczuk:2010zi,Hryczuk:2011vi}.

The tree-level DM annihilation cross section can be expanded
in the relative velocity $v_\text{rel}$ of the two annihilating particles 
($v_\text{rel}=|\boldsymbol{v}_1-\boldsymbol{v}_2|$),
\begin{align}
\label{eq:sigmav}
 \sigma_\text{ann} \, v_\text{rel} \, = \,a + b\,v_\text{rel}^2 + {\cal O}(v_\text{rel}^4)\,,
\end{align}
for non-relativistic $v_\text{rel}$.
In a previous work \cite{Beneke:2012tg}  (referred to as paper I in the
following) the calculation of the leading-order coefficient $a$ 
in (\ref{eq:sigmav}) 
was presented in analytic form for the general MSSM with neutralino LSP,
including results for all co-annihilation processes with nearly mass-degenerate
neutralinos and charginos. In this paper we complete the calculation by
providing analytic results for the subleading term in this expansion, the
coefficient $b$. 
In the non-relativistic effective theory (EFT) framework devised in paper I,
the coefficients $a,\,b$ account for the short-distance part of the neutralino
and chargino pair-annihilation processes and are written as a combination of
the absorptive parts of Wilson coefficients of local four-fermion operators.
The absorptive part of the EFT matrix element of these four-fermion operators
then gives the full neutralino and chargino pair-annihilation rates,
including the long-range Sommerfeld effect.

While the leading-order term $a$
in (\ref{eq:sigmav}) receives contributions only from $S$-wave annihilations, the subleading
term $b$ encodes both $S$- and $P$-wave annihilation contributions, which we provide separately.
This is required for a correct implementation of the Sommerfeld correction factors,
that depend on the spin and partial-wave configuration of the annihilating state. 
In recent literature which addresses the Sommerfeld effect 
including the ${\cal O}(v_\text{rel}^2)$ terms in the annihilation
cross section, only the $P$-wave contributions to the
coefficient $b$ have been computed approximately
using numerical routines at the amplitude level~\cite{Drees:2013er}, 
or it has been assumed that $b$ is entirely $P$-wave~\cite{Chen:2013bi}.

The results presented in I (and complemented in this work) also extend those from previous approaches in another relevant aspect.
The knowledge of the tree-level annihilation cross section (\ref{eq:sigmav}) for all nearly mass-degenerate
neutralino and chargino two-particle states is not sufficient for the calculation of the 
Sommerfeld corrected (co-)annihilation rates. As described in paper I, a contribution to the
full annihilation rate of an incoming $\chi_i\chi_j$ state is given by the imaginary part of the amplitude for the process 
\begin{align}
\chi_i \chi_j \to \ldots \to \chi_{e_1} \chi_{e_2} \to X_A X_B \to \chi_{e_4} \chi_{e_3} \to \ldots  \to \chi_i  \chi_j 
\ ,
\label{eq:off-diag}
\end{align}
where the transitions among the  $\chi\chi$ states are mediated by long-range
potential interactions and the short-distance annihilation into SM and light
Higgs particles ($X_A X_B$) involves the two-particle states
$\chi_{e_1} \chi_{e_2}$ and $\chi_{e_4} \chi_{e_3}$.
The case when $\chi_{e_1} \chi_{e_2}$ and $\chi_{e_4} \chi_{e_3}$ are different
states corresponds to an off-diagonal short-distance annihilation rate. In our
EFT approach the diagonal and the off-diagonal (tree-level) short-distance rates
are encoded in the absorptive part of local four-fermion operators' Wilson
coefficients, that are obtained from matching the EFT tree-level matrix
elements of the four-fermion operators to the absorptive part of the hard
(1-loop) MSSM amplitudes for the
$\chi_{e_1} \chi_{e_2} \to X_A X_B \to \chi_{e_4} \chi_{e_3}$  scattering reactions.
The off-diagonal terms have not been taken into account in the
Sommerfeld-enhanced neutralino relic abundance calculations aside from the
wino- and Higgsino-limits
\cite{Hisano:2004ds,Hisano:2006nn,Cirelli:2007xd,Hryczuk:2011vi}, and their
implementation using the numerical packages that provide the tree-level
annihilation rates has not yet been attempted.
In contrast, the analytic results presented in this work allow for a systematic 
treatment of all diagonal and off-diagonal short-distance annihilation rates at 
next-to-next-to-leading order in the non-relativistic expansion in
Sommerfeld-enhanced annihilation reactions. 

The contents of this paper are the following:
In Sec.~\ref{sec:basis} we briefly review the effective Lagrangian framework
introduced in paper I and recollect the essential notation. We then introduce
the dimension-8  four-fermion operators that encode the next-to-next-to-leading
non-relativistic corrections to the short-distance annihilation of neutralino
and chargino particle pairs. As for the case of the (leading-order) dimension-6 
operators discussed in paper I, the analytic results for the absorptive part of 
the Wilson coefficients can be obtained as the product of kinematic and coupling
factors. In the Appendix~\ref{sec:appendix}  we recall the master formula to
write down the Wilson coefficients and the rules for its implementation.
While we rely on paper I for the extraction of coupling factors also for the
Wilson coefficients presented in this work, explicit expressions
for the $P$-wave kinematic factors are given in the Appendix~\ref{sec:appendix}.
The expressions for the (rather lengthy) next-to-next-to-leading $S$-wave
Wilson coefficients are collected in a {\tt Mathematica} package attached
to this paper~\cite{dotmfile}, which also includes the $P$-wave kinematic factors and those from
the leading-order operators that were written explicitly in the appendix of
paper I. Appendix~\ref{sec:appendixsuppl} explains the notation used in this
electronic supplement. 
In Sec.~\ref{sec:gammarate} we generalise the formula for the tree-level annihilation 
rates to the case of off-diagonal annihilation processes, which is needed in
order to analyse the size of next-to-next-to-leading corrections in an
off-diagonal transition used as a case example in Sec.~\ref{sec:results}.
Apart from the latter, Sec.~\ref{sec:results} also discusses three 
selected (diagonal) processes where the role of the next-to-next-to-leading
corrections from our analytic calculation of the annihilation cross section is markedly
different. Through these examples we illustrate the importance 
of separating the different partial-wave contributions to the short-distance 
annihilation for the computation of the Sommerfeld-corrected cross sections.
Our results for the diagonal annihilation rates for these examples
are checked against the corresponding unexpanded cross sections computed 
with a numerical code.
To serve as an example of how to use the results presented in this work,
we have included in Appendix~\ref{sec:appendixexample} a step-by-step calculation
of the 
non-relativistic $\chi^+_1 \chi^-_1 \to W^+ W^-$ annihilation cross section
including up
to  ${\cal O}(v_{\rm rel}^2)$ effects for the case of pure-wino neutralino dark matter.
Analytic results for the Wilson 
coefficients needed to determine the exclusive (off-)diagonal (co-)annihilation
rates $\chi_{e_1}\chi_{e_2} \to X_A X_B \to \chi_{e_4} \chi_{e_3}$ 
in the decoupling limit of the pure-wino scenario
are also provided in that appendix.
Finally, we summarise our findings in Sec.~\ref{sec:summary}.


\section{Basis of the dimension-8 operators in $\delta \mathcal L_\text{ann}$}
\label{sec:basis}
The non-relativistic MSSM (NRMSSM) effective theory set-up of paper I is built
out of $n_0 \leq 4$ nearly on-shell non-relativistic neutralino
($\chi_i^0,\, i=1,\dots, n_0$) and $n_+ \leq 2$ nearly on-shell non-relativistic
chargino ($\chi_j^\pm,\, j=1,\dots, n_+)$ modes whose masses are nearly degenerate
with the mass $m_{\text{LSP}}$ of the lightest neutralino $\chi^0_1$.
As pair-annihilation reactions of these neutralino and chargino species into
(not non-relativistic) SM and light Higgs-particle final states take place at
distances of the order $\mathcal O(1/m_\text{LSP})$, much smaller than the
characteristic range of potential interactions between the incoming $\chi\chi$
two-particle states, we can incorporate the actual annihilation rates
in the effective theory through the absorptive part of Wilson coefficients of
local four-fermion operators ($\delta \mathcal L_\text{ann}$), in analogy to the
treatment of quarkonium annihilation in NRQCD~\cite{Bodwin:1994jh}.
In contrast to the $Q\overline{Q}$ case, here the long-range potential interactions
can lead to transitions among different two-particle states before the 
short-distance annihilation takes place:
the initially incoming $\chi_i\chi_j$ particle pair can scatter to 
any accessible (nearly on-shell) $\chi_{e_a}\chi_{e_b}$ particle pair prior to the
annihilation into SM and light Higgs two-particle final states $X_A X_B $.
Therefore we have to account for absorptive parts of  generic
$\chi_{e_1} \chi_{e_2} \to X_A X_B \to \chi_{e_4}\chi_{e_3}$
amplitudes, where $\chi_{e_1}\chi_{e_2}$ and  $\chi_{e_4}\chi_{e_3}$ 
can be different states (see Fig.~1 in paper I).

At $\mathcal O(\alpha^2_2)$, where $\alpha_2 = g^2_2/4 \pi$ with $g_2$ the $SU(2)_L$
gauge coupling in the MSSM, the absorptive part of the four-fermion
operators' Wilson coefficients are obtained by matching the absorptive
part of $\chi_{e_1}\chi_{e_2} \to \chi_{e_4}\chi_{e_3}$ 1-loop scattering amplitudes
evaluated in the MSSM with the tree-level matrix element of four-fermion
operators contained in $\delta \mathcal L_\text{ann}$ in the effective theory.
At the 1-loop level, the contribution to the absorptive part from every
individual final state $X_A X_B$ is free from infrared divergences, and can be
given separately.
At higher orders the absorptive part of the Wilson coefficients refers 
to the inclusive case, {\it i.e.} summed over all accessible final states.

The leading-order contributions to $\delta \mathcal L_\text{ann}$ are given by
dimension-6 four-fermion operators, encoding leading-order $S$-wave scattering
reactions. The corresponding operators have been given in paper I.
At next-to-next-to-leading order in the non-relativistic expansion in momenta
and mass differences, dimension-8 four-fermion operators
contribute.\footnote{Let us remark that
we do not consider next-to-leading order contributions to
$\delta \mathcal L_\text{ann}$, corresponding to dimension-7 four-fermion
operators, as these encode $^1S_0 - {}^3P_0$, $^3S_1 - {}^1P_1$ and $^3S_1 - {}^3P_1$ transitions
which will require the addition of $v_\text{rel}$-suppressed potential
interactions in the long-range part of the annihilation 
(we consider only ${\cal O}(v_\text{rel}^2)$ effects from the short-distance annihilation,
and not those arising from sub-leading non-Coulomb (non-Yukawa) potentials, in
consistency with paper I).}
Here we adopt the same notation used for $\delta \mathcal L^{d=6}_\text{ann}$ in I
in order to write the dimension-8
four-fermion operators in $\delta \mathcal L^{d=8}_\text{ann}$ as
\begin{align}
\nonumber
 \delta \mathcal L ^{d=8}_\text{ann}
  \ =& \
   \sum\limits_{\chi\chi \to \chi\chi} 
   \frac{1}{4\,M^2} ~
   f^{\chi\chi \to \chi\chi}_{\lbrace e_1 e_2 \rbrace \lbrace e_4 e_3 \rbrace}\left( ^1P_1 \right)
   ~
   \mathcal O^{\chi\chi \to \chi\chi}_{\lbrace e_4 e_3 \rbrace \lbrace e_2 e_1 \rbrace}\left( ^1P_1 \right)
 \\\nonumber
 & +
  \sum\limits_{\chi\chi \to \chi\chi} \, \sum\limits_{J=0,1,2} ~
   \frac{1}{4\,M^2} ~
   f^{\chi\chi \to \chi\chi}_{\lbrace e_1 e_2 \rbrace \lbrace e_4 e_3 \rbrace}\left( ^3P_J \right)
   ~
   \mathcal O^{\chi\chi \to \chi\chi}_{\lbrace e_4 e_3 \rbrace \lbrace e_2 e_1 \rbrace}\left( ^3P_J \right)
 \\\nonumber
 & +
  \sum\limits_{\chi\chi \to \chi\chi} \, \sum\limits_{s=0,1} ~
   \frac{1}{4\,M^2} ~
   g^{\chi\chi \to \chi\chi}_{\lbrace e_1 e_2 \rbrace \lbrace e_4 e_3 \rbrace}\left( ^{2s+1}S_s \right)
   ~
   \mathcal P^{\chi\chi \to \chi\chi}_{\lbrace e_4 e_3 \rbrace \lbrace e_2 e_1 \rbrace}\left( ^{2s+1}S_s \right)
 \\
 & +
  \sum\limits_{\chi\chi \to \chi\chi} \, \sum\limits_{s=0,1}\, \sum\limits_{i=1,2} ~
   \frac{1}{4\,M^2} ~
   h^{\chi\chi \to \chi\chi}_{i\,\lbrace e_1 e_2 \rbrace \lbrace e_4 e_3 \rbrace}\left( ^{2s+1}S_s \right)
   ~
   \mathcal Q^{\chi\chi \to \chi\chi}_{i\,\lbrace e_4 e_3 \rbrace \lbrace e_2 e_1 \rbrace}\left( ^{2s+1}S_s \right) \ .
\label{eq:basisSPwave}
\end{align}
The label $\chi\chi\to \chi\chi$ stands for all (off-)diagonal non-relativistic
scattering processes among neutralino and chargino two-particle states.
Neutral reactions involve $\chi^0\chi^0$ and $\chi^-\chi^+$ states, while
singly-charged and doubly-charged processes include $\chi^0\chi^\pm$ and 
$\chi^\pm\chi^\pm$ states, respectively.
The $f^{\chi\chi\to\chi\chi}_{\lbrace e_1 e_2 \rbrace\lbrace e_4 e_3 \rbrace}$,
$g^{\chi\chi\to\chi\chi}_{\lbrace e_1 e_2 \rbrace\lbrace e_4 e_3 \rbrace}$ and
$h^{\chi\chi\to\chi\chi}_{i\,\lbrace e_1 e_2 \rbrace\lbrace e_4 e_3 \rbrace}$ in
(\ref{eq:basisSPwave}) denote the Wilson coefficients of the corresponding
four-fermion operators $\mathcal O_{\lbrace e_4 e_3 \rbrace\lbrace e_2 e_1 \rbrace}$,
$\mathcal P_{\lbrace e_4 e_3 \rbrace\lbrace e_2 e_1 \rbrace}$ and
$\mathcal Q_{i\,\lbrace e_4 e_3 \rbrace\lbrace e_2 e_1 \rbrace}$, 
whose explicit form for
the case of $\chi_{e_1}^0\chi_{e_2}^0 \to \chi_{e_4}^0\chi_{e_3}^0$ scattering
reactions is given in Tab.~\ref{tab:operatorbasis}.\footnote{In order to ensure 
the $U(1)_\text{em}$ gauge invariance of the NRMSSM, all derivatives
$\boldsymbol{\partial}$ in dimension-8 four-fermion operators
$\mathcal O$ and $\mathcal P$ that act on chargino fields ($\eta_i$,$\zeta_i$)
have to be replaced by the corresponding covariant derivative
$\boldsymbol{D} = \boldsymbol{\partial}+i\,e\boldsymbol{A}$, where
$\boldsymbol{A}$ denotes the spatial components of the photon field $A^\mu$.}
\begin{table}[t]
\centering
\begin{tabular}{|c|c|}
\hline
 &
\\[-3ex]
   $\mathcal O^{\chi\chi \to \chi\chi}(^1P_1)$
 & $ \hspace{2ex}
    \xi^\dagger_{e_4} \left( -\frac{i}{2} ~ \overleftrightarrow{\boldsymbol{\partial}} \right) \xi^c_{e_3}
    \ \cdot \ \xi^{c \dagger}_{e_2} \left( -\frac{i}{2} ~ \overleftrightarrow{\boldsymbol{\partial}} \right) \xi^{}_{e_1}$
 \\[1ex]
   $\mathcal O^{\chi\chi \to \chi\chi}(^3P_0)$
 & $ \frac{1}{3} ~
    \xi^\dagger_{e_4} \left( -\frac{i}{2} ~ \overleftrightarrow{\boldsymbol{\partial}}\cdot\boldsymbol{\sigma} \right) \xi^c_{e_3}
    \ \cdot \ \xi^{c \dagger}_{e_2} \left( -\frac{i}{2} ~ \overleftrightarrow{\boldsymbol{\partial}}\cdot\boldsymbol{\sigma} \right) \xi^{}_{e_1}$
 \\[1ex]
   $\mathcal O^{\chi\chi \to \chi\chi}(^3P_1)$
 & $ \frac{1}{2} ~
    \xi^\dagger_{e_4} \left( -\frac{i}{2} ~ \overleftrightarrow{\boldsymbol{\partial}} \times \boldsymbol{\sigma} \right) \xi^c_{e_3}
    \ \cdot \ \xi^{c \dagger}_{e_2} \left( -\frac{i}{2} ~ \overleftrightarrow{\boldsymbol{\partial}} \times \boldsymbol{\sigma} \right) \xi^{}_{e_1}$
 \\[1ex]
   $\mathcal O^{\chi\chi \to \chi\chi}(^3P_2)$
 & $ \hspace{2ex} \xi^\dagger_{e_4} \left( -\frac{i}{2} ~ \overleftrightarrow{\boldsymbol{\partial}} ^{(i} \boldsymbol{\sigma}^{j)} \right) \xi^c_{e_3}
    \ \cdot \ \xi^{c \dagger}_{e_2} \left( -\frac{i}{2} ~ \overleftrightarrow{\boldsymbol{\partial}} ^{(i} \boldsymbol{\sigma}^{j)} \right) \xi^{}_{e_1}$
 \\[1ex]
\hline
 &
\\[-2.5ex]
   $\mathcal P^{\chi\chi \to \chi\chi}(^1S_0)$
 & $ \frac{1}{2} \hspace{0.5ex} \left[
    \xi^\dagger_{e_4}  \xi^c_{e_3}
    \ \cdot \ \xi^{c \dagger}_{e_2} \left( -\frac{i}{2} ~ \overleftrightarrow{\boldsymbol{\partial}} \right)^2 \xi^{}_{e_1}
    +
    \xi^\dagger_{e_4} \left( -\frac{i}{2} ~ \overleftrightarrow{\boldsymbol{\partial}} \right)^2 \xi^c_{e_3}
    \ \cdot \ \xi^{c \dagger}_{e_2} \xi^{}_{e_1}
   \right]$
 \\[1.5ex]
   $\mathcal P^{\chi\chi \to \chi\chi}(^3S_1)$
 & $ \frac{1}{2} ~ \left[
    \xi^\dagger_{e_4} \boldsymbol{\sigma} \, \xi^c_{e_3}
    \ \cdot \ \xi^{c \dagger}_{e_2} \, \boldsymbol{\sigma} \left( -\frac{i}{2} ~ \overleftrightarrow{\boldsymbol{\partial}} \right)^2 \xi^{}_{e_1}
    +
    \xi^\dagger_{e_4} \, \boldsymbol{\sigma} \left( -\frac{i}{2} ~ \overleftrightarrow{\boldsymbol{\partial}} \right)^2 \xi^c_{e_3}
    \ \cdot \ \xi^{c \dagger}_{e_2} \, \boldsymbol{\sigma} \, \xi^{}_{e_1}
   \right]$
 \\[2ex]
\hline
 &
\\[-2.5ex]
      $\mathcal Q_1^{\chi\chi \to \chi\chi}(^1S_0)$
 & $ \hspace{2ex}
    (\delta m\,M) \, \xi^\dagger_{e_4} \xi^c_{e_3}
    \ \cdot \ \xi^{c \dagger}_{e_2} \xi^{}_{e_1}$
\\[1ex]
   $\mathcal Q_1^{\chi\chi \to \chi\chi}(^3S_1)$
 & $ \hspace{2ex}
    (\delta m\,M) \, \xi^\dagger_{e_4} \boldsymbol{\sigma} \, \xi^c_{e_3}
    \ \cdot \ \xi^{c \dagger}_{e_2} \boldsymbol{\sigma} \, \xi^{}_{e_1}$
\\[1ex]
     $\mathcal Q_2^{\chi\chi \to \chi\chi}(^1S_0)$
 & $ \hspace{2ex}
    (\delta \overline{m}\,M) \, \xi^\dagger_{e_4} \xi^c_{e_3}
    \ \cdot \ \xi^{c \dagger}_{e_2} \xi^{}_{e_1}$
\\[1ex]
   $\mathcal Q_2^{\chi\chi \to \chi\chi}(^3S_1)$
 & $ \hspace{2ex}
    (\delta \overline{m}\,M) \, \xi^\dagger_{e_4} \boldsymbol{\sigma} \, \xi^c_{e_3}
    \ \cdot \ \xi^{c \dagger}_{e_2} \boldsymbol{\sigma} \, \xi^{}_{e_1}$
 \\[1ex]
\hline
\end{tabular}
\caption{Explicit form of the $P$-wave ($\mathcal O$) and
         next-to-next-to-leading order $S$-wave ($\mathcal P$, $\mathcal Q_i$)
         four-fermion operators contributing to
         $\chi^0_{e_1}\chi^0_{e_2} \to \chi^0_{e_4}\chi^0_{e_3}$ scattering
         reactions.
         Each index $e_i$ can take the values $e_i = 1,\ldots,n_0$.
         The $P$- and next-to-next-to-leading order $S$-wave four-fermion
         operators for the remaining neutral, charged and double-charged
         $\chi_{e_1}\chi_{e_2} \to \chi_{e_4}\chi_{e_3}$ processes
         are obtained by replacing the field operators
         $\xi_{e_i}, \, i = 1,\ldots,4$ above by those of the respective
         particle species involved.
         The quantity $\boldsymbol{\partial}$ is a 3-vector whose components are $\partial^i\equiv\partial/\partial x_i$.
         The action of $\protect\overleftrightarrow{\boldsymbol{\partial}}$ on
         the two field operators at its left and right is defined as
$\,\xi^{c\dagger}_{e_b}\protect\overleftrightarrow{\boldsymbol{\partial}}\xi^{}_{e_a}
\equiv
\xi^{c\dagger}_{e_b} \,(\boldsymbol{\partial}\xi^{}_{e_a}) -
(\boldsymbol{\partial}\xi^c_{e_b})^{\dagger}\, \xi^{}_{e_a}$.
         The symmetric traceless components of a tensor $T^{ij}$ are denoted by
         $T^{(ij)} = (T^{ij}+T^{ji})/2 - T^{kk}\delta^{ij}/3$.
         Finally, the mass scale $M$ is defined in (\protect\ref{eq:M}) and the
         mass differences
         $\delta m, \delta\overline m$ are given in (\ref{eq:deltam}).
}
\label{tab:operatorbasis}
\end{table}
The labels $e_i$ in (\ref{eq:basisSPwave}) range over $e_i = 1,\ldots,n_0$
($e_i = 1,\ldots,n_+$), if the respective field $\chi_{e_i}$ in the
$\chi_{e_1} \chi_{e_2} \to \chi_{e_4} \chi_{e_3}$ reaction refers to neutralino-
(chargino-)\linebreak species.
The factor $1/4$ in front of the operators in (\ref{eq:basisSPwave}) is
a convenient normalisation of transitions matrix elements in the effective
theory. In addition, a normalisation factor of $1/M^2$ has been factored out in
(\ref{eq:basisSPwave}), such that the next-to-next-to-leading order Wilson
coefficients have the same mass-dimension ($-2$) as the leading-order ones
presented in I.
The mass scale $M$ is equal to half the sum of the masses of the $\chi_{e_i}$
particles involved in the reaction
$\chi_{e_1} \chi_{e_2} \to \chi_{e_4} \chi_{e_3}$, {\it i.e.}
\begin{align}
\label{eq:M}
 M \ = \ \frac{1}{2} \, \sum_{i=1}^4 \, m_{e_i} \, ,
\end{align}
such that $M$ itself constitutes a process specific quantity.
The quantum numbers ${}^{2s+1}L_J$ of the operators in
$\delta \mathcal L^{d=8}_\text{ann}$
correspond to the angular-momentum configuration of the annihilating
two-particle state. Note that the operators
$\mathcal Q_i\left( ^{2s+1}S_s \right)$ have the same structure as
the dimension-6 operators $\mathcal O\left( ^{2s+1}S_s \right)$ defined in paper 
I, but are proportional to the mass differences
\begin{align}
 \delta m \ =& \ \frac{m_{e_4} - m_{e_1}}{2} \, , \hspace{7ex}
 \delta \overline m \ = \ \frac{m_{e_3} - m_{e_2}}{2} \, ,
\label{eq:deltam}
\end{align}
computed from the masses  $m_{e_i}$ in the reaction
$\chi_{e_1}\chi_{e_2} \to \chi_{e_4}\chi_{e_3}$. The mass differences
(\ref{eq:deltam}) have to be considered as ${\cal O}(v_\text{rel}^2)$ effects in the
expansion of the amplitudes according to the discussion given in Sec.~2.4 in I.
Since $\delta m=\delta \overline m=0$ for 
diagonal annihilation reactions $\chi_{e_1}\chi_{e_2} \to \chi_{e_1}\chi_{e_2}$
(where the absorptive parts of the respective amplitudes are related to
the corresponding annihilation cross section),
the $\mathcal Q_i\left( ^{2s+1}S_s \right)$ are only relevant for the computation
of the off-diagonal rates.

We note that dimension-8 operators $\mathcal P(^3S_1,{}^3D_1)$, which describe
${}^3S_1 \to {}^3D_1$ transitions, have not been included in
$\delta \mathcal L^{d=8}_\text{ann}$. In the calculation of the
tree-level annihilation cross section in the centre-of-mass frame,
contributions from these operators vanish,
while for the Sommerfeld enhanced annihilation cross section they will require
to consider a $v_\text{rel}^2$-suppressed  potential interaction in the long-range part of 
the annihilation in order to compensate for the change in orbital angular
momentum in the short-distance part, thus yielding a contribution to 
the cross section of ${\cal O}(v_\text{rel}^4)$.

As already discussed in I, we construct $\delta \mathcal L_\text{ann}$ in such a way
that it contains all redundant operators, which arise through interchanging
the single-particle field-operators at the first and second (third and fourth)
position given a specific four-fermion operator, such that several operators
describe one specific scattering reaction with a $\chi_{e_1}$ and $\chi_{e_2}$
($\chi_{e_4}$ and $\chi_{e_3}$) particle in the initial (final) state.
The respective Wilson coefficients reflect the redundancy in symmetry relations
under the exchange of the respective particle labels. Generalising from the
leading-order $S$-wave relations given in Eq.~(8)\footnote{The equation numbers
from paper I, Ref.~\cite{Beneke:2012tg}, always refer to the arXiv version.}
in I, the relations read in
case of Wilson coefficients associated with operators $\mathcal O$ and
$\mathcal P$ in (\ref{eq:basisSPwave})
\begin{align}
\nonumber
   k_{ \lbrace e_2 e_1\rbrace \lbrace e_4 e_3\rbrace }^{ \chi_{e_2} \chi_{e_1} \to \chi_{e_4} \chi_{e_3} }
                                              \left( {}^{2s+1}L_J \right)
	=
 (-1)^{s+L} \  k_{ \lbrace e_1  e_2\rbrace \lbrace e_4 e_3\rbrace }^{ \chi_{e_1} \chi_{e_2} \to \chi_{e_4} \chi_{e_3} }
                                               \left( {}^{2s+1}L_J \right)
\ ,
\\
   k_{ \lbrace e_1 e_2\rbrace \lbrace e_3 e_4\rbrace }^{ \chi_{e_1} \chi_{e_2} \to \chi_{e_3} \chi_{e_4} }
                                              \left( {}^{2s+1}L_J \right)
	=
 (-1)^{s+L} \  k_{ \lbrace e_1  e_2\rbrace \lbrace e_4 e_3\rbrace }^{ \chi_{e_1} \chi_{e_2} \to \chi_{e_4} \chi_{e_3} }
                                               \left( {}^{2s+1}L_J \right)
\ ,
\label{eq:genericWilsonCoeffSymmetry}
\end{align}
where $k = f, g$ for $P$- and next-to-next-to-leading order $S$-wave
coefficients, respectively.
Finally note, that the hermiticity property of the non-relativistic Lagrangian
leads to the relation
\begin{align}
   k_{ \lbrace e_1 e_2\rbrace \lbrace e_4 e_3\rbrace }^{ \chi \chi  \to \chi \chi } \left( {}^{2s+1}L_J \right)
	=
  \  \left[ k_{ \lbrace e_4  e_3\rbrace \lbrace e_1 e_2\rbrace }^{ \chi \chi \to \chi \chi } \left( {}^{2s+1}L_J \right) \right]^*
\label{eq:hermiticitySP}
\end{align}
for Wilson-coefficients $k = f, g$ associated with the $P$- and
next-to-next-to-leading order $S$-wave operators $\mathcal O$ and $\mathcal P$,
in analogy to the respective relation given in Eq.~(13) in I.
Similar relations as (\ref{eq:genericWilsonCoeffSymmetry},
\ref{eq:hermiticitySP}) above apply for the Wilson coefficients $h_i$, where
however an additional exchange of the particles in the definition of the mass
differences $\delta m,\delta\overline m$ in front of the corresponding operators
$\mathcal Q_i$ has to be taken into account.

\section{(Off-)diagonal annihilation rates $\Gamma^{\chi_{e_1}\chi_{e_2} \to X_A
    X_B \to \chi_{e_4} \chi_{e_3}}$}
\label{sec:gammarate}
In order to assess the importance of the non-relativistic corrections computed
in this work, we shall compare in Sec.~\ref{sec:results} the EFT and full
results for the tree-level annihilation rates for some selected processes. 
To extend this analysis to the case of off-diagonal annihilation rates, we
generalise the definition of the rates in the following way.
We  define the centre-of-mass frame tree-level annihilation rate $\Gamma$
associated with the (off-) diagonal
$\chi_{e_1}\chi_{e_2} \to X_A X_B \to \chi_{e_4} \chi_{e_3}$ scattering reaction as
the product of the $\chi_{e_1} \chi_{e_2} \to X_A X_B$ tree-level annihilation
amplitude with the complex conjugate of the tree-level amplitude for the
$\chi_{e_4} \chi_{e_3} \to X_A X_B$ annihilation reaction, integrated over the
final $X_A X_B$ particles' phase space\footnote{\label{fn:symmetryfactor}
The product of tree-level annihilation amplitudes has to be multiplied with an
additional symmetry factor of $1/2$ if the final state particles are
identical, $X_A = X_B$.}
and averaged over the spin states of the respective incoming particles
$\chi_{e_i},\,i = 1,\ldots,4$.
In the latter spin-average it is assumed that the $\chi_{e_1}\chi_{e_2}$ and
$\chi_{e_4}\chi_{e_3}$ pair reside in the same spin state.\footnote{In the
calculation of Sommerfeld enhanced $\chi_i \chi_j \to X_A X_B$ pair-annihilation
rates through the imaginary part of the
$\chi_i \chi_j \to \ldots \to \chi_{e_1} \chi_{e_2} \to X_A X_B \to
\chi_{e_4} \chi_{e_3} \to \ldots \to \chi_i \chi_j$ forward scattering reaction,
the assumption that the incoming and outgoing particle pairs in the
$\chi_{e_1}\chi_{e_2} \to X_A X_B \to \chi_{e_4}\chi_{e_3}$ short-distance
annihilation part have the same spin state implies that just leading-order
potential interactions in the 
$\chi_i \chi_j \to \ldots \to \chi_{e_1} \chi_{e_2}$ and
$\chi_i \chi_j \to \ldots \to \chi_{e_4} \chi_{e_3}$ scattering reactions are
considered, since the long-range potentials are spin-diagonal only at
leading order and hence pass the spin-configuration of the incoming
$\chi_i \chi_j$ pair to the $\chi_{e_1} \chi_{e_2}$ and $\chi_{e_4} \chi_{e_3}$
pairs.}
The external $\chi_{e_a}\chi_{e_b}$ states are further taken to be
non-relativistic normalised in order to match with the definition of the
annihilation cross section times relative velocity in case of diagonal
reactions $\chi_{e_1}\chi_{e_2}\to X_A X_B \to \chi_{e_1}\chi_{e_2}$.
In terms of the  Wilson coefficients of the four-fermion operators in
$\delta\mathcal L_\text{ann}$, the expansion of the annihilation
rate $\Gamma$ in the non-relativistic momenta and in the
mass differences $\delta m$, $\delta \overline m$ is then given by
\begin{eqnarray}
\label{eq:res_Gammarate}
\Gamma^{\chi_{e_1} \chi_{e_2} \to X_A X_B \to \chi_{e_4} \chi_{e_3}}
& =& 
 \hat f(^1S_0) + 3~\hat f(^3S_1)
\\
&&
\nonumber
\hspace*{-2.5cm}
      +\, \frac{\delta m}{M} ~\left( \hat h_1(^1S_0) + 3~\hat h_1(^3S_1) \right)
      +\, \frac{\delta\overline m}{M} ~\left( \hat h_2(^1S_0) + 3~\hat h_2(^3S_1)
                                     \right)
\\
&&
\nonumber
\hspace*{-2.5cm} +\, \frac{~\boldsymbol p \cdot \boldsymbol{p}^{\,\prime}}{M^2} ~ \Bigl(
        \hat f(^1P_1) + \frac{1}{3}~\hat f(^3P_0) + \hat f(^3P_1)
      + \frac{5}{3}~\hat f(^3P_2) \Bigr)
\\
&&
\nonumber
\hspace*{-2.5cm} + \, \frac{~\boldsymbol p^{\,2} + \,\boldsymbol p^{\,\prime\,2}}{2\,M^2}~
           \Bigl( \hat g(^1S_0) + 3~\hat g(^3S_1) \Bigr)
\\
&&
\nonumber
\hspace*{-2.5cm} + \ \mathcal O\left(\,
                       (\boldsymbol p^{~2} + \boldsymbol p^{~\prime\,2})^2,~
                       (\boldsymbol{p}\cdot\boldsymbol{p}^{~\prime})^2,~
                       \boldsymbol p^{\,(\prime)2} \delta m,~
                       \boldsymbol p^{\,(\prime)2} \delta\overline m,~
                       \delta m \, \delta\overline m \,\right) \ , \quad
\end{eqnarray}
where $\boldsymbol{p}$ and $\boldsymbol{p}^{\,\prime}$ correspond to the momenta of the $\chi_{e_1}$ and
$\chi_{e_4}$ particle, respectively, in the centre-of-mass frame of the reaction.
To shorten the notation we have suppressed in (\ref{eq:res_Gammarate}) the label
``$\chi_{e_1} \chi_{e_2} \to X_A X_B \to \chi_{e_4} \chi_{e_3}$'' on the Wilson
coefficients $\hat f, \hat g$ and $\hat h_i$.
As we study annihilation rates of non-relativistic $\chi_{e_a}\chi_{e_b}$ particle
pairs, the mass differences $\delta m$ and $\delta\overline m$ have to be
(at most) of the order of the $\chi_{e_a} \chi_{e_b}$  non-relativistic
kinetic energy, as argued in 
Sec.~2.4 of paper I.
Note that the non-relativistic expansion (\ref{eq:res_Gammarate}) in\-corporates
this convention and assumes that $\delta m,\, \delta \overline m \sim {\cal O}(\boldsymbol{p}^2/M)$.
In case of diagonal $\chi_{e_1} \chi_{e_2} \to X_A X_B \to \chi_{e_1} \chi_{e_2}$
scattering reactions, the definition of the corresponding 
annihilation rate $\Gamma$ obviously coincides with the definition of the
spin-averaged centre-of-mass frame tree-level $\chi_{e_1}\chi_{e_2} \to X_A X_B$
annihilation cross section times relative velocity,
$\sigma^{\chi_{e_1}\chi_{e_2}\to X_A X_B}\,v_\text{rel}$, and the 
expansion in (\ref{eq:res_Gammarate}), with $\boldsymbol{p}^{\,\prime}=\boldsymbol{p}$,
reduces to the non-relativistic expansion 
of $\sigma^{\chi_{e_1}\chi_{e_2}\to X_A X_B}\,v_\text{rel}$ as given in Eq.~(27) of paper I
(see also Eqs.~(\ref{eq:res_sigmavrel}--\ref{eq:res_bcoeff2}) below).

\section{Results}
\label{sec:results}
In paper I we have presented several examples for the numeric comparison of the 
non-relativistic approximation to the tree-level centre-of-mass frame
annihilation cross-sec\-tion $ \sigma^{\chi_{e_1} \chi_{e_2} \to X_A X_B}$ times
relative velocity $ v_\text{rel}=\vert \vec v_{e_1} - \vec v_{e_2} \vert$,
\begin{align}
 \sigma^{\chi_{e_1} \chi_{e_2} \to X_A X_B}~ v_\text{rel}
  \ = \
 a ~ + ~ b~v_\text{rel}^2 \ + \ \mathcal O(v_\text{rel}^4) \ ,
\label{eq:res_sigmavrel}
\end{align}
with the corresponding unexpanded result obtained with {\sc{MadGraph}}~\cite{Alwall:2011uj}.
The coefficient $a$ in
the expansion (\ref{eq:res_sigmavrel}) is expressed in terms of the leading
order $S$-wave Wilson coefficients as
\begin{align}
\label{eq:res_acoeff}
 a \ =& \ \hat f(^1S_0) + 3~\hat f(^3S_1) \ ,
\end{align}
and the coefficient $b$ can be written as the sum $b = b_P + b_S$,
where
\begin{align}
\label{eq:res_bcoeff1}
 b_P \ =& \ 
 \frac{\mu_{e_1 e_2}^2}{ M^2} 
      \left(  \hat f(^1P_1) + \frac{1}{3}~\hat f(^3P_0)
             +\hat f(^3P_1) + \frac{5}{3}~\hat f(^3P_2) \right) \ ,
\\
 b_S \ =& \
 \frac{\mu_{e_1 e_2}^2}{ M^2}~ 
      \Bigl( ~\hat g(^1S_0) + 3~\hat g(^3S_1) \Bigr) \ ,
\label{eq:res_bcoeff2}
\end{align}
and
\begin{align}
\label{eq:redmass}
 \mu_{e_1 e_2} \ =& \ 
 \frac{m_{e_1} m_{e_2}}{ m_{e_1} + m_{e_2}}  
\end{align}
is the reduced mass of the $\chi_{e_1}\chi_{e_2}$ two-particle state.
We have again suppressed in (\ref{eq:res_acoeff}--\ref{eq:res_bcoeff2}) the labels on the Wilson coefficients
$\hat  f$ and $\hat g$  that indicate the specific $\chi_{e_1}\chi_{e_2} \to X_A X_B \to \chi_{e_1} \chi_{e_2}$
reaction under consideration to simplify the notation. The prefactor $(\mu_{e_1 e_2}/M)^2$ in front of the Wilson coefficients
in (\ref{eq:res_bcoeff1}--\ref{eq:res_bcoeff2}) is needed to translate
the cross section's expansion in
$\boldsymbol{p}^{2}$, Eq.~(\ref{eq:res_Gammarate}) with $\boldsymbol{p}^{\,\prime}=\boldsymbol{p}$,
to the $v_\text{rel}^2$ expansion used in
(\ref{eq:res_sigmavrel}).

From the comparison of the non-relativistic approximation to 
$\sigma^{\chi_{e_1}\chi_{e_2} \to X_A X_B}~v_\text{rel}$  with the full result
from {\sc{MadGraph}}, it was shown in paper I that the non-relativistic
approximation reproduces the behaviour of the exact tree-level cross section times
relative velocity within a percent level
deviation up to $v_\text{rel}/c\sim 0.6$.\footnote{In case of processes with
vanishing $S$-wave contributions the agreement between {\sc{MadGraph}}
and the non-relativistic approximation
is a bit worse, but still with an accuracy at the level of $\sim 6\%$ for
$v_\text{rel}/c\sim 0.4$.}
The numeric extraction of the coefficients $a$ and $b$ from
{\sc{MadGraph}} data by means of a parabola fit to
$\sigma^{\chi_{e_1}\chi_{e_2} \to X_A X_B}~v_\text{rel}$ in the non-relativistic regime
thus provides a useful numeric check for the sum 
of leading-order $S$-wave Wilson coefficients (\ref{eq:res_acoeff}), 
as well as for the sum of next-to-next-to-leading order $S$-wave and
$P$-wave  Wilson coefficients in (\ref{eq:res_bcoeff1},\ref{eq:res_bcoeff2}).
However, a splitting of the numerically extracted coefficients $a$ and $b$ into 
their constituting partial-wave contributions is not straightforward from
publicly available numeric codes, as this requires manipulations at the
amplitude level. The separate knowledge of the different $^{2s+1}L_J$ partial
wave contributions to the tree-level (co-)annihilation rates is essential
for a precise determination of Sommerfeld enhanced neutralino
(co-)annihilation cross sections, because
the Sommerfeld enhancements depend both on the spin- and orbital angular
momentum quantum numbers of the annihilating particle pair. Therefore a
consistent treatment of the Sommerfeld enhancement including $P$-wave effects
requires the separate knowledge of all relevant (off-)diagonal tree-level
$^1S_0$ and $^3S_1$ partial-wave annihilation rates both at leading and
next-to-next-to-leading order, as well as the individual (off-)diagonal
tree-level $^1P_1$ and $^3P_J$ partial-wave annihilation rates.
In the latter case, the knowledge of the (spin-weighted) sum over the three
different $^3P_0$, $^3P_1$ and $^3P_2$ partial-wave Wilson coefficients,
\begin{align}
 \hat f(^3P_{\cal J})
\ = \
 \frac{1}{3} \hat f(^3P_0) +  \hat f(^3P_1) + \frac{5}{3} \hat f(^3P_2) \ ,
\end{align}
is sufficient, as long as only leading-order non-relativistic potential interactions between
the neutralino and chargino states are taken
into account in the full annihilation amplitudes. 
This is because the leading-order potential interactions depend
on the spin ($s=0,1$)
of the
$\chi_{e_a}\chi_{e_b}$ particle pairs taking part in the
 $\chi_i\chi_j \to \ldots \to \chi_{e_1}\chi_{e_2} \to X_A X_B \to  \chi_{e_4} \chi_{e_3} \to \ldots \to \chi_i \chi_j$
scattering process, but do not discriminate among the three spin-1
$P$-wave states $^3P_J$ with different total angular momentum $J = 0,1,2$.

Recently, Sommerfeld corrections including $P$-wave effects
have been subject of study at 1-loop \cite{Drees:2013er} and with full resummation
\cite{Chen:2013bi}.
In these studies, the next-to-next-to-leading order contributions in the
expansion of the relevant (co-)annihilation rates  were assumed to be 
given only by $P$-waves.
While such reasoning is justified when the leading-order $S$-wave
contributions to the annihilation rates are strongly suppressed with respect to 
the next-to-next-to-leading order coefficients
in (\ref{eq:res_Gammarate}, \ref{eq:res_sigmavrel}), it does
not hold for the general case. In particular, $P$- and next-to-next-to-leading
order $S$-wave terms can come with differing signs, such that a
partial compensation of different next-to-next-to-leading order
contributions to the annihilation rates may occur.

\begin{figure}[t]
\includegraphics[width=0.49\textwidth]{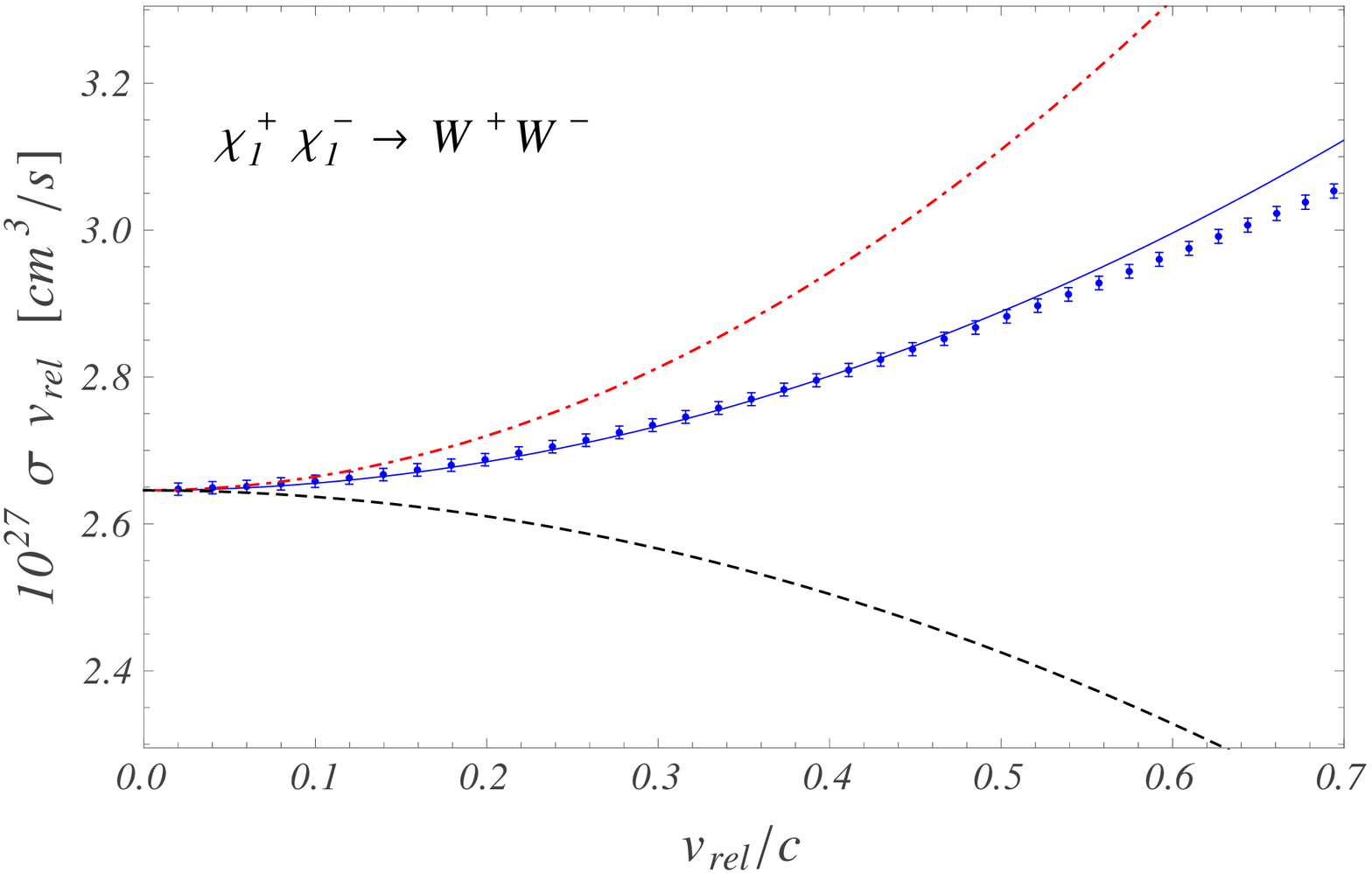} \ \ 
\includegraphics[width=0.49\textwidth]{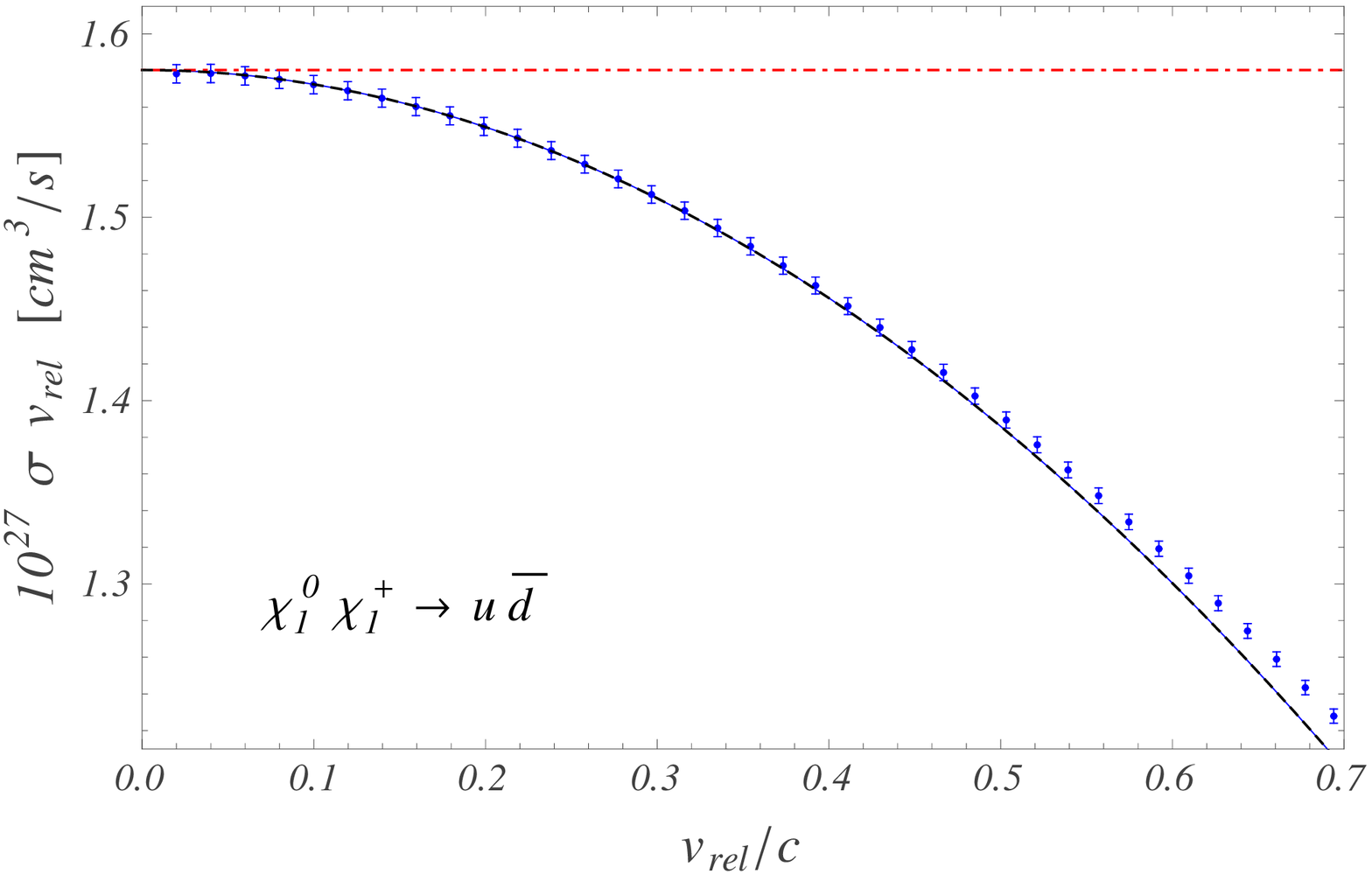}
\caption{Numeric comparison of the non-relativistic approximation 
(solid lines) to the tree-level annihilation cross section times relative
velocity, $\sigma\, v_{\rm rel}$, for the $\chi^+_1 \chi^-_1 \to W^+ W^-$ process
(left) and for the $S$-wave dominated reaction
$\chi^0_1 \chi^+_1 \to u \overline d$ (right) with the corresponding unexpanded
annihilation cross sections produced with {\sc MadGraph}.
Numeric errors on the {\sc MadGraph} data are given by
$\sigma\, v_{\rm rel}/\sqrt{N}$, where $N = 10^5$ gives the number of
events used in the {\sc MadGraph} calculation of each cross section value.
The dash-dotted red (dashed black) curves represent the constant leading-order 
term in the non-relativistic expansion of the cross section plus the
$P$-wave (next-to-next-to-leading order $S$-wave) contribution,
$a+b_{P}\,v_\text{rel}^2$ ($a + b_S\,v_\text{rel}^2$).}
\label{fig:xsectionsMG_1}
\end{figure}

In order to illustrate the different behaviour of the $P$- and
next-to-next-to-leading order $S$-wave contributions to the tree-level
annihilation cross sections, we show  in Figs.~\ref{fig:xsectionsMG_1}
and \ref{fig:xsectionsMG_2} results for the tree-level
annihilation cross section times relative velocity,
$\sigma^{\chi_{e_1}\chi_{e_2}\to X_A X_B}~v_\text{rel}$,
for three different processes. The plots refer to the same  SUSY
spectrum that was used in paper I, which contains a wino-like
neutralino LSP with mass $m_{\chi^0_1} = 2748.92\,$GeV, and an
almost mass-degenerate wino-like chargino partner with
$m_{\chi^+_1} = 2749.13$\,GeV. The next-to-lightest chargino state has 
a mass $m_{\chi^+_2} = 3073.31\,$GeV.

\subsection{Example 1: $\chi^+_1 \chi^-_1 \to W^+ W^-$}
\label{subsec:x1+x1-w+w-}

The plot on the left hand side in Fig.~\ref{fig:xsectionsMG_1} shows the
$\chi^+_1 \chi^-_1 \to W^+ W^-$ tree-level annihilation rate, a relevant
co-annihilation rate in the neutralino LSP relic density computation.
The solid blue line corresponds to the
non-relativistic approximation to the tree-level annihilation cross section,
$\sigma^{\chi^+_1 \chi^-_1 \to W^+ W^-}\,v_\text{rel}$,
and the points correspond to the full tree-level result obtained with
{\sc{MadGraph}}. The deviation
between our approximation and the {\sc{MadGraph}} data is at one percent level
for $v_\text{rel}/c \sim 0.6$ and in the permille regime for smaller relative
velocities. Further, the composition of the non-relativistic approximation
to $\sigma^{\chi^+_1 \chi^-_1 \to W^+ W^-}~v_\text{rel}$ out of $P$- and
next-to-next-to-leading order $S$-wave contributions can be read off 
from Fig.~\ref{fig:xsectionsMG_1}: the dash-dotted red line represents the
contribution $a + b_P~v_\text{rel}^2$ to (\ref{eq:res_sigmavrel}), while the
dashed black line is $a + b_S~v_\text{rel}^2$.
While both $b_P$ and $b_S$ are roughly of the same order of magnitude, the
summed $P$-wave contributions enter with a positive sign
($b_P\,c^2 = 1.86\cdot10^{-27}\,$cm$^3$\,s$^{-1}$), whereas the summed next-to-next-to-leading
order $S$-wave contributions come with a negative weight,
$b_S\,c^2 = -0.88\cdot10^{-27}\,$cm$^3$\,s$^{-1}$.
It is worth noting that the sum of next-to-next-to-leading order 
corrections in the $\chi^+_1 \chi^-_1 \to W^+ W^-$ tree-level cross section times
relative velocity gives a $\sim 6\%$ correction to the leading-order
approximation for $v_\text{rel}/c \sim 0.4$.
For this relative velocity, the corrections to the leading-order approximation
from $P$-waves only amount to $\sim 11\%$, while those from
next-to-next-to-leading order $S$-wave contributions amount to $\sim -5\%$.
Hence, in the light of the expected future experimental precision on the
measured dark matter density, it is crucial to take these corrections into 
account. Further, as generically the Sommerfeld enhancements for each of the
contributing partial waves are different, it will be needed to
investigate the Sommerfeld enhanced annihilation cross section including
$P$- and next-to-next-to-leading order $S$-wave enhancements separately.
This study within our formalism is postponed to \cite{paperIII}.

The fact that the $P$-wave terms in the example of Fig.~\ref{fig:xsectionsMG_1}
contribute with positive sign is generic: the sum of all $^{2s+1}P_J$
partial-wave contributions to any $\chi_{e_1}\chi_{e_2} \to X_A X_B$
annihilation cross section has to be positive, as it
results from the  absolute square  of the coefficient of the
${\cal O}(\boldsymbol p)$ terms in the expansion of the annihilation amplitude. 
Moreover, the separate $^{2s+1}P_J$ partial-wave contributions must also be
positive, since different $^{2s+1}P_J$-wave amplitudes do not interfere in the
absolute square of the annihilation amplitude due to total angular-momentum
conservation and the additional conservation of spin in the non-relativistic
regime.
The next-to-next-to-leading order $S$-wave contributions to the
$\chi_{e_1} \chi_{e_2} \to X_A X_B$ annihilation cross section, however, result
from the product of leading-order and next-to-next-to-leading order $S$-wave
contributions in the expansion of the $\chi_{e_1}\chi_{e_2} \to X_A X_B$
amplitude. There is a priori no reason why this product should be positive, and
hence negative next-to-next-to-leading order $S$-wave contributions to the
cross section can occur, as can be explicitly seen in the examples presented
in this section.

\subsection{Example 2: $\chi^0_1 \chi^+_1 \to u \overline d$ }
\label{subsec:n1x1+ubard}

The right  plot in Fig.~\ref{fig:xsectionsMG_1} shows results for the $S$-wave
dominated tree-level $\chi^0_1 \chi^+_1 \to u \overline d$ annihilation process,
also of importance in the neutralino relic abundance
calculation including co-annihilations.
The dashed black line, representing the $a + b_S~v_\text{rel}^2$ contribution
to the non-relativistic expansion of the annihilation rate
with $b_S\,c^2 = -0.78\cdot10^{-27}\,$cm$^3$\,s$^{-1}$, basically coincides
with the solid blue line, which corresponds to the complete non-relativistic
approximation (\ref{eq:res_sigmavrel}). Data produced with {\sc{MadGraph}}
for the $\chi^0_1 \chi^+_1 \to u \overline d$ tree-level annihilation rate
are shown in addition, illustrating once again the nice agreement of the
non-relativistic approximation with the unexpanded tree-level cross section
results for relative velocities up to $v_\text{rel}/c \sim 0.6$.
It is worthwhile to understand the suppression of $P$-waves with respect to the
next-to-next-to-leading order $S$-wave contributions in the
$\chi^0_1\chi^+_1\to u \overline d$ process as well as the composition of the
coefficient $b_S$ out of its $^1S_0$ and $^3S_1$ partial-wave contributions:
First note, that in the case of vanishing final state masses, $m_u = m_d = 0$,
the contributions to both $a$ and $b_S$ can be attributed solely to $^3S_1$
partial waves.
The absence (or more generally the suppression in $m_q/M$, $q=u,d$) of $^1S_0$
partial-wave contributions both in the leading-order coefficient $a$ and in
$b_S$ is a helicity suppression effect.
The helicity suppression argument applies to all $^{2s+1}L_J$ partial-wave
reactions with $J = 0$, as the final state of a massless (left-handed) quark and
a massless (right-handed)
anti-quark in its centre-of-mass system cannot build a total angular-momentum
state $J = 0$. Hence both $^1S_0$ as well as $^3P_0$ partial-wave contributions
are helicity suppressed.

The suppression of $^1P_1$, ${}^3P_1$ and ${}^3P_2$ partial-wave
contributions that proceed through single $s$-channel gauge-boson or
Higgs exchange is related to either factors of
$\Delta_m = (m_{\chi^0_1} -  m_{\chi^+_1})/(m_{\chi^0_1} + m_{\chi^+_1})$ 
or to vertex couplings that vanish in the exact $SU(2)_L$ symmetric
limit.
Similarly, contributions from $t$-channel exchange amplitudes introduce $\Delta_m$
factors or coupling factor combinations that lead to vanishing contributions in
the $SU(2)_L$ symmetric theory (case of $^1P_1$ waves), or are additionally suppressed 
(as it is the case of $^3P_1$ and $^3P_2$ partial-wave configurations) by
the masses of $t$-channel exchanged sfermions, since the mass scale of the
latter is above  $5\,$TeV  in the  MSSM scenario considered.
Consequently, as the initial two particle state in the reaction
$\chi^0_1\chi^+_1 \to u \overline d$ consists of two wino-like particles
with $|\Delta_m| \sim 4 \cdot 10^{-5}$, the $^1P_1$, $^3P_1$ and $^3P_2$ partial
waves give suppressed contributions to the tree-level annihilation rate.

\begin{figure}[t]
\includegraphics[width=0.490\textwidth]{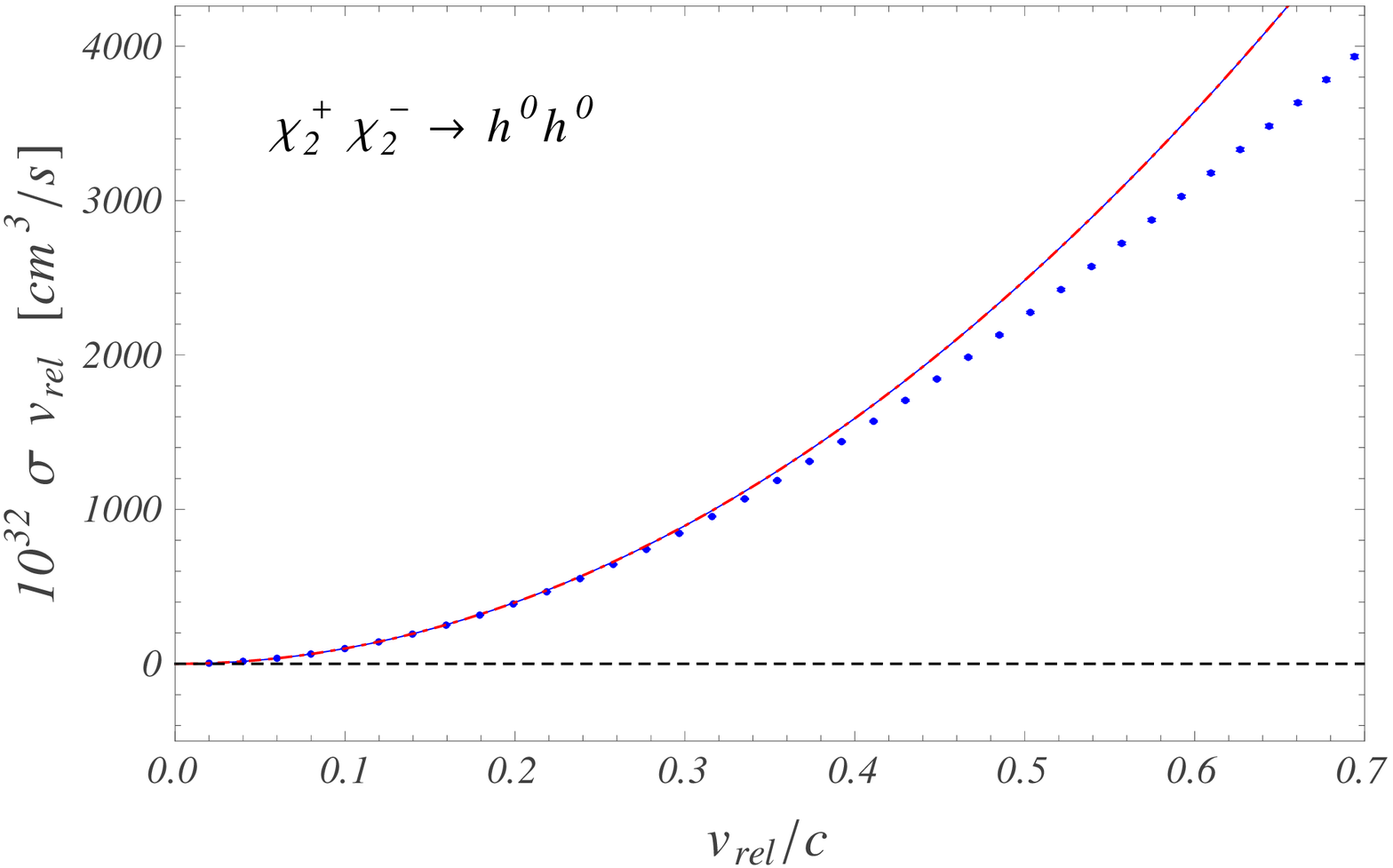} \ \ 
\includegraphics[width=0.490\textwidth]{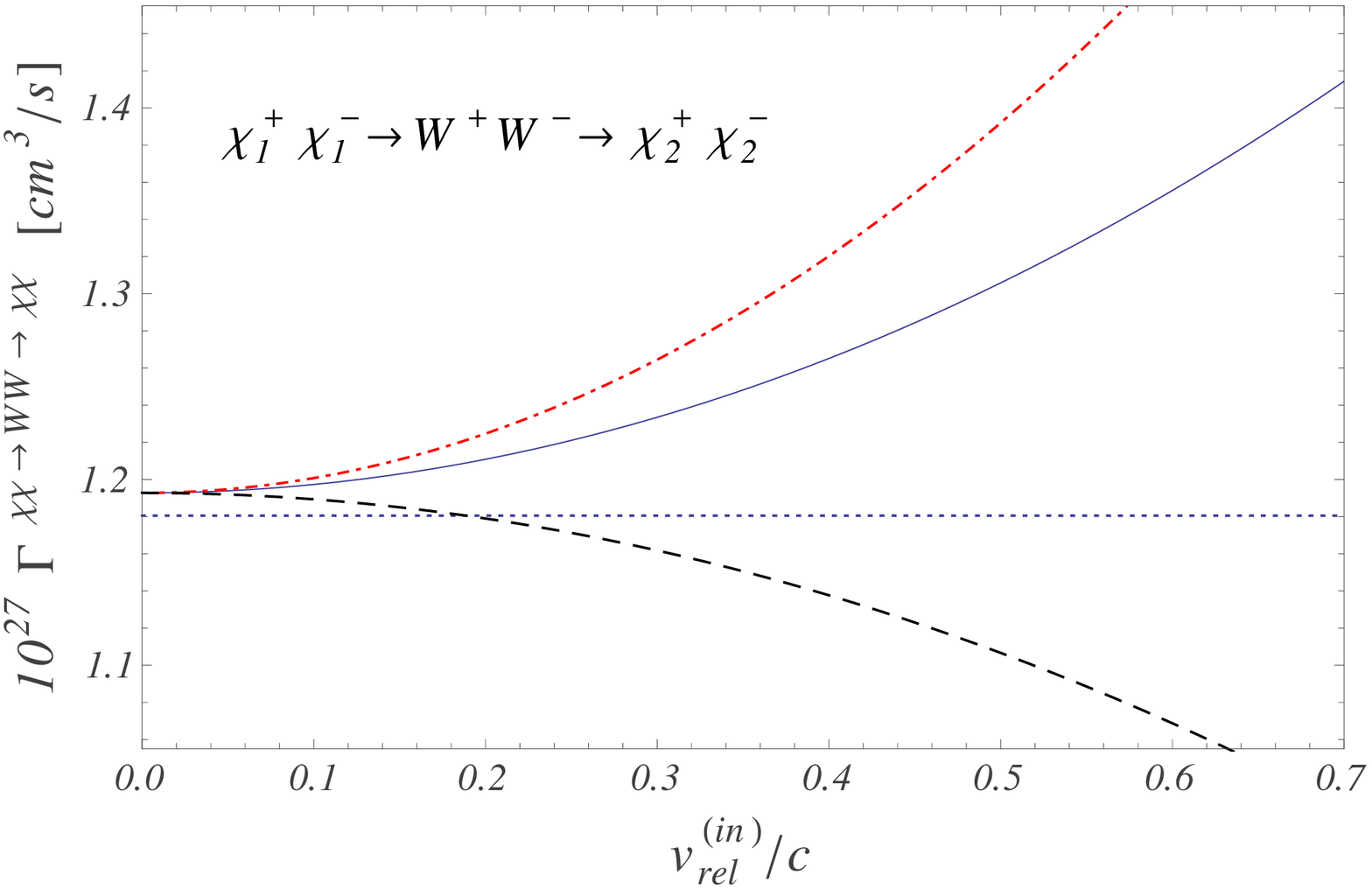}
\caption{Left plot: Numeric comparison of the non-relativistic approximation
(solid blue curve) to the tree-level annihilation cross section times relative
velocity, $\sigma\, v_{\rm rel}$, for the $P$-wave dominated
$\chi^+_2 \chi^-_2 \to h^0 h^0$ reaction to data for the corresponding unexpanded
annihilation cross section produced with {\sc MadGraph}.
Numeric errors on the {\sc MadGraph} data are taken to
be $\sigma\, v_{\rm rel}/\sqrt{N}$, where $N = 10^5$ gives the number of
events used in the {\sc MadGraph} calculation of each cross section value.
The dash-dotted red and dashed black lines represent the constant leading-order
term plus the
$P$-wave or the next-to-next-to-leading order $S$-wave contribution,
$a + b_{P}\,v_\text{rel}^2$ or $a + b_S\,v_\text{rel}^2$, respectively.
Note that the $a + b_{P}\,v_\text{rel}^2$ contribution and the non-relativistic
approximation coincide, as there are no $S$-wave contributions in this
particular annihilation reaction.
Right plot: Off-diagonal annihilation rate $\Gamma$ for the
reaction $\chi^+_1 \chi^-_1 \to W^+ W^- \to \chi^+_2 \chi^-_2$. 
The solid line includes all contributions to $\Gamma$ up to
next-to-next-to-leading order in the non-relativistic expansion. It is
obtained from (\ref{eq:res_Gammarate}) assuming that $\boldsymbol{p}$ and
$\boldsymbol{p}^\prime$ are parallel to each other. 
The constant dotted blue line gives the leading-order approximation to
$\Gamma$. Summing the $P$- or the (momentum-dependent) next-to-next-to-leading
order $S$-wave contributions to the constant $S$-wave terms (given by the leading order plus the 
terms proportional to $\delta m$ and  $\delta\overline m$) yields
the dash-dotted red or the dashed black line, respectively.
The curves are plotted against the relative velocity 
$v_{\rm rel}^{(\rm in)}$
of the incoming state $\chi^+_1 \chi^-_1$.
}
\label{fig:xsectionsMG_2}
\end{figure}

\subsection{Example 3: $\chi^+_2 \chi^-_2 \to h^0 h^0$}
\label{subsec:x2+x2-h0h0}

An example of a $P$-wave dominated process is provided in 
the left plot of Fig.~\ref{fig:xsectionsMG_2}. It corresponds to the
tree-level $\chi^+_2 \chi^-_2 \to h^0 h^0$ annihilation, wherein $S$-wave
contributions vanish, such that the process is purely $P$-wave mediated in the
non-relativistic regime (the coefficient $b_P\,c^2$ is given by
$9.94\cdot10^{-29}\,$cm$^3$\,s$^{-1}$).
The absence of $S$-wave contributions can be
explained by $CP$ and total angular-momentum conservation in the
$\chi^+_2 \chi^-_2 \to h^0 h^0$ reaction.\footnote{The following reasoning
applies to all possible $\chi^+_a \chi^-_a \to X_A X_B$ annihilation reactions
with two $CP$-even MSSM Higgs particles in the final state,
$X_A X_B = h^0 h^0, h^0 H^0, H^0 H^0$. Note that $CP$ is conserved in these
reactions if the mixing matrices in the chargino sector are real, which 
is the case for the scenario we consider.}
The $CP$ quantum number of the final two-particle state $h^0 h^0$ 
is given by $CP = (-1)^L=(-1)^J$, as the total angular
momentum of a $h^0h^0$ state coincides with its orbital angular momentum 
and the parity of such a state is given by $P = (-1)^L$, while its charge
conjugation is $C = 1$.
In case of the annihilating $\chi^+_a \chi^-_a$ two-particle state the $J^{PC}$
quantum numbers are $0^{-+}$ for a $^1S_0$ partial-wave configuration and $1^{--}$
for a $^3S_1$ partial-wave state. Hence, for the $\chi^+_a \chi^-_a$ state,
$CP = -1$ is realised in case of $S$-waves for the  $J = 0$ configuration, and
$CP = +1$ for  $J = 1$, which are opposite to the $CP$ quantum
numbers of a $h^0 h^0$ final state with the same total angular momentum.
The same reasoning explains the absence of $^3P_1$ annihilations
in any of the processes $\chi^+_a\chi^-_a\to X_A X_B$ with
$X_A X_B = h^0 h^0, h^0 H^0, H^0 H^0$, as the $J^{PC}$ quantum numbers of the
$^3P_1$ partial-wave configuration of the incoming $\chi^+_a\chi^-_a$ states are
$1^{++}$, hence $CP = +1$ for $J = 1$.
This is opposite to the $CP$ quantum number of the two $CP$-even Higgs boson
final state with total angular momentum $J = 1$.

Let us finally note that there are also no contributions from $^1P_1$
partial waves in the process
shown in the left plot in Fig.~{\ref{fig:xsectionsMG_2}}. This feature
is generic to 
$\chi^+_a\chi^-_b\to X_A X_B$ annihilations with identical
scalar particles in the final state, $X_A X_B= h^0h^0, H^0H^0$.
The argument relies on the statistics of the final state identical bosons,
and applies to all $\chi_a^+ \chi_b^-$ incoming states and not only to
particle-anti-particle states $\chi_a^+ \chi_a^-$: Bose statistics forbids the
two identical final state scalars to be in a $J = L = 1$ state, as the
corresponding two-particle wave-function for odd total angular momentum $J$
would be anti-symmetric.
This argument can also be used to explain the absence of the $J = 1$ $^3S_1$
and $^3P_1$ states in a $\chi_a^+ \chi_b^- \to h^0h^0, H^0 H^0$ annihilation
reaction.

\subsection{Example 4: $\chi^+_1 \chi^-_1 \to W^+ W^- \to \chi^+_2 \chi^-_2$}
\label{subsec:x1+x1-w+w-x2+x2-}

Let us finally turn to the case of an off-diagonal annihilation rate. The
right plot in Fig.~\ref{fig:xsectionsMG_2} shows the off-diagonal
annihilation rate $\Gamma$ associated with the process
$\chi^+_1 \chi^-_1 \to W^+ W^- \to \chi^+_2 \chi^-_2$, which is relevant, for
instance, in the calculation of the Sommerfeld enhanced
$\chi^0_1 \chi^0_1\to W^+ W^-$ and $\chi^+_1 \chi^-_1\to W^+ W^-$
(co-)annihilation cross sections. The mass splitting between the $\chi^\pm_1$ and
$\chi^\pm_2$ charginos is given by $324.18\,$GeV in the MSSM scenario considered,
which results in rather large mass differences, namely
$\delta m = \delta\overline m = 162.09\,$GeV.
In this case, the Wilson coefficients $h_1$ and $h_2$, that are proportional to
$\delta m$ and $\delta \overline m$, lead to a $1\%$ positive correction to the
constant leading-order rate. This positive shift corresponds to the difference
between the leading-order approximation to the annihilation rate $\Gamma$
(first line in (\ref{eq:res_Gammarate}), dotted blue line in the right plot in
Fig.~\ref{fig:xsectionsMG_2}), and the complete non-relativistic result for
$\Gamma$ including next-to-next-to-leading corrections (solid blue line) at
zero momentum. The corrections induced by the terms proportional to 
$\delta m,\,\delta\overline m$ turn out to be somewhat
smaller than the naive expectation $\delta m/M = \delta\overline m/M = 2.78\%$,
but represent nevertheless the dominant next-to-next-to-leading order correction
up to $v_\text{rel}/c \sim 0.16$.
For larger relative velocities, the $P$- and next-to-next-to-leading
order $S$-wave terms provide larger contributions to the absorptive part of the
$\chi^+_1 \chi^-_1 \to W^+ W^- \to \chi^+_2 \chi^-_2$ scattering amplitude.
This is indicated by the dash-dotted red and dashed black curves, which result 
from the addition of the
constant $S$-wave contributions (first two lines in (\ref{eq:res_Gammarate}))
and 
the $P$-wave contributions (third line in (\ref{eq:res_Gammarate}))
or the momentum-dependent $S$-wave next-to-next-to-leading
terms (fourth line in (\ref{eq:res_Gammarate})), respectively.
The correction to the leading-order $\Gamma$ rate
due to the $P$- and next-to-next-to-leading order $S$-wave terms amounts to a $7\%$ for $v_\text{rel}/c = 0.4$.

Note that no comparison with public numeric codes providing results for
(tree-level) $\chi\chi \to X_A X_B$ annihilation rates is available for the
off-diagonal annihilation rates.
We emphasise that the calculation of the partial-wave decomposed
off-diagonal annihilation rates therefore constitutes one of the main results
presented in paper I and in this work.
The relevance of off-diagonal annihilation rates in the calculation of
Sommerfeld enhanced (co-)annihilation amplitudes in context of the $\chi^0_1$
relic abundance calculation was in particular pointed out in Sec.~4.2 of I, and
will be further investigated in subsequent work \cite{paperIII}.

\section{Summary}
\label{sec:summary}
With this work we finish the presentation of results associated to the short-distance
annihilation rates, that are prerequisites for
a refined study of Sommerfeld enhancements in neutralino dark matter
(co-)annihilation processes in the MSSM including $P$- and
next-to-next-to-leading order $S$-wave contributions.
Our analysis can be applied to a set of nearly mass-degenerate non-relativistic
neutralino and chargino states with masses around the TeV scale and
excludes accidental mass-degeneracies with further supersymmetric or Higgs
particles.

A factorisation between the short- and long-distance contributions in the pair annihilation
of non-relativistic neutralino and chargino pairs is possible given the large separation between the 
associated scales.
Paper I~\cite{Beneke:2012tg} introduced
an effective field theory set-up (the NRMSSM),
that provides the basis for a systematic study of radiative corrections in
(co-)annihilation processes of non-relativistic neutralinos and charginos,
applicable to both neutralino DM freeze-out in the Early
Universe as well as to neutralino pair-annihilations today.
In the EFT approach, the tree-level (co-)annihilation rates 
of neutral, single and double charged neutralino/chargino pairs
into SM and Higgs two-particle final states $X_A X_B$,
related to the absorptive parts of the 1-loop scattering reactions
$\chi_{e_1}\chi_{e_2} \to X_A X_B \to \chi_{e_4} \chi_{e_3}$, are
encoded in the absorptive parts of the Wilson coefficients of
four-fermion operators.
As the first step in the construction of the  NRMSSM, paper I provided the basis
of dimension-6 four-fermion operators, which describe leading-order $S$-wave
annihilation processes. The absorptive parts of the corresponding four-fermion
operators' Wilson coefficients are provided in analytic form in the appendix of paper I.
In the present work, we extend the results from paper I and provide
the operator basis for dimension-8 four-fermion operators, contributing at
next-to-next-to-leading order in the non-relativistic expansion of
$\chi_{e_1}\chi_{e_2} \to X_A X_B \to \chi_{e_4} \chi_{e_3}$ annihilation rates,
and present analytic results for the absorptive
parts of the corresponding $^1P_1$-wave Wilson coefficients as well as the
spin-averaged sum of spin-1 $P$-wave ($^3P_{\cal J}$) Wilson coefficients in the
Appendix~\ref{sec:appendix}.
An electronic supplement~\cite{dotmfile} to this paper contains analytic results for all kinematic
factors that are needed in the construction of the absorptive part of
partial-wave separated (next-to-next-to-)leading order $S$- and $P$-wave Wilson 
coefficients, relevant for the determination of the $\mathcal O(v_\text{rel}^2)$ 
approximation to any (off-)diagonal tree-level (co-)annihilation rate.
Our results apply to neutralino and chargino states with arbitrary composition and
include the full mass dependence of the final state SM and Higgs particles. 
As a straightforward application of our work, we have provided in
Appendix~\ref{sec:appendixexample}
the previously unknown $\mathcal O(v_\text{rel}^2)$ corrections to the exclusive (off-)diagonal (co-)annihilation
rates for the same  pure-wino neutralino dark matter scenario
considered in earlier works~\cite{Hisano:2004ds,Hisano:2006nn,Cirelli:2007xd}.

While there are situations where the main part of the
$\mathcal O(v_\text{rel}^2)$ corrections to a given annihilation rate can be
attributed to a specific partial wave
(for instance when $CP$-conservation or helicity-suppression forbids or
suppresses annihilation reactions from other partial-wave states),
we have presented two examples in
Sec.~\ref{sec:results}, where the $\mathcal O(v_\text{rel}^2)$ $P$- and
next-to-next-to-leading order $S$-wave
contributions in the annihilation rates are roughly of the same order of
magnitude and enter with differing signs.
A proper treatment of Sommerfeld enhanced annihilation rates beyond leading
order $S$-wave annihilations therefore generally requires the knowledge of each 
separate partial-wave contribution, which is now available with the analytic
results given in the appendices of paper I and II as well as those collected
in the electronic supplement.
In particular a numeric extraction of the $\mathcal O(v_\text{rel}^2)$
contributions in the non-relativistic expansion of the annihilation rates
without a separation of the different constituting $P$- and
next-to-next-to-leading order $S$-waves will, in general, not be sufficient in 
a rigorous analysis of the Sommerfeld effect beyond leading-order $S$-wave
enhancements.

In addition, it is important to stress that
our work allows for a consistent treatment of off-diagonal annihilation rates,
required for the accurate description of Sommerfeld enhanced annihilation
reactions, as the potential exchange of electroweak gauge-bosons and light
Higgses prior to the actual annihilation can change the incoming neutralino or
chargino two-particle state to another nearly on-shell two-particle state.
This implies that the annihilation process itself is generally described by a
non-diagonal hermitian matrix in the space of neutralino and chargino
two-particle states.
Apart from the usual expansion in non-relativistic momenta, a consistent
treatment of off-diagonal reactions within the NRMSSM requires an additional
expansion in the mass differences between initial
and final state particles in the off-diagonal annihilation rates
$\chi_{e_1} \chi_{e_2} \to X_A X_B \to \chi_{e_4} \chi_{e_3}$.
The respective contributions count as
next-to-next-to-leading order in the non-relativistic expansion.
Consequently, we account for the corresponding set of four-fermion operators in
the basis of the dimension-8 four-fermion operators given in this paper and
include the results for their Wilson coefficients in the electronic supplement.

With the above results at hand, the study of the long-range effects in the 
annihilation of non-relativistic neutralino and chargino pairs
as well as their impact on the neutralino relic-abundance calculation in
selected examples
will be the subject of a forthcoming publication \cite{paperIII}.

\subsubsection*{Acknowledgements}
We would like to thank M.~Beneke for very useful discussions and suggestions regarding
the preparation of this paper.
This work is supported in part by the Gottfried Wilhelm Leibniz programme of the
Deutsche Forschungsgemeinschaft (DFG) and the DFG
Sonder\-for\-schungs\-bereich/Trans\-regio~9 ``Computergest\"utzte
Theoreti\-sche Teilchenphysik''.
C.H. would like to thank the ``Deutsche Telekom Stiftung'' for its support
while the main part of this work was done.
The work of P.~R. is partially supported by MEC (Spain) under grants FPA2007-60323 and FPA2011-23778 and by the
Spanish Consolider-Ingenio 2010 Programme CPAN (CSD2007-00042).

\appendix

\section{Absorptive parts of   
Wilson coefficients of 
dim\-ension-8 operators in $\delta \mathcal L_\text{ann}$}
\label{sec:appendix}
We provide in this appendix analytic expressions for the kinematic factors
related to the  absorptive part of the $P$-wave Wilson coefficients
in $\delta\mathcal L^{d=8}_\text{ann}$, Eq.~(\ref{eq:basisSPwave}).
The corresponding expressions for  the kinematic factors of the
next-to-next-to-leading order $S$-wave Wilson coefficients,
$\hat g(^{2s+1}S_s)$ and $\hat h_i(^{2s+1}S_s)$ with $s=0,1$ are quite lengthy.
Therefore we have collected the latter in an electronic
supplement~\cite{dotmfile} attached to this paper, that also contains the
kinematic factors associated with the absorptive part of the $P$-wave Wilson
coefficients ($\hat f(^1P_1)$, $\hat f(^3P_J)$, $J = 0,1,2$) and those
corresponding to the leading-order $S$-wave Wilson coefficients
($\hat f(^1S_0)$, $\hat f(^3S_1)$ in $\delta \mathcal L^{d=6}_\text{ann}$), which
were written in the appendix of paper I.
Details on the nomenclature used in the electronic supplement can be
found in Appendix \ref{sec:appendixsuppl}.

We aim at the description of Sommerfeld enhanced annihilation rates, and will
in a forthcoming publication \cite{paperIII} consider the potentials which
are responsible for the long-range Sommerfeld corrections at leading order. 
Despite that the leading-order
potential interactions cannot change the spin of the incoming two-particle state, they depend
on the spin ($s=0,1$) of the latter.
Consequently, as far as our study of Sommerfeld enhancements is
concerned, 
the separate knowledge of the different $\hat f(^3P_J)$
coefficients, which share the same orbital angular-momentum and spin but
different total angular momentum ($J=0,1,2$), is not needed.
It suffices to consider the combination of spin-1 $P$-wave 
Wilson coefficients $\hat f(^3P_{\cal J})$ entering the short-distance part, Eqs.~(\ref{eq:res_Gammarate}) and (\ref{eq:res_bcoeff1}),
\begin{align}
 \hat f(^3P_{\cal J}) \ = \
 \frac{1}{3}~\hat f(^3P_0) +  \hat f(^3P_1) + \frac{5}{3}~\hat f(^3P_2) \,
\end{align}
that will be multiplied by the $P$-wave Sommerfeld correction factor 
computed with potentials for spin-1 scattering states.
Hence we give in this appendix analytic expressions for the kinematic factors
corresponding to $\hat f(^1P_1)$ Wilson coefficients as well as the kinematic
factors associated with the combination $\hat f(^3P_{\cal J})$.
For completeness, the kinematic factors for the separate
$\hat f(^3P_J),\, J = 0,1,2,$ Wilson coefficients 
can be found in the electronic attachment~\cite{dotmfile},
together with those of the combination $\hat f(^3P_{\cal J})$.

\subsection{Master formula to build the Wilson coefficients}

The results for the Wilson coefficients $\hat{f}$, 
$\hat{g}$ and $\hat{h}_i$  $(i = 1,2)$
at $\mathcal O(\alpha^2_2)$ are obtained
through  matching of the EFT tree-level matrix element of four-fermion operators
in $\delta \mathcal L_\text{ann}$ with the absorptive part of the MSSM 1-loop
$\chi_{e_1}\chi_{e_2} \to X_A X_B \to \chi_{e_4}\chi_{e_3}$
scattering amplitude with states $\chi_{e_1}\chi_{e_2}$ and $\chi_{e_4}\chi_{e_3}$ 
in a ${}^{2s+1}L_J$ partial-wave configuration. 
In case of $s=0$, the total angular momentum $J$ of the $^{2s+1}L_J$ state
takes the value $J=L$, while for $s=1$,
$J=|L-1|,\dots,L+1$.
As tree-level annihilation processes
are free from infrared
divergences, the individual contributions to the Wilson coefficients
from exclusive final states $X_A X_B$ at $\mathcal O(\alpha_2^2)$ can be given separately.
Our results cover separately all possible exclusive SM and light Higgs
two-particle final states $X_A X_B$ in neutral ($\chi^0\chi^0$, $\chi^-\chi^+$), 
single-charged ($\chi^0\chi^+$, $\chi^0\chi^-$) and double-charged
($\chi^+\chi^+$, $\chi^-\chi^-$) chargino and neutralino pair-annihilation
reactions, where the $X_A X_B$ states are conveniently classified to be of
vector-vector $(V V)$, vector-scalar $(V S)$, scalar-scalar $(SS)$,
fermion-fermion $(ff)$ or ghost-anti-ghost $(\eta\bar{\eta})$ type, see Tab.~3 
in paper I.

In paper I we have provided a master formula to obtain the
absorptive part $\hat f(^{2s+1}L_J)$ of a given Wilson coefficient 
from its constituent parts, the kinematic and coupling factors. 
For the sake of clarity, we write the formula here as well and briefly comment
on its structure. It reads
\begin{eqnarray}
\nonumber
&&
\hspace*{-2cm} 
  \hat{f}^{\, \chi_{e_1}\chi_{e_2} \to X_A  X_B\to \chi_{e_4}\chi_{e_3}}_{\lbrace e_1 e_2 \rbrace \lbrace e_4 e_3 \rbrace}(^{2s+1}L_J)
\\[0.2cm]\nonumber
 \ = \ 
 \frac{\pi \alpha_2^2}{M^2} ~
 \Biggl(
& &
 \sum\limits_{n} \sum\limits_{i_1, i_2}
       b^{\, \chi_{e_1}\chi_{e_2} \to X_A X_B\to \chi_{e_4}\chi_{e_3} }_{n, \, i_1 i_2} \
       B^{\, X_A X_B }_{n, \, i_1 i_2} (^{2s + 1}L_J)
\\\nonumber
& &+ 
 \sum\limits_{\alpha = 1}^{4} \sum\limits_{n} \sum\limits_{i_1, i_2}
       c^{(\alpha) \, \chi_{e_1}\chi_{e_2} \to X_A X_B\to \chi_{e_4}\chi_{e_3} }_{n, \, i_1 i_2} \
       C^{(\alpha) \,  X_A X_B }_{n, \, i_1 i_2} (^{2s + 1}L_J)
\\
& &+
 \sum\limits_{\alpha = 1}^{4} \sum\limits_{n} \sum\limits_{i_1, i_2}
       d^{(\alpha) \, \chi_{e_1}\chi_{e_2} \to X_A X_B\to \chi_{e_4}\chi_{e_3} }_{n, \, i_1 i_2} \
       D^{(\alpha) \, X_A X_B }_{n, \, i_1 i_2} (^{2s + 1}L_J)
 \Biggr)
 \ .
\label{eq:genericstructureWilson}
\end{eqnarray}
Formula (\ref{eq:genericstructureWilson})  also applies to the next-to-next-to-leading order $S$-wave Wilson coefficients denoted with
$g$ and $h_i$ $(i = 1,2)$ in (\ref{eq:basisSPwave}), 
with $\hat f$ being replaced by
$\hat g$ or $\hat h_i$.
However in the  discussion that follows, we will generically
refer to the absorptive part of any four-fermion operator's Wilson coefficient
as $\hat f$.
An exclusive final state contribution is indicated with the label
$\chi_{e_1} \chi_{e_2} \to X_A X_B \to \chi_{e_4} \chi_{e_3}$ on $\hat f$ in
(\ref{eq:genericstructureWilson}),
while the actual absorptive part of the Wilson coefficient $\hat f(^{2s+1}L_J)$
is given by the inclusive annihilation rate, summed over all accessible final
states.
Note that we have used $\alpha_2 = g_2^2/4\pi$, where $g_2$ denotes the $SU(2)_L$
gauge coupling. 

The first line on the right-hand side of (\ref{eq:genericstructureWilson})
collects all contributions from
$\chi_{e_1}\chi_{e_2} \to X_A X_B \to \chi_{e_4}\chi_{e_3}$ MSSM selfenergy
amplitudes, while the second and third lines give the triangle and box
amplitudes' contributions, respectively.
Quantities $ B^{}_{n, \, i_1 i_2}$, $C^{(\alpha)}_{n, \, i_1 i_2}$ and
$D^{(\alpha)}_{n, \, i_1 i_2}$ in (\ref{eq:genericstructureWilson}) denote the
kinematic factors, which encode the $^{2s+1}L_J$ partial-wave specific
information on the process.\footnote{In order to distinguish the kinematic factors associated with 
the leading and
$\mathcal O(v_\text{rel}^2)$ $S$-wave
Wilson coefficients $\hat f$ and $\hat g, \hat h_i$, we write the
partial-wave state label $^{2s+1}L_J$ in the latter as $^{2s+1}L_J = {}^1S_0, {}^3S_1$ for kinematic factors related to
$\hat f$, and $^{2s+1}L_J = {}^1S_0^{(p^2)}, {}^3S_1^{(p^2)}$
as well as $^{2s+1}L_J = {}^1S_0^{(\delta m)}, {}^3S_1^{(\delta m)}, {}^1S_0^{(\delta\overline m)}, {}^3S_1^{(\delta\overline m)}$
if the kinematic factors are related to $\hat g$ or $\hat h_i$,
respectively.}
They are obtained from a generic
$\chi_{e_1}\chi_{e_2} \to X_A X_B \to \chi_{e_4}\chi_{e_3}$ 1-loop scattering
reaction with generic external Majorana fermions and generic final state
particles $X_A X_B$ of the type vector-vector ($VV$), vector-scalar ($VS$), 
scalar-scalar ($SS$), fermion-fermion ($ff$) or ghost-anti-ghost ($\eta\overline\eta$),
respectively, and can be applied to any $\chi\chi\to X_A X_B \to \chi\chi$
annihilation reaction with external Majorana or Dirac fermions by appropriate
construction of the corresponding process-specific coupling factors, denoted
with lowercase letters ($b^{}_{n, \, i_1 i_2}$, $c^{(\alpha)}_{n, \, i_1 i_2}$,
$d^{(\alpha)}_{n, \, i_1 i_2} $) in (\ref{eq:genericstructureWilson}).
\begin{figure}[t!]
\begin{center}
\includegraphics[width=0.3\textwidth]{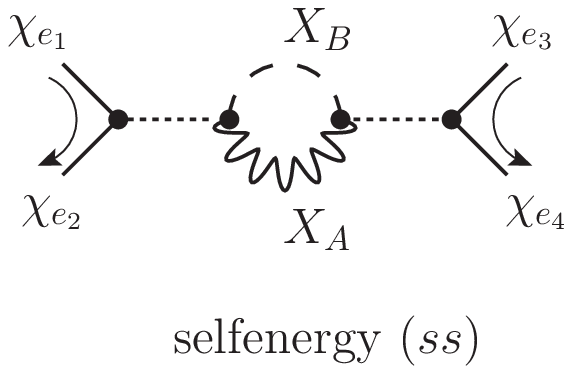}\\
 \vspace*{0.2cm}
\includegraphics[width=0.9\textwidth]{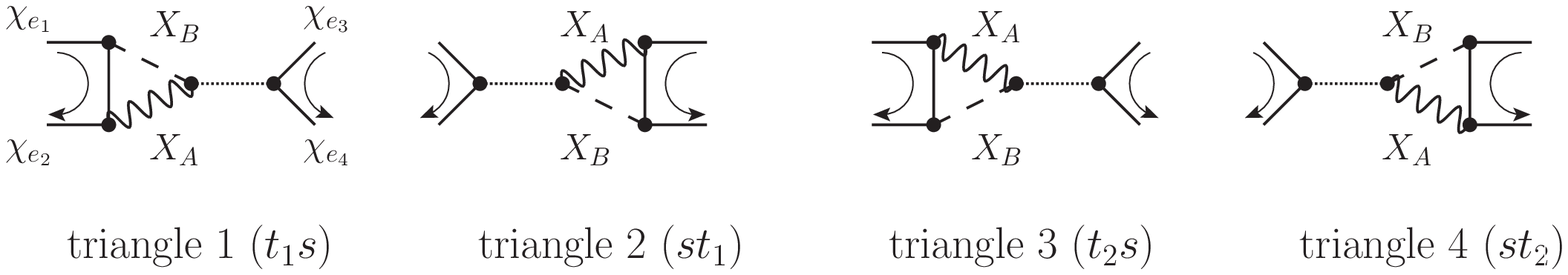}\\
 \vspace*{0.4cm}
\includegraphics[width=0.85\textwidth]{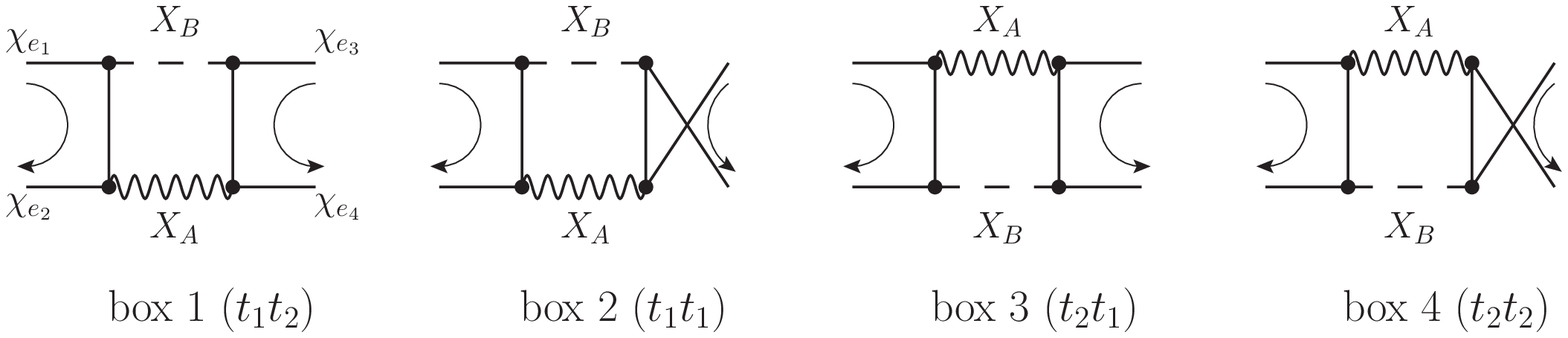}\\
 \vspace*{0.4cm}
\includegraphics[width=0.85\textwidth]{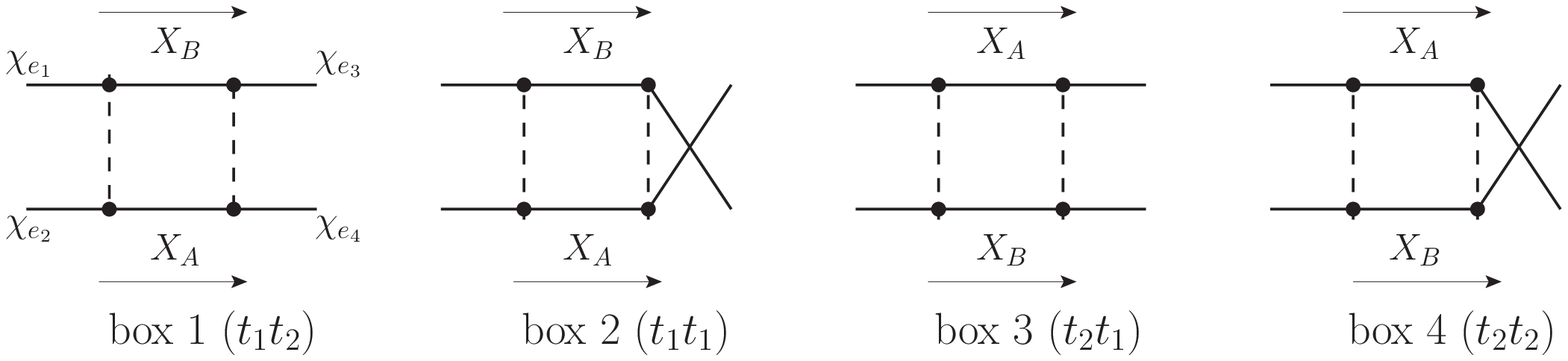}
\caption{ Generic selfenergy-, triangle- and box-diagrams in $\chi\chi \to X_A X_B \to \chi\chi$
          reactions, with $X_A$ and $X_B$ representing any two-body 
          final state of SM and Higgs particles.
          The box-amplitudes in the third line refer to $X_A X_B = VV, VS, SS$
          while the box-amplitudes in the last line apply to $X_A X_B = ff$.
          The shorthand $a \tilde a$ notation, with
          $a,\,\tilde a=s,\,t_1,\,t_2$, indicates the tree-level diagrams
          $a$ and $\tilde a$ in the $\chi_{e_1}\chi_{e_2}\to X_A X_B$ and
          $\chi_{e_4}\chi_{e_3}\to X_A X_B$ processes, respectively, to which the
          coupling factors in a specific reaction are related
          (see Figs.~9,~10 in paper I for details on the latter).}
\label{fig:genericamplitudes}
\end{center}
\end{figure}
%
The index $\alpha$ enumerates the
expressions related to the four different triangle and box amplitudes, as 
shown in Fig.~\ref{fig:genericamplitudes}.
Depending on the type of the particles $X_A$ and $ X_B$ as well as on the
topology, there is a fixed number of coupling-factor expressions that result
from all possible combinations of (axial-)vector or (pseudo-)scalar vertex
factors at the four vertices of a given MSSM
$\chi_{e_1}\chi_{e_2}\to X_A X_B\to\chi_{e_4}\chi_{e_3}$ 1-loop amplitude. These
combinations are labelled with the index $n$ in
(\ref{eq:genericstructureWilson}).
Finally, in each of the processes there is a certain set of
particle species that can be exchanged in the $s$- or the $t$-channels of
the contributing amplitudes. These are labelled with the indices $i_1$ 
and $i_2$.
The general recipe on how to derive the coupling factors is given in 
Appendix~A.2 of paper I. 
Note that the coupling
factors do not depend on the kinematics and hence are the same for Wilson
coefficients with different quantum numbers $^{2s+1}L_J$.
With the purpose to illustrate how to obtain the annihilation rates
from our results, we build in Appendix~\ref{sec:appendixexample} the Wilson coefficients
needed to describe up to  
next-to-next-to-leading non-relativistic corrections
for the case of pure-wino neutralino dark matter.

\subsection{Kinematic factors}
\label{sec:app_kinematicfactors}
Let us first collect from paper I the relevant notation that enters the formulae
for the kinematic factors. The kinematic factors for a given 
$\chi_{e_1} \chi_{e_2} \to X_A X_B \to \chi_{e_4} \chi_{e_3}$ 
scattering reaction depend on the external particles' 
masses, which are rewritten in terms of two (in principle) different
reference mass scales $m$, $\overline m$ and two mass differences $\delta m$, $\delta\overline m$  as
\begin{align}
\nonumber
 m_{e_1} \ =& \ m -\delta m \, , \hspace{7ex}
 m_{e_2} \ = \ \overline m -\delta \overline m \, ,
\\
 m_{e_4} \ =& \ m +\delta m \, , \hspace{7ex}
 m_{e_3} \ = \ \overline m +\delta \overline m \, ,
\label{eq:mei}
\end{align}
such that
\begin{align}
\nonumber
 m \ =& \ \frac{m_{e_1} + m_{e_4}}{2} \ ,
&
 \overline m \ = \ \frac{m_{e_2} + m_{e_3}}{2} \ ,
\\
 \delta m \ =& \ \frac{m_{e_4 }- m_{e_1}}{2} \ ,
&
 \delta \overline m \ = \ \frac{m_{e_3} - m_{e_2}}{2} \ .
\label{eq:mmbardef}
\end{align}
The mass differences $\delta m$, 
$\delta \overline m$ vanish for the diagonal reactions, $\chi_{e_1} \chi_{e_2} \to \chi_{e_1} \chi_{e_2}$,
while they have to be considered as ${\cal O}(v_\text{rel}^2)$ corrections
in the non-relativistic expansion for the off-diagonal
amplitudes.
The convention established by (\ref{eq:mei}) implies that particles
1 and 4 have masses closer to the reference mass scale $m$, while particles 2
and 3 share the reference scale $\overline{m}$. 
Introducing two distinct mass scales  for
the particle species allows our results for the absorptive part of the Wilson coefficients
to cover both the cases of a set of particles nearly mass-degenerate with the neutralino LSP ($m\sim  \overline{m}$)
and that of a set of non-relativistic hydrogen-like neutralino and chargino systems ($m\gg \overline{m}$ or $m\ll \overline{m}$).
If in a process $\chi_i\chi_j \to \chi_l \chi_k$, the mass $m_k$($m_l$) is
actually closer to the mass $m_i$($m_j$) and the mass scales $m$
and $\overline{m}$ differ beyond $\mathcal O(v_\text{rel}^2)$,
the results for the kinematic
factors presented below and collected in the electronic attachment cannot
directly be used to determine the corresponding $\hat f$ expressions, as the
mass differences $\delta m,\delta\overline m$ related to this reaction are not
necessarily small.
However the
absorptive part of the Wilson coefficients for the reaction with particles 3 and 4
exchanged,
$\chi_i\chi_j \to \chi_k \chi_l$, can be obtained from the kinematic factors presented in this work
if for that case the particle masses obey 
(\ref{eq:mei}) with mass differences $\delta m,\delta\overline m$ of $\mathcal O(v_\text{rel}^2)$.
The symmetry relations given in (\ref{eq:genericWilsonCoeffSymmetry}) then
allow to relate the obtained result for the $\hat f$ in
$\chi_i\chi_j \to \chi_k \chi_l$ rates to the $\hat f$ for the
$\chi_i\chi_j \to \chi_l \chi_k$ reactions.

We use the hat notation $\widehat m_a$ to denote the mass $m_a$ rescaled by the mass scale
$M=m + \overline m$, {\it i.e.}
\begin{align}
 \rescm m_a \ = \ \frac{m_a}{M} \ ,
\end{align}
and define the dimensionless quantities
\begin{align}
\nonumber
 \Delta_m \ =& \ ~\widehat m - \widehat{\overline m} \ ,
\\
\nonumber
 \Delta_{AB} \ =& \ ~\widehat m_A^2 - \widehat m_B^2 \ ,
\\
 \beta \ =& \
  ~\sqrt{ 1 - 2 ~ (\widehat m_A^2 + \widehat m_B^2) + \Delta_{AB}^2 } \ \ ,
\label{eq:parameters}
\end{align}
where  $m_A$ and $m_B$ are the masses of the particles $X_A$ and $X_B$, and
for $X_A=X_B$ $\beta$ corresponds to the relative velocity of the $X_A$ and $X_B$
particle at leading order in the expansion in the non-relativistic
3-momenta and the mass
differences of the $\chi_{e_i}$.\footnote{For a set of nearly mass-degenerate
particles $\chi_{e_i}$, $\Delta_m$ will be of the order of
the mass differences $\delta m$ and $\delta\overline m$, and thus yield
terms beyond ${\cal O}(v_\text{rel}^2)$.
The exact dependence on $\Delta_m$ is however kept in our results, which in
particular allows us to cover the case of annihilation reactions in
hydrogen-like $\chi\chi$ systems as well.}
Performing the same expansion for 
the single $s$-channel (gauge or Higgs boson $X_i$) exchange
propagators, we obtain the following denominator-structure at leading order:
\begin{align}
 P^s_i \ = \ 1 - \widehat m_i^2 \ .
\end{align}
The corresponding leading-order expansion of $t$- and $u$-channel
chargino, neutralino and sfermion propagators gives
\begin{align}
\nonumber
 P_{i\,AB} \ =& \
         \rescm{m} ~ \rescm{\overline m} + \rescm m_i^2
       - \rescm m ~ \rescm m_A^2
       - \rescm{\overline m} ~ \rescm m_B^2 \ ,
\\
 P_{i\,BA} \ =& \
         P_{i\,AB} \ \vert_{ A \leftrightarrow B} \ .
\label{eq:Pti}
\end{align}
The index $i$ in (\ref{eq:Pti}) refers to the $t$-channel exchanged particle 
species and the labels $A$ and $B$ are related to the final state particles
$X_A$ and $X_B$ in the actual $\chi \chi \to X_A X_B$ annihilation reaction.

It is convenient to write the kinematic factors  for the 
Wilson coefficients of dimension-8 operators by pulling out factors of the
leading-order propagator and $(\widehat{m} ~ \widehat{\overline{m}})$, as well
as the factor $\beta$ arising from the phase-space integration.
For instance, for the kinematic factors related to dimension-8 four-fermion
operators, that derive from the selfenergy topology we define
\begin{align}
B^{\, X_A  X_B}_{n, \, i_1 i_2}(^{2s+1}L_J) \ = \
  \frac{ \beta}
       { (\widehat{m} ~ \widehat{\overline{m}})^2 ~ P^s_{i_1} \, P^s_{i_2}} \ 
  \tilde B^{\, X_A X_B}_{n, \, i_1 i_2}(^{2s+1}L_J) \ ,
\label{eq:app_B_generic}
\end{align}
where the labels $i_1$ and $i_2$ refer to the particle species that are 
exchanged in the left and right $s$-channel propagator of the selfenergy
diagram. As generically either gauge-boson~($V$) or Higgs~($S$) $s$-channel
exchange occurs in the processes under consideration, the combination $i_1 i_2$ 
is given by $i_1 i_2 = VV, VS, SV, SS$. Note that with respect to the
definitions in the first paper, there are additional normalisation factors in
the prefactor's denominator of (\ref{eq:app_B_generic}).
Likewise, kinematic factors of dimension-8 Wilson coefficients arising from the 
triangle-topologies are rewritten as
\begin{align}
\nonumber
C^{(\alpha)\, X_A  X_B}_{n, \, i_1 X}(^{2s+1}L_J) \ =& \
  \frac{ \beta}{ \widehat{m} ~ \widehat{\overline{m}} ~ P_{i_1 AB} \, P^s_{X}} \ 
  \tilde C^{(\alpha)\, X_A X_B}_{n, \, i_1 X}(^{2s+1}L_J)  
\qquad \alpha=1,2 \quad ,
\\
C^{(\alpha)\, X_A  X_B}_{n, \, i_1 X}(^{2s+1}L_J) \ =& \
  \frac{ \beta}{ \widehat{m} ~ \widehat{\overline{m}} ~ P_{i_1 BA} \, P^s_{X}} \ 
  \tilde C^{(\alpha)\, X_A X_B}_{n, \, i_1 X}(^{2s+1}L_J) 
\qquad \alpha=3,4 \quad ,
\label{eq:app_C_generic}
\end{align}
where the index $i_1$ is now related to the $t$- or $u$-channel
exchanged particle species, and the subscript-index $X$ indicates the type
of exchanged particle ($X = V,S$) in the $s$-channel.
Finally, the kinematic factors associated with the box topologies 
are written in the same form as the correspondent expressions related to
leading dimension-6 operators:
\begin{align}
\nonumber
D^{(1)\, X_A  X_B}_{n, \, i_1 i_2}(^{2s+1}L_J) \ = \
  \frac{ \beta}{ P_{i_1 AB} \, P_{i_2 BA}} \ 
  \tilde D^{(1)\, X_A X_B}_{n, \, i_1 i_2}(^{2s+1}L_J) \ ,
 \\\nonumber
D^{(2)\, X_A X_B}_{n, \, i_1 i_2}(^{2s+1}L_J) \ = \
  \frac{ \beta}{ P_{i_1 AB} \, P_{i_2 AB}} \ 
  \tilde D^{(2)\, X_A X_B}_{n, \, i_1 i_2}(^{2s+1}L_J) \ ,
 \\\nonumber
D^{(3)\, X_A X_B}_{n, \, i_1 i_2}(^{2s+1}L_J) \ = \
  \frac{ \beta}{ P_{i_1 BA} \, P_{i_2 AB}} \ 
  \tilde D^{(3)\, X_A X_B}_{n, \, i_1 i_2}(^{2s+1}L_J) \ ,
 \\
D^{(4)\, X_A X_B}_{n, \, i_1 i_2}(^{2s+1}L_J) \ = \
  \frac{ \beta}{ P_{i_1 BA} \, P_{i_2 BA}} \ 
  \tilde D^{(4)\, X_A X_B}_{n, \, i_1 i_2}(^{2s+1}L_J) \ .
\label{eq:app_D_generic}
\end{align}
In (\ref{eq:app_D_generic}) the indices $i_1$ and $i_2$ refer to the exchanged 
particle species in the left and
right $t$- and $u$-channels of the 1-loop box amplitudes, respectively.

Finally, let us recall the conventions for the label $n$ established in paper I.
Each entry for the index $n$ in 
(\ref{eq:app_B_generic}--\ref{eq:app_D_generic}) is given by a character 
string with a length equal to the number of vertices  that involve fermions
in the underlying 1-loop amplitude.
In case of $X_A X_B = VV, VS, SS$ or $\eta\overline\eta$, the string $n$ hence
has 2,3 and 4 characters for the selfenergy, triangle and box amplitudes,
respectively. If $X_A X_B = ff$, the string $n$ has always 4 characters.
The $i$th element in a string $n$ indicates if the coupling factor at the $i$th 
vertex of the respective 1-loop amplitude is of vector/scalar ($r$) or
axialvector/pseudoscalar ($q$) type.
We enumerate the vertices of box amplitudes according to the respective attached
external particles $\chi_{e_i}, i = 1,\ldots,4$ in ascending order.
In case of selfenergy and triangle diagrams with inner vertices without an
attached external $\chi_{e_i}$ our convention to enumerate the vertices is from
top to bottom and from left to right.
Only those kinematic factors with a given label $n$ that are non-vanishing
are quoted in the following.

\subsubsection{$P$-wave kinematic factors for $X_A X_B = V V$}
The only non-vanishing kinematic factor $\tilde B^{VV}_{n,\, i_1 i_2}$
in case of $^1P_1$ partial-wave reactions is given by
\begin{align}
\label{eq:B_VV_first}
\tilde B^{VV}_{qq,\, VV }(^1P_1) \ =& \
 \frac{\diffmmbar^2}{24} \left( 8~\beta^2 - 3~\Delta_{AB}^2 - 27 \right)
 \ ,
\end{align}
while for the combined $^3P_{\cal J}$ waves the non-vanishing kinematic factors
read
\begin{align}
\tilde B^{VV}_{rr,\, VV }(^3P_{\cal J}) \ =& \ 
 -~\frac{\diffmmbar^2}{8}~ \left( \beta^2 - 6~\Delta_{AB}^2 \right) \ ,
\\
\tilde B^{VV}_{qq,\, VV }(^3P_{\cal J}) \ =& \
 \frac{1}{12} \left( 8~\beta^2 - 3~\Delta_{AB}^2 - 27 \right) \ ,
\\
\tilde B^{VV}_{rr,\, VS }(^3P_{\cal J}) \ =& \ \tilde B^{VV}_{rr,\, SV }(^3P_{\cal J}) \ = \
 -\frac{3}{4}~\widehat{m}_W~\diffmmbar~\Delta_{AB} \ ,
\\
\tilde B^{VV}_{rr,\, SS }(^3P_{\cal J}) \ =& \
   \widehat{m}_W^2 \ .
\end{align}
In the case $X_A X_B = VV$, there are relations among the $\alpha=1 (2)$ and
$\alpha=3 (4)$ kinematic factors for the triangle and box topologies which are 
fulfilled for any $^{2s+1}L_J$ configuration (in particular also for the
kinematic factors associated with the absorptive part of the
next-to-next-to-leading order $S$-wave Wilson coefficients, $\hat g(^{2s+1}S_s)$
and $\hat h_i(^{2s+1}S_s)$).
These can be found in paper I, but are repeated here for completeness:
\begin{align}
\nonumber
\tilde C^{(3)\, VV}_{n, \, i_1 V}(^{2s+1}L_J) \ =& \ 
 - \tilde C^{(1)\, VV}_{n, \, i_1 V}(^{2s+1}L_J) \
                             \vert_{A \leftrightarrow B} \ ,
\\\nonumber
\tilde C^{(4)\, VV}_{n, \, i_1 V}(^{2s+1}L_J) \ =& \ 
 - \tilde C^{(2)\, VV}_{n, \, i_1 V}(^{2s+1}L_J) \
                             \vert_{A \leftrightarrow B} \ ,
\\\nonumber
\tilde C^{(3)\, VV}_{n, \, i_1 S}(^{2s+1}L_J) \ =& \ 
 \tilde C^{(1)\, VV}_{n, \, i_1 S}(^{2s+1}L_J) \
                             \vert_{A \leftrightarrow B} \ ,
\\\nonumber
\tilde C^{(4)\, VV}_{n, \, i_1 S}(^{2s+1}L_J) \ =& \ 
 \tilde C^{(2)\, VV}_{n, \, i_1 S}(^{2s+1}L_J) \
                             \vert_{A \leftrightarrow B} \ ,
\\\nonumber
\tilde D^{(3)\, VV}_{n, \, i_1 i_2}(^{2s+1}L_J) \ =& \ \tilde D^{(1)\, VV}_{n, \, i_1 i_2}(^{2s+1}L_J) \
                             \vert_{A \leftrightarrow B} \ ,
\\
\tilde D^{(4)\, VV}_{n, \, i_1 i_2}(^{2s+1}L_J) \ =& \ \tilde D^{(2)\, VV}_{n, \, i_1 i_2}(^{2s+1}L_J) \
                             \vert_{A \leftrightarrow B} \ .
\label{eq:app_12_34_relation_VV}
\end{align}
The minus sign in the relation for the triangle coefficients
$\tilde C^{(\alpha)\, VV}_{n,\, i_1 V}$ is a consequence of interchanging the two
gauge bosons $X_A$ and $X_B$ at the internal three-gauge boson vertex.
By virtue of the relations (\ref{eq:app_12_34_relation_VV}), we only need to
give the kinematic factors for diagram-topologies $\alpha=1,2$ for both
the cases of triangle and box diagram kinematic factors. Starting with 
the expressions $\tilde C^{(\alpha) \, VV}_{n,\,i_1V}$ for $^1P_1$ partial waves we
have
\begin{align}
\nonumber
\tilde C^{(1)\,VV}_{rqq,\, i_1 V}(^1P_1) \ =& \
 \frac{3~\mihat}{4~\mmbar}~\diffmmbar
 + \frac{\beta^2~\diffmmbar}{12~P_{i_1 AB}}
        \left( \diffmmbar - 6~\mihat + 2~\Delta_{AB} \right)
 \\ & \
 + \frac{\diffmmbar}{24~\mmbar}
        \left( 6~\diffmmbar^2~\Delta_{AB}-\diffmmbar( 5~\beta^2 - 3~\Delta_{AB}^2)
              -3~\Delta_{AB} \right) \ ,
\\
\tilde C^{(2)\,VV}_{qqr,\, i_1 V}(^1P_1) \ =& \ 
\tilde C^{(1)\,VV}_{rqq,\, i_1 V}(^1P_1) \ ,
\end{align}
whereas for the combined $^3P_{\cal J}$ quantum numbers we find
\begin{align}
\nonumber
\tilde C^{(1)\,VV}_{rrr,\, i_1 V}(^3P_{\cal J}) \ =& \
 \frac{3~\mihat}{4~\mmbar}~\diffmmbar~\Delta_{AB}
 + \frac{\beta^2~\diffmmbar}{12~P_{i_1 AB}}
        \left( \diffmmbar + 2~\Delta_{AB} \right)
 \\ & \
 - \frac{\diffmmbar}{8~\mmbar}
        \left( 2~\diffmmbar^2\Delta_{AB} - \diffmmbar~(\beta^2 - 3~\Delta_{AB}^2)
            +\Delta_{AB} \right) \ ,
\\
\tilde C^{(2)\,VV}_{rrr,\, i_1 V}(^3P_{\cal J}) \ =& \ \tilde C^{(1)\,VV}_{rrr,\, i_1 V}(^3P_{\cal J}) \ ,
\\\nonumber
\tilde C^{(1)\,VV}_{rqq,\, i_1 V}(^3P_{\cal J}) \ =& \
 \frac{3~\mihat}{2~\mmbar}~\diffmmbar
 -\frac{\beta^2}{2~P_{i_1 AB}}
 \\ & \
 - \frac{1}{12~\mmbar}
        \left( 5~\beta^2 - 9 + 9~\diffmmbar^2
               - 3~\Delta_{AB}~(\diffmmbar + \Delta_{AB})\right) \ ,
\\
\tilde C^{(2)\,VV}_{qqr,\, i_1 V}(^3P_{\cal J}) \ =& \ \tilde C^{(1)\,VV}_{rqq,\, i_1 V}(^3P_{\cal J}) \ .
\end{align}
The coefficients $\tilde C^{(\alpha)\,VV}_{n,\,i_1 S}(^1P_1)$, corresponding to triangles with a Higgs
particle exchanged in the $s$-channel, vanish for all $n$.
The corresponding expressions related to $^3P_{\cal J}$ reactions read for diagram
topologies $\alpha = 1,2$
\begin{align}
\tilde C^{(\alpha)\,VV}_{rrr,\, i_1 S}(^3P_{\cal J}) \ =& \
 -\frac{\widehat{m}_W ~ \mihat}{\mmbar}
 -\frac{\beta^2~\widehat{m}_W}{6~P_{i_1 AB}}
 +\frac{\widehat{m}_W}{4~\mmbar} \left( \diffmmbar~\Delta_{AB} + 1 \right) \ .
\end{align}
All the remaining non-vanishing kinematic factors
$\tilde C^{(\alpha)\,VV}_{n,\,i_1 X}$ associated with $^1P_1$ and $^3P_{\cal J}$
scattering reactions with both $X = V,S$ are related to the above expressions by
\begin{align}
\nonumber
\tilde C^{(1)\,VV}_{qqr,\, i_1 X}(^{2s+1}P_J) \ =& \
 \tilde C^{(2)\,VV}_{rqq,\, i_1 X}(^{2s+1}P_J) \ = \
 \tilde C^{(1)\,VV}_{rrr,\, i_1 X}(^{2s+1}P_J)\vert_{m_{i_1} \to -m_{i_1}} \ ,
\\
\tilde C^{(1)\,VV}_{qrq,\, i_1 X}(^{2s+1}P_J) \ =& \
 \tilde C^{(2)\,VV}_{qrq,\, i_1 X}(^{2s+1}P_J) \ = \
 \tilde C^{(1)\,VV}_{rqq,\, i_1 X}(^{2s+1}P_J)\vert_{m_{i_1} \to -m_{i_1}} \ ,
\label{eq:CtildeXrelations}
\end{align}
where these relations hold in particular in case of separate $^3P_J, J = 0,1,2$,
partial-wave configurations and hence trivially for the combined $^3P_{\cal J}$
waves.
Finally, the terms related to box diagrams give rise to the following
non-vanishing coefficients
\begin{align}
\nonumber
\lefteqn{\tilde D^{(1)\,VV}_{rrrr,i_1 i_2}(^1P_1)} &\\\nonumber
=& \
 - \frac{\mihat\mjhat}{4~(\mmbar)^2}
 - \frac{\mihat}{4~(\mmbar)^2} \left( \diffmmbar~\Delta_{AB} - 1 \right)
 \\\nonumber & \
 - \frac{1}{48~(\mmbar)^2}
        \left( \diffmmbar^2~(2~\beta^2 - 3~\Delta_{AB}^2) + 3 \right)
 - \frac{\beta^4}{12~P_{i_1 AB}~P_{i_2 BA}}
 \\ & \
 - \frac{\beta^2}{12~\mmbar~P_{i_1 AB}}
        \left( \diffmmbar~\Delta_{AB} + 2~\mjhat - 1 \right)
 + \Bigl\lbrace A \leftrightarrow B, i_1 \leftrightarrow i_2 \Bigr\rbrace \ ,
\\[3ex]\nonumber
\lefteqn{\tilde D^{(2)\,VV}_{rrrr,i_1 i_2}(^1P_1)} &\\\nonumber
=& \
 \frac{\mihat\mjhat}{4~(\mmbar)^2}
 - \frac{\mihat}{4~(\mmbar)^2} \left( \diffmmbar~\Delta_{AB} + 1 \right)
 \\\nonumber & \
 - \frac{1}{48~(\mmbar)^2}
        \left( \diffmmbar^2~(2~\beta^2 - 3~\Delta_{AB}^2)
               - 6~\diffmmbar\Delta_{AB} - 3 \right)
 + \frac{\beta^4}{12~P_{i_1 AB}~P_{i_2 AB}}
 \\ & \
 - \frac{\beta^2}{12~\mmbar~P_{i_1 AB}}
        \left( \diffmmbar~\Delta_{AB} - 2~\mjhat + 1 \right)
 + \sumij \ ,
\\[3ex]\nonumber
\lefteqn{\tilde D^{(1)\,VV}_{rqqr,i_1 i_2}(^1P_1)} & \\\nonumber
=& \
 \frac{\mihat\mjhat}{4~(\mmbar)^2}
 - \frac{1}{48~(\mmbar)^2}
        \left( 12~\diffmmbar^4 - \diffmmbar^2~(12 - 4~\beta^2 + 3~\Delta_{AB}^2)
               +3 \right)
 \\\nonumber &
 - \frac{\beta^2}{12~\mmbar~P_{i_1 AB}}
        \left( 2~\diffmmbar^2 + \diffmmbar~(2~\mihat - \Delta_{AB}) - 1 \right)
 \\\nonumber &
 - \frac{\beta^2}{12~P_{i_1 AB} P_{i_2 BA}}
        \left( 2~\diffmmbar^2 - 2~\diffmmbar~(\mihat+\mjhat)
               + \beta^2 + 8~\mihat\mjhat \right.
 \\ & \phantom{  - \frac{\beta^2}{12~P_{i_1 AB} P_{i_2 BA}} } \hspace{2ex}
        \left. + 2~(\mihat - \mjhat)~\Delta_{AB}
               - 2~\Delta_{AB}^2 \right)
 + \sumABij \ ,
\\[3ex]\nonumber
\lefteqn{\tilde D^{(2)\,VV}_{rqqr,i_1 i_2}(^1P_1)} & \\\nonumber
=& \
 - \frac{\mihat\mjhat}{4~(\mmbar)^2}
 - \frac{\beta^2}{12~\mmbar~P_{i_1 AB}}
        \left( 2~\diffmmbar^2 - \diffmmbar~(2~\mihat - \Delta_{AB}) - 1 \right)
 \\\nonumber & \
 - \frac{1}{48~(\mmbar)^2}
        \left( 12~\diffmmbar^3~(\diffmmbar + \Delta_{AB})
               - \diffmmbar^2~(12 + 4~\beta^2 - 3~\Delta_{AB}^2) \right.
 \\\nonumber & \ \phantom{ - \frac{1}{48~(\mmbar)^2} } \hspace{2ex}
        \left. -~6~\diffmmbar~\Delta_{AB} + 3 \right)
 \\\nonumber & \
 + \frac{\beta^2}{12~P_{i_1 AB}~P_{i_2 AB}}
      \left( 2~\diffmmbar^2 - 2~\diffmmbar~(\mihat+\mjhat - 2~\Delta_{AB})\right.
 \\\nonumber & \ \phantom{ + \frac{\beta^2}{12~P_{i_1 AB}~P_{i_2 AB}} } \hspace{3ex}
      \left.  - 2~\Delta_{AB}~(\mihat + \mjhat - \Delta_{AB})
              - \beta^2 + 8~\mihat\mjhat \right)
 \\ & \
 + \sumij \ .
\end{align}
In case of combined $^3P_{\cal J}$ waves we have
\begin{align}
\nonumber
\lefteqn{\tilde D^{(1)\,VV}_{rrrr,i_1 i_2}(^3P_{\cal J})} &
\\\nonumber
=& \
 \frac{\mihat\mjhat}{2~(\mmbar)^2} \left( 1 - \diffmmbar^2 \right)
 + \frac{\mihat}{4~(\mmbar)^2}
        \left( 2~\diffmmbar^2 - \diffmmbar~\Delta_{AB} - 1 \right)
 \\\nonumber & \
 + \frac{1}{48~(\mmbar)^2}
        \left( 18~\diffmmbar^4 + 3~\diffmmbar^2~(\beta^2 - 2~\Delta_{AB}^2 - 10)
               -4~\beta^2 + 6~\Delta_{AB}^2 + 12 \right)
 \\\nonumber & \
 + \frac{\beta^2}{12~\mmbar~P_{i_1 AB}}
        \left( 5~\diffmmbar^2 + \diffmmbar~\Delta_{AB}
               + 2~(2~\mihat + \mjhat - 2) \right)
 \\\nonumber & \
 + \frac{\beta^2}{12~P_{i_1 AB}~P_{i_2 BA}}
        \left( 3~\diffmmbar^2 + 4~(\beta^2 - 3~\mihat\mjhat)
               - 3~\Delta_{AB}^2 \right)
 \\ & \
 + \sumABij \ ,
\\[3ex]\nonumber
\lefteqn{\tilde D^{(2)\,VV}_{rrrr,i_1 i_2}(^3P_{\cal J})} &
\\\nonumber
=& \
 \frac{\mihat\mjhat}{2~(\mmbar)^2} \left( 1 + \diffmmbar^2 \right)
 - \frac{\mihat}{4~(\mmbar)^2}
        \left( 2~\diffmmbar^2 + 3~\diffmmbar~\Delta_{AB} + 1 \right)
 \\\nonumber & \
 + \frac{1}{48~(\mmbar)^2}
        \left( 18~\diffmmbar^4 + 12~\diffmmbar^3~\Delta_{AB}
               -3~\diffmmbar^2~( \beta^2 - 2~\Delta_{AB}^2 + 6 )
               + 12~\diffmmbar~\Delta_{AB} \right.
 \\\nonumber & \ \phantom{  + \frac{1}{48~(\mmbar)^2} } \hspace{3ex}
        \left.  -~4~\beta^2 + 6~\Delta_{AB}^2 + 12 \right)
 \\\nonumber & \
 + \frac{\beta^2}{12~\mmbar~P_{i_1 AB}}
        \left( 3~\diffmmbar^2 - \diffmmbar~\Delta_{AB} 
               + 2~(2~\mihat + \mjhat - 2) \right)
 \\\nonumber & \
 - \frac{\beta^2}{36~P_{i_1 AB}~P_{i_2 AB}}
        \left( 9~\diffmmbar^2  + 18~\diffmmbar~\Delta_{AB}
               - 12~(\beta^2 - 3~\mihat\mjhat) + 9~\Delta_{AB}^2 \right)
 \\ & \
 + \sumij \ ,
\\[3ex]\nonumber
\lefteqn{\tilde D^{(1)\,VV}_{rqqr,i_1 i_2}(^3P_{\cal J})} &
\\\nonumber
=& \
 \frac{\mihat\mjhat}{2~(\mmbar)^2}~\diffmmbar^2
 + \frac{2~\mihat}{\mmbar}~\diffmmbar
 \\\nonumber & \
 + \frac{1}{48~(\mmbar)^2}
        \left( 6~\diffmmbar^4 + 3~\diffmmbar^2~(\beta^2-6)
               - 8~\beta^2 + 6~\Delta_{AB}^2 + 6 \right)
 \\\nonumber & \
 + \frac{\beta^2}{12~\mmbar~P_{i_1 AB}}
        \left( \diffmmbar^2 + \diffmmbar~(2~\mihat - 4~\mjhat - \Delta_{AB})
               -2 \right)
 \\\nonumber & \
 + \frac{\beta^2}{12~P_{i_1 AB}~P_{i_2 BA}}
        \left( 3~\diffmmbar^2 - 6~\diffmmbar~(\mihat + \mjhat)
               + 2~(\beta^2 + 6~\mihat\mjhat) \right.
 \\ & \ \phantom{  + \frac{\beta^2}{36~P_{i_1 AB}~P_{i_2 BA}} } \hspace{3ex}
         \left. +~6~(\mihat - \mjhat)~\Delta_{AB}
                - 3~\Delta_{AB}^2 \right)
 + \sumABij \ ,
\\[3ex]\nonumber
\lefteqn{\tilde D^{(2)\,VV}_{rqqr,i_1 i_2}(^3P_{\cal J})} &
\\\nonumber
=& \
 - \frac{\mihat\mjhat}{2~(\mmbar)^2}~\diffmmbar^2
 - \frac{2~\mihat}{\mmbar}~\diffmmbar
 \\\nonumber & \
 - \frac{1}{48~(\mmbar)^2}
        \left( 6~\diffmmbar^4 - 3~\diffmmbar^2~(\beta^2+2)
               + 12~\diffmmbar~\Delta_{AB}-8~\beta^2 + 6~\Delta_{AB}^2 + 6 \right)
 \\\nonumber & \
 - \frac{\beta^2}{12~\mmbar~P_{i_1 AB}}
        \left( 3~\diffmmbar^2 - \diffmmbar~(2~\mihat+4~\mjhat - \Delta_{AB})
               - 2 \right)
 \\\nonumber & \
 + \frac{\beta^2}{12~P_{i_1 AB}~P_{i_2 AB}}
        \left( 3~\diffmmbar^2 - 6~\diffmmbar~(\mihat+\mjhat-\Delta_{AB})
               - 2~(\beta^2 - 6~\mihat\mjhat) \right.
 \\ & \ \phantom{ + \frac{\beta^2}{12~P_{i_1 AB}~P_{i_2 BA}} } \hspace{3ex}
        \left. -~6~(\mihat+\mjhat)~\Delta_{AB} + 3~\Delta_{AB}^2 \right)
 + \sumij \ .
\end{align}
The remaining non-vanishing kinematic factors $\tilde D^{(\alpha)\,VV}_{n,\, i_1 i_2}$ 
for diagram topologies $\alpha = 1,2$ are related to the expressions given above
by
\begin{align}
\nonumber
\tilde D^{(\alpha)\,VV}_{qqqq,i_1 i_2}(^{2s+1}L_J) \ =& \ 
 \tilde D^{(\alpha)\,VV}_{rrrr,i_1 i_2}(^{2s+1}L_J)\vert_{m_{i_{1,2}} \to -m_{i_{1,2}}}\ ,
\\
\nonumber
\tilde D^{(\alpha)\,VV}_{rrqq,i_1 i_2}(^{2s+1}L_J) \ =& \
 \tilde D^{(\alpha)\,VV}_{rrrr,i_1 i_2}(^{2s+1}L_J)\vert_{m_{i_{2}} \to -m_{i_{2}}}\ ,
\\
\nonumber
\tilde D^{(\alpha)\,VV}_{qqrr,i_1 i_2}(^{2s+1}L_J) \ =& \
 \tilde D^{(\alpha)\,VV}_{rrrr,i_1 i_2}(^{2s+1}L_J)\vert_{m_{i_{1}} \to
   -m_{i_{1}}}\ ,
\\
\nonumber
\tilde D^{(\alpha)\,VV}_{qrrq,i_1 i_2}(^{2s+1}L_J) \ =& \ 
 \tilde D^{(\alpha)\,VV}_{rqqr,i_1 i_2}(^{2s+1}L_J)\vert_{m_{i_{1,2}} \to -m_{i_{1,2}}}\ ,
\\
\nonumber
\tilde D^{(\alpha)\,VV}_{rqrq,i_1 i_2}(^{2s+1}L_J) \ =& \
 \tilde D^{(\alpha)\,VV}_{rqqr,i_1 i_2}(^{2s+1}L_J)\vert_{m_{i_{2}} \to -m_{i_{2}}}\ ,
\\
\tilde D^{(\alpha)\,VV}_{qrqr,i_1 i_2}(^{2s+1}L_J) \ =& \
 \tilde D^{(\alpha)\,VV}_{rqqr,i_1 i_2}(^{2s+1}L_J)\vert_{m_{i_{1}} \to -m_{i_{1}}}\ ,
\end{align}
where these relations hold for the kinematic factors related to any
$^{2s+1}L_J$ partial-wave reaction.

\subsubsection{$P$-wave kinematic factors for $X_A X_B = VS$}
The only non-vanishing kinematic factor expression associated with $^1P_1$
partial-wave reactions and related to selfenergy diagrams reads
\begin{align}
\nonumber
 \tilde B^{VS}_{qq,\,VV}(^1P_1) \ =& \ \frac{\widehat{m}_W^2}{4}~\diffmmbar^2 \ .
\end{align}
In case of combined $^3P_{\cal J}$ waves we have
\begin{align}
\tilde B^{VS}_{rr,\, VV }(^3P_{\cal J}) \ =& \ -\frac{\widehat{m}_W^2}{4}~\diffmmbar^2 \ ,
\\
\tilde B^{VS}_{qq,\, VV }(^3P_{\cal J}) \ =& \ \frac{\widehat{m}_W^2}{2} \ ,
\\
\tilde B^{VS}_{rr,\, VS }(^3P_{\cal J}) \ =& \ \tilde B^{VS}_{rr,\, SV }(^3P_{\cal J}) \ = \
  \frac{\widehat{m}_W}{8}~\diffmmbar~(3 - \Delta_{AB}) \ ,
\\
\tilde B^{VS}_{rr,\, SS }(^3P_{\cal J}) \ =& \ 
 \frac{1}{16}~(\beta^2 -9 + 6~\Delta_{AB} - \Delta_{AB}^2) \ .
\end{align}
The non-vanishing kinematic factors $\tilde C^{(\alpha)\, VS}_{n,\, i_1 V}$ related to the
four generic triangle topologies with gauge-boson exchange $V$ in the
single $s$-channel read
\begin{align}
\tilde C^{(1)\,VS}_{rqq,\, i_1 V}(^1P_1) \ =& \
 \frac{\widehat{m}_W\,\mihat}{4~\mmbar}\diffmmbar^2
 + \frac{\widehat{m}_W\,\diffmmbar}{8~\mmbar}~
        \left(\diffmmbar^2 + \diffmmbar - 1 + \Delta_{AB} \right)
 + \frac{\beta^2~\widehat{m}_W\diffmmbar}{12~P_{i_1 AB}} \ ,
\\
\tilde C^{(2)\,VS}_{qqr,\, i_1 V}(^1P_1) \ =& \
 \tilde C^{(1)\,VS}_{rqq,\, i_1 V}(^1P_1) \ ,
\\
\tilde C^{(3)\,VS}_{rqq,\, i_1 V}(^1P_1) \ =& \
 - \frac{\widehat{m}_W\,\mihat}{4~\mmbar}\diffmmbar^2
 - \frac{\widehat{m}_W\,\diffmmbar}{8~\mmbar}~
        \left(\diffmmbar^2 - \diffmmbar - 1 + \Delta_{AB} \right)
 - \frac{\beta^2~\widehat{m}_W\diffmmbar}{12~P_{i_1 BA}}  ,
\\
\tilde C^{(4)\,VS}_{qqr,\, i_1 V}(^1P_1) \ =& \
 \tilde C^{(3)\,VS}_{rqq,\, i_1 V}(^1P_1) \ .
\end{align}
In case of combined $^3P_{\cal J}$ wave reactions the kinematic factors
$\tilde C^{(\alpha)\, VS}_{n,\, i_1 V}$ read
\begin{align}
\tilde C^{(1)\,VS}_{rrr,\, i_1 V}(^3P_{\cal J}) \ =& \ \tilde C^{(2)\,VS}_{rrr,\, i_1 V}(^3P_{\cal J})\ =\
 -~\tilde C^{(1)\,VS}_{rqq,\, i_1 V}(^1P_1) \ ,
\\
\tilde C^{(3)\,VS}_{rrr,\, i_1 V}(^3P_{\cal J}) \ =& \ \tilde C^{(4)\,VS}_{rrr,\, i_1 V}(^3P_{\cal J})\ =\
 -~\tilde C^{(3)\,VS}_{rqq,\, i_1 V}(^1P_1)\vert_{\mihat \to - \mihat} \ ,
\\
\tilde C^{(1)\,VS}_{rqq,\, i_1 V}(^3P_{\cal J}) \ =& \ \tilde C^{(2)\,VS}_{qqr,\, i_1 V}(^3P_{\cal J})\ =\
 \frac{\widehat{m}_W\,\mihat}{2~\mmbar}
 + \frac{\widehat{m}_W}{4~\mmbar} \left( \diffmmbar~\Delta_{AB} + 1 \right)
 - \frac{\beta^2 ~ \widehat{m}_W}{6~P_{i_1 AB}} \ ,
\\\nonumber
\tilde C^{(3)\,VS}_{rqq,\, i_1 V}(^3P_{\cal J}) \ =& \ \tilde C^{(4)\,VS}_{qqr,\, i_1 V}(^3P_{\cal J})
\\ \ =& \
 - \frac{\widehat{m}_W\,\mihat}{2~\mmbar}
 - \frac{\widehat{m}_W}{4~\mmbar} \left( \diffmmbar~\Delta_{AB} - 1 \right)
 - \frac{\beta^2 ~ \widehat{m}_W}{6~P_{i_1 BA}} \ .
\end{align}
Turning to $\tilde C^{(\alpha)\,VS}_{n,\,i_1 S}$ factors we find that all kinematic
factors corresponding to the $^1P_1$ configuration vanish. Kinematic factors
$\tilde C^{(\alpha)\,VS}_{n,\, i_1 S}$ in combined $^3P_{\cal J}$ partial-wave reactions read
\begin{align}
\nonumber
\tilde C^{(1)\,VS}_{rrr,\, i_1 S}(^3P_{\cal J}) \ =& \
 \frac{\mihat\,\diffmmbar}{8~\mmbar} \left( 3 - \Delta_{AB} \right)
 + \frac{\beta^2}{24~P_{i_1 AB}} ~ (\diffmmbar + 3 + 2~\mihat)
\\ & \
 + \frac{1}{16~\mmbar}
        \left( \beta^2 - 3 + (4 - \Delta_{AB})~\Delta_{AB}
              + ( \diffmmbar^2 + \diffmmbar )( 3 - \Delta_{AB} ) \right) ,
\\
\tilde C^{(2)\,VS}_{rrr,\, i_1 S}(^3P_{\cal J}) \ =& \
 \tilde C^{(1)\,VS}_{rrr,\, i_1 S}(^3P_{\cal J}) \ ,
\\
\nonumber
\tilde C^{(3)\,VS}_{rrr,\, i_1 S}(^3P_{\cal J}) \ =& \
 \frac{\mihat\,\diffmmbar}{8~\mmbar} \left( 3 - \Delta_{AB} \right)
 + \frac{\beta^2}{24~P_{i_1 BA}} ~ (\diffmmbar - 3 - 2~\mihat)
 \\ & \
 - \frac{1}{16~\mmbar}
        \left( \beta^2 - 3 + (4 - \Delta_{AB})~\Delta_{AB}
               + (\diffmmbar^2 - \diffmmbar)(3 - \Delta_{AB}) \right) ,
\\
\tilde C^{(4)\,VS}_{rrr,\, i_1 S}(^3P_{\cal J}) \ =& \
 \tilde C^{(3)\,VS}_{rrr,\, i_1 S}(^3P_{\cal J}) \ .
\end{align}
The additional non-vanishing kinematic factor expressions
$\tilde C^{(\alpha)VS}_{n,\,i_1 X}$ with $X=V,S$ are related to the above given
expressions via
\begin{align}
\nonumber
\tilde C^{(1)\,VS}_{qqr,\, i_1 X}(^{2s+1}P_J) \ =& \ \tilde C^{(2)\,VS}_{rqq,\, i_1 X}(^{2s+1}P_J) \ = \
 -~\tilde C^{(1)\,VS}_{rrr,\, i_1 X}(^{2s+1}P_J)
 \vert_{\widehat m_{i_1} \to - \widehat m_{i_1}} \ ,
\\\nonumber
\tilde C^{(3)\,VS}_{qqr,\, i_1 X}(^{2s+1}P_J) \ =& \ \tilde C^{(4)\,VS}_{rqq,\, i_1 X}(^{2s+1}P_J)  \ = \
 \tilde C^{(3)\,VS}_{rrr,\, i_1 X}(^{2s+1}P_J)
 \vert_{\widehat m_{i_1} \to - \widehat m_{i_1}} \ ,
\\\nonumber
\tilde C^{(1)\,VS}_{qrq,\, i_1 X}(^{2s+1}P_J) \ =& \ \tilde C^{(2)\,VS}_{qrq,\, i_1 X}(^{2s+1}P_J) \ = \
 -~\tilde C^{(1)\,VS}_{rqq,\, i_1 X}(^{2s+1}P_J)
 \vert_{\widehat m_{i_1} \to - \widehat m_{i_1}} \ ,
\\
\tilde C^{(3)\,VS}_{qrq,\, i_1 X}(^{2s+1}P_J) \ =& \ \tilde C^{(4)\,VS}_{qrq,\, i_1 X}(^{2s+1}P_J) \ = \
 \tilde C^{(3)\,VS}_{rqq,\, i_1 X}(^{2s+1}P_J)
 \vert_{\widehat m_{i_1} \to - \widehat m_{i_1}} \ .
\label{eq:CtildeVSrelations}
\end{align}
The above relations hold for kinematic factors
$\tilde C^{(\alpha)\,VS}_{n,\, i_1 X}(^{2s+1}L_J)$ associated with the absorptive part
of Wilson coefficients $f(^{2s+1}L_J)$ and $g(^{2s+1}S_s)$ in
$\delta\mathcal L^{d=6}_\text{ann}$ and $\delta\mathcal L^{d=8}_\text{ann}$.

Finally, kinematic factors $\tilde D^{(\alpha)}_{n,\,i_1 i_2}$ for $^1P_1$ partial
wave reactions are given by
\begin{align}
\nonumber
\lefteqn{\tilde D^{(\alpha)\,VS}_{rrrr,\, i_1 i_2}(^1P_1) \ = \
 \frac{\beta^2}{24~(\mmbar)^2} \ , } &
\\[2ex]
\nonumber
\lefteqn{\tilde D^{(1)\,VS}_{rqqr,\, i_1 i_2}(^1P_1)} &
\\\nonumber =& \
 - \frac{\diffmmbar}{8~(\mmbar)^2}
        \left( (\diffmmbar^2 -1 + \Delta_{AB})~(\mihat + \mjhat)
               - \diffmmbar (\mihat - \mjhat) \right)
 \\\nonumber & \
 - \frac{\mihat\mjhat}{4~(\mmbar)^2}~\diffmmbar^2
 - \frac{1}{48~(\mmbar)^2}
        \left( \diffmmbar^2~( 3\,\diffmmbar^2 - 9 + 6\Delta_{AB}) \right.
 \\\nonumber & \ \phantom{  - \frac{\mihat\mjhat}{4~(\mmbar)^2}~\diffmmbar^2
                            - \frac{1}{48~(\mmbar)^2} } \hspace{3ex}
        \left.  - \beta^2 + 3 - 3~( 2-\Delta_{AB})\Delta_{AB} \right)
 \\\nonumber & \
  - \frac{\beta^2}{24~\mmbar~P_{i_1 AB}}
         \left( \diffmmbar (\diffmmbar + 2~\mjhat - 2) -1 - 2~\mihat \right)
 \\ & \
 \nonumber
 - \frac{\beta^2}{24~\mmbar~P_{i_2 BA}}
        \left( \diffmmbar (\diffmmbar + 2~\mihat + 2) -1 + 2~\mjhat \right)
 \\ & \
 \nonumber
 - \frac{\beta^2}{12~P_{i_1 AB} P_{i_2 BA}}
        \left( \diffmmbar (\diffmmbar + 2~\mihat + 2~\mjhat) + \beta^2
               + 4~\mihat \mjhat \right.
 \\ & \ \phantom{  - \frac{\beta^2}{12~P_{i_1 AB} P_{i_2 BA}} } \hspace{3ex}
         \left. - (2~\mihat -2~\mjhat + \Delta_{AB})~\Delta_{AB} \right) \ ,
\\[2ex]
 \nonumber
\lefteqn{\tilde D^{(2)\,VS}_{rqqr,\, i_1 i_2}(^1P_1)} &
\\\nonumber =& \
 \frac{\mihat \diffmmbar}{8~(\mmbar)^2}
      \left( \diffmmbar^2 + \diffmmbar - 1 + \Delta_{AB} \right)
 + \frac{\mihat\mjhat}{8~(\mmbar)^2}~\diffmmbar^2
 \\ & \
 \nonumber
 + \frac{1}{96~(\mmbar)^2}
        \left(\, 3\,(\diffmmbar^2+\diffmmbar-1)\,
                    (\diffmmbar^2+\diffmmbar-1+2~\Delta_{AB})
             -\beta^2 + 3~\Delta_{AB}^2 \right)
 \\ & \ 
 \nonumber
 + \frac{\beta^2}{24~\mmbar~P_{i_1 AB}}
        \left( \diffmmbar\,(\diffmmbar + 2~\mjhat) - 2~\mihat - 1 \right)
 \\ & \
 \nonumber
 - \frac{\beta^2}{24~P_{i_1 AB} P_{i_2 AB}}
        \Bigl( (\diffmmbar + 2~(\mihat+\mjhat+\Delta_{AB}) )~\diffmmbar
                -\beta^2 +\,4~\mihat\mjhat
 \\ & \ \phantom{  - \frac{\beta^2}{24~P_{i_1 AB} P_{i_2 AB}} } \hspace{3ex}
               +( 2~(\mihat+\mjhat) +\Delta_{AB}) \, \Delta_{AB} \Bigr) 
 \ + \ \sumij  \ ,
\\[2ex]
\lefteqn{\tilde D^{(3)\,VS}_{rqqr,\, i_1 i_2}(^1P_1) \ = \
 \tilde D^{(1)\,VS}_{rqqr,\, i_1 i_2}(^1P_1)
        \vert_{\widehat{m}_{i_{1}} \leftrightarrow \,\widehat{m}_{i_{2}}} \ ,} &
\\
\lefteqn{\tilde D^{(4)\,VS}_{rqqr,\, i_1 i_2}(^1P_1) \ = \
 \tilde D^{(2)\,VS}_{rqqr,\, i_1 i_2}(^1P_1) 
        \vert_{\, \widehat{m} \leftrightarrow \widehat{\overline{m}}, ~
                  \widehat{m}_{i_{1,2}} \rightarrow -\widehat{m}_{i_{1,2}} } \ . } &
\end{align}
The kinematic factor expressions $\tilde D^{(\alpha)}_{n,\, i_1 i_2}$ related to
$^3P_{\cal J}$ partial-wave reactions read
\begin{align}
\nonumber
\lefteqn{\tilde D^{(1)\,VS}_{rrrr,\, i_1 i_2}(^3P_{\cal J})} &
\\\nonumber =& \
 \frac{\diffmmbar}{8~(\mmbar)^2}
      \left(   (\mihat - \mjhat)\,(\diffmmbar^2 + \Delta_{AB} - 1)
             - (\mihat + \mjhat)\,\diffmmbar \right)
 - \frac{\mihat\mjhat}{4(\mmbar)^2}~\diffmmbar^2
 \\\nonumber & \
 +\frac{1}{48~(\mmbar)^2}
       \left( (3~\diffmmbar^2 + 4~\beta^2 - 9 + 6~\Delta_{AB})\,\diffmmbar^2
               - 3\,(\beta^2 - (1-\Delta_{AB})^2\,) \right)
 \\\nonumber & \
 - \frac{\beta^2}{24~\mmbar~P_{i_1 AB}}
        \left( (\diffmmbar+2\,( 2~\mihat+\mjhat + \Delta_{AB} + 1)) \diffmmbar
               + 2~\mihat + 1 \right)
 \\\nonumber & \
 - \frac{\beta^2}{24~\mmbar~P_{i_2 BA}}
        \left( (\diffmmbar-2\,( 2~\mjhat+\mihat + \Delta_{AB} + 1)) \diffmmbar
               + 2~\mjhat + 1 \right)
 \\\nonumber & \
 + \frac{\beta^2}{12~P_{i_1 AB}~P_{i_2 BA}}
        \left( 3\,\diffmmbar\,(\diffmmbar + 2~(\mihat-\mjhat) )
               - 12~\mihat\mjhat +\beta^2 \right.
 \\ & \ \phantom{ + \frac{\beta^2}{12~P_{i_1 AB}~P_{i_2 BA}} } \hspace{3ex}
        \left. - 3~\Delta_{AB}\,(2~(\mihat+\mjhat)+\Delta_{AB}) \right) \ ,
\\[2ex]
\nonumber
\lefteqn{\tilde D^{(2)\,VS}_{rrrr,\, i_1 i_2}(^3P_{\cal J})} &
 \\\nonumber =& \
 - \frac{\mihat \diffmmbar}{8~(\mmbar)^2}
        \left( \diffmmbar (\diffmmbar+1) + \Delta_{AB} - 1 \right)
 - \frac{\mihat\mjhat}{8~(\mmbar)^2}~\diffmmbar^2
 \\\nonumber & \
 -\frac{1}{96~(\mmbar)^2}
       \Bigl( (3~\diffmmbar^2 + 6~\diffmmbar - 4~\beta^2 - 3 + 6~\Delta_{AB})
              \,\diffmmbar^2
 \\\nonumber & \ \phantom{  -\frac{1}{96~(\mmbar)^2} } \hspace{3ex}
               +\,6~\diffmmbar (\Delta_{AB}-1)
               - 3\,(\beta^2 - (1-\Delta_{AB})^2\,) \Bigr)
 \\\nonumber & \
 - \frac{\beta^2}{24~\mmbar~P_{i_1 AB}}
        \left( \diffmmbar\,(3~\diffmmbar + 2\,(2~\mihat + \mjhat + \Delta_{AB}))
               -2~\mihat - 1 \right)
 \\\nonumber & \
 + \frac{\beta^2}{24~P_{i_1 AB}~P_{i_2 AB}}
        \Bigl( 3~\diffmmbar (\diffmmbar + 2~(\mihat + \mjhat + \Delta_{AB}))
               + 12~\mihat\mjhat - \beta^2
 \\ & \ \phantom{ + \frac{\beta^2}{24~P_{i_1 AB}~P_{i_2 BA}} } \hspace{3ex}
               + 3~\Delta_{AB}\,(2~(\mihat+\mjhat)+\Delta_{AB}) \Bigr) 
 \ + \ \sumij \ ,
\\[2ex] &
\label{eq:D3VSrrrr3Prepls}
 \tilde D^{(3)\,VS}_{rrrr,\, i_1 i_2}(^3P_{\cal J}) \ =\ \tilde D^{(1)\,VS}_{rrrr,\, i_1 i_2}(^3P_{\cal J})
           \vert_{\widehat{m}_{i_{1}} \leftrightarrow ~\widehat{m}_{i_{2}}} \ ,
\\ &
\label{eq:D4VSrrrr3Prepls}
 \tilde D^{(4)\,VS}_{rrrr,\, i_1 i_2}(^3P_{\cal J}) \ =\ \tilde D^{(2)\,VS}_{rrrr,\, i_1 i_2}(^3P_{\cal J})
           \vert_{\, \widehat{m} \leftrightarrow \widehat{\overline{m}}} \ ,
\\[2ex]
\nonumber
\lefteqn{\tilde D^{(1)\,VS}_{rqqr,\, i_1 i_2}(^3P_{\cal J})} &
 \\\nonumber =& \
 - \frac{1}{4~(\mmbar)^2}
        \left( (\mihat + \mjhat)~\diffmmbar\,\Delta_{AB} -\mihat + \mjhat \right)
 - \frac{\mihat\mjhat}{2~(\mmbar)^2}
 \\\nonumber & \
 + \frac{1}{24~(\mmbar)^2}
        \left( \diffmmbar^2 (\beta^2 - 3~\Delta_{AB}^2) + 3 \right)
 + \frac{\beta^2}{12~\mmbar~P_{i_1 AB}}
        \left( \diffmmbar\,\Delta_{AB} + 2~\mjhat - 1 \right)
 \\ & \
 - \frac{\beta^2}{12~\mmbar~P_{i_2 BA}}
        \left( \diffmmbar~\Delta_{AB} + 2~\mihat + 1 \right)
 + \frac{\beta^4}{6~~P_{i_1 AB}~P_{i_2 BA}} \ ,
\\[2ex]
\nonumber
\lefteqn{\tilde D^{(2)\,VS}_{rqqr,\, i_1 i_2}(^3P_{\cal J})} &
 \\\nonumber =& \
 \frac{\mihat}{4~(\mmbar)^2} \left(\diffmmbar\,\Delta_{AB} + 1 \right)
 + \frac{\mihat\mjhat}{4~(\mmbar)^2}
 \\\nonumber & \
 - \frac{1}{48~(\mmbar)^2}
        \left( \diffmmbar^2\,(\beta^2 - 3~\Delta_{AB}^2)
                - 6~\diffmmbar\,\Delta_{AB} - 3 \right)
 + \frac{\beta^4}{12~P_{i_1 AB} P_{i_2 AB}}
 \\ & \
 - \frac{\beta^2}{12~\mmbar~P_{i_1 AB}}
        \left( \diffmmbar\,\Delta_{AB} + 2~\mjhat + 1 \right) 
 + \sumij \ , 
\\[2ex] &
\label{eq:D3VSrqqr3Prepls}
 \tilde D^{(3)\,VS}_{rqqr,\, i_1 i_2}(^3P_{\cal J}) \ =\ \tilde D^{(1)\,VS}_{rqqr,\, i_1 i_2}(^3P_{\cal J})
           \vert_{\widehat{m}_{i_{1}} \leftrightarrow~ \widehat{m}_{i_{2}}} \ ,
\\ &
\label{eq:D4VSrqqr3Prepls}
 \tilde D^{(4)\,VS}_{rqqr,\, i_1 i_2}(^3P_{\cal J}) \ =\ \tilde D^{(2)\,VS}_{rqqr,\, i_1 i_2}(^3P_{\cal J})
           \vert_{\, \widehat{m} \leftrightarrow \widehat{\overline{m}}, ~
                   \widehat{m}_{i_{1,2}} \to~ -\widehat{m}_{i_{1,2}}} \ .
\end{align}
Note that relation (\ref{eq:D3VSrrrr3Prepls}) implies that the denominator
structures $P_{i_1 AB}$ and $P_{i_2 BA}$ in the kinematic factor corresponding
to diagram topology $\alpha = 1$ have to be replaced by $P_{i_2 AB}$ and
$P_{i_1 BA}$ respectively, in order to arrive at the kinematic factor related
to diagram topology $\alpha = 3$. Likewise, in (\ref{eq:D4VSrrrr3Prepls}) the
replacement rule for the kinematic factor for diagram-topology $\alpha = 2$
implies the replacement of $P_{i_1 AB}$ and $P_{i_2 AB}$ by $P_{i_1 BA}$ and
$P_{i_2 BA}$, respectively.
Similar replacements are needed to obtain the $\alpha = 3,4$ 
kinematic factors from the $\alpha = 1,2$ expressions with 
$n = rqqr$ using
(\ref{eq:D3VSrqqr3Prepls}) and (\ref{eq:D4VSrqqr3Prepls}).
The relations among kinematic factors in 
(\ref{eq:D3VSrrrr3Prepls}--\ref{eq:D4VSrrrr3Prepls}) and
(\ref{eq:D3VSrqqr3Prepls}--\ref{eq:D4VSrqqr3Prepls}) also
hold for the individual kinematic factors
related to $^3P_J$ partial-wave reactions with $J = 0,1,2$.

The remaining non-vanishing kinematic factors $\tilde D^{(\alpha)}_{n,\, i_1 i_2}$
for diagram topologies $\alpha = 1,2$ derive from the above given expressions in
the following way:
\begin{align}
\nonumber
\tilde D^{(\alpha)\,VS}_{qqqq,\, i_1 i_2}(^{2s+1}L_J) \ =& \ 
 (-1)^{\,\alpha} ~ \tilde D^{(\alpha)\,VS}_{rrrr,\, i_1 i_2}(^{2s+1}L_J)
 \vert_{\ \widehat m_{\,i_{1,2}}  \rightarrow \ - \widehat m_{\,i_{1,2}}} \ ,
\\\nonumber
\tilde D^{(\alpha)\,VS}_{rrqq,\, i_1 i_2}(^{2s+1}L_J) \ =& \ 
 (-1)^{\,\alpha+1} ~ \tilde D^{(\alpha)\,VS}_{rrrr,\, i_1 i_2}(^{2s+1}L_J)
 \vert_{\ \hat m_{\,i_{2}}  \rightarrow \ - \hat m_{\,i_{2}}} \ ,
\\\nonumber
\tilde D^{(\alpha)\,VS}_{qqrr,\, i_1 i_2}(^{2s+1}L_J) \ =& \ 
 - \tilde D^{(\alpha)\,VS}_{rrrr,\, i_1 i_2}(^{2s+1}L_J)
 \vert_{\ \hat m_{\,i_{1}}  \rightarrow \ - \hat m_{\,i_{1}}} \ ,
\\\nonumber
\tilde D^{(\alpha)\,VS}_{qrrq,\, i_1 i_2}(^{2s+1}L_J) \ =& \ 
 (-1)^{\,\alpha} ~ \tilde D^{(\alpha)\,VS}_{rqqr,\, i_1 i_2}(^{2s+1}L_J)
  \vert_{\ \hat m_{\,i_{1,2}}  \rightarrow \ - \hat m_{\,i_{1,2}}} \ ,
\\\nonumber
\tilde D^{(\alpha)\,VS}_{rqrq,\, i_1 i_2}(^{2s+1}L_J) \ =& \ 
 (-1)^{\,\alpha+1} ~ \tilde D^{(\alpha)\,VS}_{rqqr,\, i_1 i_2}(^{2s+1}L_J)
  \vert_{\ \hat m_{\,i_{2}}  \rightarrow \ - \hat m_{\,i_{2}}} \ ,
\\
\tilde D^{(\alpha)\,VS}_{qrqr,\, i_1 i_2}(^{2s+1}L_J) \ =& \ 
 - \tilde D^{(\alpha)\,VS}_{rqqr,\, i_1 i_2}(^{2s+1}L_J)
  \vert_{\ \hat m_{\,i_{1}}  \rightarrow \ - \hat m_{\,i_{1}}} \ .
\label{eq:app_D_VS_12}
\end{align}
Similarly, in case of diagram topologies $\alpha = 3,4$ we find
\begin{align}
\nonumber
\tilde D^{(\alpha)\,VS}_{qqqq,\, i_1 i_2}(^{2s+1}L_J) \ =& \ 
 (-1)^{\,\alpha} ~ \tilde D^{(\alpha)\,VS}_{rrrr,\, i_1 i_2}(^{2s+1}L_J)
 \vert_{\ \widehat m_{\,i_{1,2}}  \rightarrow \ - \widehat m_{\,i_{1,2}}} \ ,
\\\nonumber
\tilde D^{(\alpha)\,VS}_{rrqq,\, i_1 i_2}(^{2s+1}L_J) \ =& \ 
 (-1)^{\,\alpha} ~ \tilde D^{(\alpha)\,VS}_{rrrr,\, i_1 i_2}(^{2s+1}L_J)
 \vert_{\ \hat m_{\,i_{2}}  \rightarrow \ - \hat m_{\,i_{2}}} \ ,
\\\nonumber
\tilde D^{(\alpha)\,VS}_{qqrr,\, i_1 i_2}(^{2s+1}L_J) \ =& \ 
 \tilde D^{(\alpha)\,VS}_{rrrr,\, i_1 i_2}(^{2s+1}L_J)
 \vert_{\ \hat m_{\,i_{1}}  \rightarrow \ - \hat m_{\,i_{1}}} \ ,
\\\nonumber
\tilde D^{(\alpha)\,VS}_{qrrq,\, i_1 i_2}(^{2s+1}L_J) \ =& \ 
 (-1)^{\,\alpha} ~ \tilde D^{(\alpha)\,VS}_{rqqr,\, i_1 i_2}(^{2s+1}L_J)
  \vert_{\ \hat m_{\,i_{1,2}}  \rightarrow \ - \hat m_{\,i_{1,2}}} \ ,
\\\nonumber
\tilde D^{(\alpha)\,VS}_{rqrq,\, i_1 i_2}(^{2s+1}L_J) \ =& \ 
 (-1)^{\,\alpha} ~ \tilde D^{(\alpha)\,VS}_{rqqr,\, i_1 i_2}(^{2s+1}L_J)
  \vert_{\ \hat m_{\,i_{2}}  \rightarrow \ - \hat m_{\,i_{2}}} \ ,
\\
\tilde D^{(\alpha)\,VS}_{qrqr,\, i_1 i_2}(^{2s+1}L_J) \ =& \ 
 \tilde D^{(\alpha)\,VS}_{rqqr,\, i_1 i_2}(^{2s+1}L_J)
  \vert_{\ \hat m_{\,i_{1}}  \rightarrow \ - \hat m_{\,i_{1}}} \ .
\label{eq:app_D_VS_34}
\end{align}
The relations in (\ref{eq:app_D_VS_12})--(\ref{eq:app_D_VS_34}) are
valid for $\tilde D^{(\alpha)\,VS}_{n,\, i_1 i_2}(^{2s+1}L_J)$ expressions related to
any $^{2s+1}L_J$ partial-wave reaction.

\subsubsection{$P$-wave kinematic factors for $X_A X_B = S S$}
In case of $X_A X_B = SS$ the only non-vanishing kinematic factor
$\tilde B^{SS}_{n,\, i_1 i_2}$ in $^1P_1$ partial-wave scattering reactions reads
\begin{align}
\tilde B^{SS}_{qq,\, VV }(^1P_1) \ =& \
 \frac{\beta^2}{12}~\diffmmbar^2 \ ,
\end{align}
while the corresponding kinematic factors for combined $^3P_{\cal J}$ reactions
read
\begin{align}
\tilde B^{SS}_{rr,\, VV}(^3P_{\cal J}) \ =& \ \frac{1}{4}~\diffmmbar^2 ~ \Delta_{AB}^2 \ ,
\\
\tilde B^{SS}_{qq,\, VV}(^3P_{\cal J}) \ =& \ \frac{\beta^2}{6} \ ,
\\
\tilde B^{SS}_{rr,\, VS}(^3P_{\cal J}) \ =& \ \tilde B^{SS}_{rr,\, SV}(^3P_{\cal J}) \ = \
 \frac{\widehat{m}_W}{4}~\diffmmbar~\Delta_{AB} \ ,
\\
\tilde B^{SS}_{rr,\, SS}(^3P_{\cal J}) \ =& \
 \frac{\widehat{m}_W^2}{4} \ .
\end{align}
The kinematic factors for diagram topologies $\alpha = 3 (4)$ and
$\alpha = 1 (2)$ obey in both the cases of triangle and box diagrams
certain relations,
\begin{align}
\nonumber
\tilde C^{(3)\, SS}_{n, \, i_1 V}(^{2s+1}L_J) \ =& \ 
 -~\tilde C^{(1)\, SS}_{n, \, i_1 V}(^{2s+1}L_J) \
                             \vert_{A \leftrightarrow B} \ ,
\\\nonumber
\tilde C^{(4)\, SS}_{n, \, i_1 V}(^{2s+1}L_J) \ =& \ 
 -~ \tilde C^{(2)\, SS}_{n, \, i_1 V}(^{2s+1}L_J) \
                             \vert_{A \leftrightarrow B} \ ,
\\\nonumber
\tilde C^{(3)\, SS}_{n, \, i_1 S}(^{2s+1}L_J) \ =& \ 
 \tilde C^{(1)\, SS}_{n, \, i_1 S}(^{2s+1}L_J) \
                             \vert_{A \leftrightarrow B} \ ,
\\\nonumber
\tilde C^{(4)\, SS}_{n, \, i_1 S}(^{2s+1}L_J) \ =& \ 
 \tilde C^{(2)\, SS}_{n, \, i_1 S}(^{2s+1}L_J) \
                             \vert_{A \leftrightarrow B} \ ,
\\\nonumber
\tilde D^{(3)\, SS}_{n, \, i_1 i_2}(^{2s+1}L_J) \ =& \ \tilde D^{(1)\, SS}_{n, \, i_1 i_2}(^{2s+1}L_J) \
                             \vert_{A \leftrightarrow B} \ ,
\\
\tilde D^{(4)\, SS}_{n, \, i_1 i_2}(^{2s+1}L_J) \ =& \ \tilde D^{(2)\, SS}_{n, \, i_1 i_2}(^{2s+1}L_J) \
                             \vert_{A \leftrightarrow B} \ ,
\end{align}
that generically apply for the respective kinematic factors related to a given
$^{2s+1}L_J$ partial-wave configuration, including kinematic factors related to
coefficients $\hat g(^{2s+1}S_s)$ and $\hat h_i(^{2s+1}S_s)$
(see also Eq.~(111) in paper I).

In case of $^1P_1$ waves we find the following expressions for kinematic factors
$\tilde C^{(\alpha)\,SS}_{n,\, i_1 V}$ and diagram topologies $\alpha = 1,2$:
\begin{align}
\tilde C^{(1)\,SS}_{rqq,\, i_1 V}(^1P_1) \ =& \ \tilde C^{(2)\,SS}_{qqr,\, i_1 V}(^1P_1)
 \ = \
   \frac{\beta^2}{24~\mmbar}~\diffmmbar^2
 - \frac{\beta^2~\diffmmbar}{12~P_{i_1 AB}}
        \left( \diffmmbar + 2~\mihat + \Delta_{AB} \right) \ .
\end{align}
In case of combined $^3P_{\cal J}$ reactions the corresponding expressions read
\begin{align}
\nonumber
\tilde C^{(1)\,SS}_{rrr,\, i_1 V}(^3P_{\cal J}) \ =& \ \tilde C^{(2)\,SS}_{rrr,\, i_1 V}(^3P_{\cal J})
 \\ =& \
 - \frac{\mihat}{4~\mmbar}~\diffmmbar\,\Delta_{AB}
 - \frac{\diffmmbar\,\Delta_{AB}}{8~\mmbar}\left(\diffmmbar\,\Delta_{AB}+1 \right)
 +\frac{\beta^2~\diffmmbar\,\Delta_{AB}}{12~P_{i_1 AB}} \ ,
\\
\tilde C^{(1)\,SS}_{rqq,\, i_1 V}(^3P_{\cal J}) \ =& \ \tilde C^{(2)\,SS}_{qqr,\, i_1 V}(^3P_{\cal J})
 \ = \
 \frac{\beta^2}{12~\mmbar} \ .
\end{align}
Turning to kinematic factors $\tilde C^{(\alpha)\,SS}_{n,\, i_1 S}$ with
$\alpha = 1,2$ we find
\begin{align}
\nonumber
\tilde C^{(1)\,SS}_{rrr,\, i_1 S}(^3P_{\cal J}) \ =& \ \tilde C^{(2)\,SS}_{rrr,\, i_1 S}(^3P_{\cal J})
\\ =& \
 -\frac{\widehat{m}_W}{8~\mmbar}~\left( \diffmmbar\,\Delta_{AB} + 1 \right)
 - \frac{\widehat{m}_W}{4~\mmbar}~\mihat
 +\frac{\beta^2}{12~P_{i_1 AB}}~\widehat{m}_W \ ,
\end{align}
and, as in the case of leading-order $^1S_0$ and $^3S_1$ kinematic
factors (see Eq.(115) in paper I), the remaining non-vanishing expressions for
$\tilde C^{(\alpha)\,SS}_{n,\, i_1 X}$ with both $X = V,S$ and $\alpha = 1,2$
that are associated with $^1P_1$ and $^3P_{\cal J}$ (as well as the separate
$^3P_J, J = 0,1,2$) partial-wave configurations,
derive from the above given expressions in the
following way:
\begin{align}
\nonumber
\tilde C^{(1)\,SS}_{qqr,\, i_1 X}(^{2s+1}P_J) \ =& \ 
\tilde C^{(2)\,SS}_{rqq,\, i_1 X}(^{2s+1}P_J) \ = \
 -~\tilde C^{(1)\,SS}_{rrr,\, i_1 X}(^{2s+1}P_J)
 \vert_{\widehat m_{i_1} \to - \widehat m_{i_1}} \ ,
\\
\tilde C^{(1)\,SS}_{qrq,\, i_1 X}(^{2s+1}P_J) \ =& \ 
\tilde C^{(2)\,SS}_{qrq,\, i_1 X}(^{2s+1}P_J)  \ = \
 -~\tilde C^{(1)\,SS}_{rqq,\, i_1 X}(^{2s+1}P_J)
 \vert_{\widehat m_{i_1} \to - \widehat m_{i_1}} \ .
\end{align}
Finally, the box-diagram related kinematic factors
$\tilde D^{(\alpha)\,SS}_{n,\, i_1 i_2}$ for diagram topologies $\alpha = 1,2$ are
given by
\begin{align}
\nonumber
\tilde D^{(1)\, SS}_{rqqr,\, i_1 i_2}(^1P_1) \ =& \
 -\frac{\beta^2~\diffmmbar^2}{96~(\mmbar)^2}
 +\frac{\beta^2~\diffmmbar}{24~\mmbar~P_{i_1 AB}}
      \left( \diffmmbar + 2~\mihat + \Delta_{AB} \right)
 \\ & \
 \nonumber
 -\frac{\beta^2}{24~P_{i_1 AB} P_{i_2 BA}}
       \left( \diffmmbar^2+ 2~\diffmmbar~(\mihat+\mjhat) \right.
 \\ & \ \phantom{  -\frac{\beta^2}{24~P_{i_1 AB} P_{i_2 BA}} } \hspace{3ex}
 \nonumber
       \left. + (2~\mihat + \Delta_{AB})~(2~\mjhat - \Delta_{AB}) \right)
 \\ & \
 + \sumABij  \ ,
\\
\nonumber
\tilde D^{(2)\, SS}_{rqqr,\, i_1 i_2}(^1P_1) \ =& \
 \phantom{-} \frac{\beta^2~\diffmmbar^2}{96~(\mmbar)^2}
 -\frac{\beta^2~\diffmmbar}{24~\mmbar~P_{i_1 AB}}
       \left( \diffmmbar + 2~\mihat + \Delta_{AB} \right)
 \\ & \
\nonumber
 +\frac{\beta^2}{24~P_{i_1 AB} P_{i_2 AB}}
       \left( \diffmmbar^2 + 2~\diffmmbar~ (\mihat+\mjhat+ \Delta_{AB}) \right.
 \\ & \ \phantom{  -\frac{\beta^2}{24~P_{i_1 AB} P_{i_2 BA}} } \hspace{3ex}
 \nonumber
        \left. + (2~\mihat + \Delta_{AB})~(2~\mjhat + \Delta_{AB}) \right)
 \\ & \
 + \sumij \ .
\end{align}
For the combined $^3P_{\cal J}$ reactions we have
\begin{align}
\nonumber
\tilde D^{(1)\, SS}_{rrrr,\, i_1 i_2}(^3P_{\cal J}) \ =& \
   \frac{\mihat\mjhat}{8~(\mmbar)^2}
 - \frac{\mihat}{8~(\mmbar)^2}
        \left( \diffmmbar\,\Delta_{AB} - 1 \right)
 \\ & \
 \nonumber
 + \frac{\beta^2}{24~\mmbar~P_{i_1 AB}}
        \left( \diffmmbar \Delta_{AB} - 2~\mjhat - 1 \right)
 + \frac{\beta^4}{24~P_{i_1 AB} P_{i_2 BA}}
 \\ & \
 - \frac{1}{32~(\mmbar)^2} \left( \diffmmbar^2\Delta_{AB}^2 -1 \right)
 + \sumABij \ ,
\\
\nonumber
\tilde D^{(2)\, SS}_{rrrr,\, i_1 i_2}(^3P_{\cal J}) \ =& \
   \frac{\mihat\mjhat}{8~(\mmbar)^2}
 + \frac{\mihat}{8~(\mmbar)^2}
        \left( \diffmmbar\,\Delta_{AB} + 1 \right)
 \\ & \
 \nonumber
 - \frac{\beta^2}{24~\mmbar~P_{i_1 AB}}
        \left( \diffmmbar\,\Delta_{AB} + 2~\mjhat + 1 \right)
 + \frac{\beta^4}{24~P_{i_1 AB} P_{i_2 AB}}
 \\ & \
 + \frac{1}{32~(\mmbar)^2} \left( \diffmmbar\,\Delta_{AB} +1 \right)^2
 + \sumij \ ,
\\
\tilde D^{(\alpha)\, SS}_{rqqr,\, i_1 i_2}(^3P_{\cal J}) \ =& \
 (-1)^\alpha ~ \frac{\beta^2}{24~(\mmbar)^2} \ .
\end{align}
The remaining non-vanishing kinematic factors can be related to the above given 
expressions by making use of the following relations among
$\tilde D^{(\alpha)\, SS}_{n,\, i_1 i_2}$ kinematic factors with different labels $n$:
\begin{align}
\nonumber
\tilde D^{(\alpha)\,SS}_{qqqq,\, i_1 i_2}(^{2s+1}L_J) \ =& \ 
  \tilde D^{(\alpha)\,SS}_{rrrr,\, i_1 i_2}(^{2s+1}L_J)
  \vert_{\ \hat m_{\,i_{1,2}}  \rightarrow \ - \hat m_{\,i_{1,2}}} \ ,
\\\nonumber
\tilde D^{(\alpha)\,SS}_{rrqq,\, i_1 i_2}(^{2s+1}L_J) \ =& \
 - \tilde D^{(\alpha)\,SS}_{rrrr,\, i_1 i_2}(^{2s+1}L_J)
 \vert_{\ \hat m_{\,i_{2}}  \rightarrow \ - \hat m_{\,i_{2}}} \ ,
\\\nonumber
 \tilde D^{(\alpha)\,SS}_{qqrr,\, i_1 i_2}(^{2s+1}L_J) \ =& \
 - \tilde D^{(\alpha)\,SS}_{rrrr,\, i_1 i_2}(^{2s+1}L_J)
 \vert_{\ \hat m_{\,i_{1}}  \rightarrow \ - \hat m_{\,i_{1}}} \ ,
\\\nonumber
\tilde D^{(\alpha)\,SS}_{qrrq,\, i_1 i_2}(^{2s+1}L_J) \ =& \ 
 \tilde D^{(\alpha)\,SS}_{rqqr,\, i_1 i_2}(^{2s+1}L_J) \ ,
 \vert_{\ \hat m_{\,i_{1,2}}  \rightarrow \ - \hat m_{\,i_{1,2}}} \ ,
\\\nonumber
\tilde D^{(\alpha)\,SS}_{rqrq,\, i_1 i_2}(^{2s+1}L_J) \ =& \ 
 - \tilde D^{(\alpha)\,SS}_{rqqr,\, i_1 i_2}(^{2s+1}L_J)
 \vert_{\ \hat m_{\,i_{2}}  \rightarrow \ - \hat m_{\,i_{2}}} \ ,
\\
\tilde D^{(\alpha)\,SS}_{qrqr,\, i_1 i_2}(^{2s+1}L_J) \ =& \ 
 - \tilde D^{(\alpha)\,SS}_{rqqr,\, i_1 i_2}(^{2s+1}L_J)
 \vert_{\ \hat m_{\,i_{1}}  \rightarrow \ - \hat m_{\,i_{1}}} \ .
\end{align}
Note that these relations hold among the kinematic factors
associated with any of the Wilson coefficients $\hat f(^{2s+1}L_J)$,
$\hat g(^{2s+1}S_s)$ and $\hat h_i(^{2s+1}S_s)$.

\subsubsection{$P$-wave kinematic factors for $X_A X_B = ff$}
The relevant kinematic factors $\tilde B^{ff}_{n,\,i_1 i_2}$, related to the
selfenergy diagram topology with a fermion-fermion final state, read
\begin{align}
\label{eq:app_ff_self_first}
\tilde B^{ff}_{qqqq,\, VV}(^1P_1) \ =& \
 \frac{\diffmmbar^2}{12}
      \left( \beta^2 + 3 - 12~\mAhat\mBhat - 3~\Delta_{AB}^2 \right) \ ,
\end{align}
for the $^1P_1$ partial-wave configuration, and
\begin{align}
\tilde B^{ff}_{rrrr,\, VV}(^3P_{\cal J}) \ =& \
 -\frac{\diffmmbar^2}{4}
       \left( \beta^2 - 1 + 4~\mAhat\mBhat + \Delta_{AB}^2 \right) \ ,
\\
\tilde B^{ff}_{rrrr,\, VS}(^3P_{\cal J}) \ =& \ \tilde B^{ff}_{rrrr,\, SV}(^3P_{\cal J}) \ = \
 -\frac{\diffmmbar}{2}
       \left( \mAhat - \mBhat - (\mAhat + \mBhat)~\Delta_{AB} \right) ,
\\
\tilde B^{ff}_{rrrr,\, SS}(^3P_{\cal J}) \ =& \
 \frac{1}{4} \left( \beta^2 + 1 - 4~\mAhat\mBhat - \Delta_{AB}^2 \right) \ ,
\\
\tilde B^{ff}_{qqqq,\, VV}(^3P_{\cal J}) \ =& \
 \frac{1}{6} \left( \beta^2 + 3 - 12~\mAhat\mBhat - 3~\Delta_{AB}^2 \right) \ ,
\label{eq:app_ff_self_last}
\end{align}
for the $^3P_{\cal J}$ case. In the case that the $s$-channel
exchanged particles are of the same type ($i_1 i_2 = VV, SS$),
the additional
non-vanishing kinematic factors are related to the expressions 
(\ref{eq:app_ff_self_first})--(\ref{eq:app_ff_self_last}) as
\begin{align}
\tilde B^{ff}_{rqqr,\, i_1 i_2 }(^{2s+1}P_J) \ =& \
 \tilde B^{ff}_{rrrr,\, i_1 i_2 }(^{2s+1}P_J)
 \vert_{\widehat m_A \widehat m_B \to - \widehat m_A \widehat m_B} \ ,
\\
\tilde B^{ff}_{qrrq,\, i_1 i_2 }(^{2s+1}P_J) \ =& \
 \tilde B^{ff}_{qqqq,\, i_1 i_2 }(^{2s+1}P_J)
 \vert_{\widehat m_A \widehat m_B \to - \widehat m_A \widehat m_B} \ ,
\end{align}
where the notation for the replacement rule applies to the term 
$\widehat m_A \widehat m_B$, but all other occurrences of $\widehat m_A$ or $\widehat m_B$
shall be left untouched. Similarly, in case of $s$-channel
particles of different type ($i_1 i_2 = VS, SV$), the additional
non-vanishing $\tilde B^{ff}_{n,\, i_1 i_2}$ terms are given by
\begin{align}
\tilde B^{ff}_{rqqr,\, i_1 i_2 }(^{2s+1}P_J) \ =& \
 -\tilde B^{ff}_{rrrr,\, i_1 i_2 }(^{2s+1}P_J)
 \vert_{\widehat m_A \to - \widehat m_A} \ ,
\\
\tilde B^{ff}_{qrrq,\, i_1 i_2 }(^{2s+1}P_J) \ =& \
 -\tilde B^{ff}_{qqqq,\, i_1 i_2 }(^{2s+1}P_J)
 \vert_{\widehat m_A \to - \widehat m_A} \ .
\end{align}
There are relations among kinematic factors for diagram topologies
$\alpha = 3 (4)$ and diagram topologies $\alpha = 1 (2)$ for both the cases of
box and triangle diagrams, that are given by ($X = V,S$)
\begin{align}
\nonumber
C^{(3)\, ff}_{n, \, i_1 X}(^{2s+1}L_J) \ =& \ 
  C^{(1)\, ff}_{n, \, i_1 X}(^{2s+1}L_J) \
  \vert_{A \leftrightarrow B} \ ,
\\\nonumber
C^{(4)\, ff}_{n, \, i_1 X}(^{2s+1}L_J) \ =& \ 
  C^{(2)\, ff}_{n, \, i_1 X}(^{2s+1}L_J) \
  \vert_{A \leftrightarrow B} \ ,
\\\nonumber
D^{(3)\, ff}_{n, \, i_1 i_2}(^{2s+1}L_J) \ =& \ D^{(1)\, ff}_{n, \, i_1 i_2}(^{2s+1}L_J) \
                             \vert_{A \leftrightarrow B} \ ,
\\
D^{(4)\, ff}_{n, \, i_1 i_2}(^{2s+1}L_J) \ =& \ D^{(2)\, ff}_{n, \, i_1 i_2}(^{2s+1}L_J) \
                             \vert_{A \leftrightarrow B} \ .
\end{align}
Note that these relations are valid among kinematic factors
associated with any $^{2s+1}L_J$ partial wave (in particular also for
$\hat g(^{2s+1}S_s)$ and $\hat h_i(^{2s+1}S_s)$ associated kinematic factors).

The structures $C^{(\alpha)\, ff}_{n, \, i_1 V}(^{2s+1}P_J)$ that we obtain for
diagram topologies $\alpha=1,2$ read
\begin{align}
\nonumber
C^{(\alpha)\, ff}_{qqqq, \, i_1 X}(^1P_1) \ =& \
 \frac{\diffmmbar}{48~\mmbar}
      \Bigl(  6~(\mAhat + \mBhat) \Delta_{AB} - 6~(\mAhat - \mBhat)
 \\\nonumber & \
 \phantom{ \frac{\diffmmbar}{48~\mmbar} \hspace{1em}}
             - \diffmmbar(\beta^2 + 3 - 12~\mAhat\mBhat - 3~\Delta_{AB}^2)
       \Bigr)
 \\ & \
  + \frac{\beta^2 \diffmmbar}{12~P_{i_1 AB}}
         \left( \mAhat - \mBhat - \Delta_{AB} \right) \ ,
\end{align}
and, for the combined $^3P_{\cal J}$ partial-wave reactions,
\begin{align}
\nonumber
C^{(\alpha)\, ff}_{rrrr, \, i_1 X}(^3P_{\cal J}) \ =& \
 \frac{\diffmmbar}{16~\mmbar}
      \Bigl( - 2~(\mAhat + \mBhat) \Delta_{AB} + 2~(\mAhat - \mBhat)
 \\\nonumber & \
 \phantom{ \frac{\diffmmbar}{16~\mmbar} \hspace{1em}}
             + \diffmmbar(\beta^2 - 1 + 4~\mAhat\mBhat + \Delta_{AB}^2)
       \Bigr)
 \\ & \
  - \frac{\beta^2 \diffmmbar}{12~P_{i_1 AB}}
         \left( \mAhat - \mBhat + \Delta_{AB} \right) \ ,
\\\nonumber
C^{(1)\, ff}_{rqrq, \, i_1 X}(^3P_{\cal J}) \ =& \
 \frac{1}{24~\mmbar}
      \Bigl( 6\,(\mAhat + \mBhat - (\mAhat - \mBhat)~\Delta_{AB})~\diffmmbar
 \\\nonumber & \
 \phantom{ \frac{1}{24~\mmbar} \hspace{1em}}
            + \beta^2 + 3 + 12~\mAhat\mBhat - 3~\Delta_{AB}^2
      \Bigr)
 \\ & \
 - \frac{\beta^2}{6~P_{i_1 AB}} \left( 1 - \mAhat + \mBhat \right) \ ,
\\
C^{(2)\, ff}_{qrqr, \, i_1 X}(^3P_{\cal J}) \ =& \ C^{(1)\, ff}_{rqrq, \, i_1 X}(^3P_{\cal J}) \ .
\end{align}
The following relations for the additional non-vanishing
$C^{(\alpha)\,ff}_{n,\, i_1 V}$ hold:
\begin{align}
\nonumber
\tilde C^{(1)\,ff}_{rqrq,\, i_1 V}(^{1}P_1) \ =& \
 \tilde C^{(2)\,ff}_{qrqr,\, i_1 V}(^{1}P_1) \ = \
 -~\tilde C^{(\alpha)\,ff}_{qqqq,\, i_1 V}(^{1}P_1)
 \vert_{m_B \to - m_B} \ ,
\\\nonumber
\tilde C^{(\alpha)\,ff}_{qrrq,\, i_1 V}(^{1}P_1) \ =& \
 \tilde C^{(\alpha)\,ff}_{qqqq,\, i_1 V}(^{1}P_1)
 \vert_{m_A \to - m_A} \ ,
\\\nonumber
\tilde C^{(1)\,ff}_{rrqq,\, i_1 V}(^1P_1) \ =& \
 \tilde C^{(2)\,ff}_{qqrr,\, i_1 V}(^1P_1) \ = \
 -~\tilde C^{(\alpha)\,ff}_{qqqq,\, i_1 V}(^1P_1)
 \vert_{m_{A,B} \to - m_{A,B}} \ ,
\\
\nonumber
\tilde C^{(1)\,ff}_{qqrr,\, i_1 V}(^3P_{\cal J}) \ =& \
 \tilde C^{(2)\,ff}_{rrqq,\, i_1 V}(^3P_{\cal J}) \ = \
 -~\tilde C^{(1)\,ff}_{rrrr,\, i_1 V}(^3P_{\cal J})
 \vert_{m_{A,B} \to - m_{A,B}} \ ,
\\
\tilde C^{(\alpha)\,ff}_{qrrq,\, i_1 V}(^3P_{\cal J}) \ =& \
 -\tilde C^{(1)\,ff}_{rqrq,\, i_1 V}(^3P_{\cal J})
  \vert_{m_{A,B} \to - m_{A,B}} \ .
\end{align}
Turning to the expressions $C^{(\alpha)\,ff}_{n,\, i_1 S}$, we find that all kinematic
factors in case of $^1P_1$ reactions vanish, as it has to be due to total
angular-momentum conservation. 
The non-vanishing kinematic factors in combined $^3P_{\cal J}$ reactions read
($\alpha = 1,2$)
\begin{align}
\nonumber
\tilde C^{(\alpha)\,ff}_{rrrr,\, i_1 S}(^3P_{\cal J}) \ =& \
 \frac{1}{16~\mmbar}
      \Bigl( 2\left(-\mAhat + \mBhat + (\mAhat + \mBhat)~\Delta_{AB}
              \right)\diffmmbar
 \\\nonumber & \
 \phantom{  \frac{1}{16~\mmbar} } \hspace{1em}
             + \beta^2 + 1 - 4~\mAhat\mBhat - \Delta_{AB}^2 \Bigr)
 \\ & \
 - \frac{\beta^2}{12~P_{i_1 AB}} \left( 1 + \mAhat + \mBhat \right) \ ,
\\
\tilde C^{(1)\,ff}_{qqrr,\, i_1 S}(^3P_{\cal J}) \ =& \
 \tilde C^{(2)\,ff}_{rrqq,\, i_1 S}(^3P_{\cal J}) \ = \
 \tilde C^{(1)\,ff}_{rrrr,\, i_1 S}(^3P_{\cal J}) \vert_{m_{A,B} \to - m_{A,B}} \ .
\end{align}
In case of $^1P_1$ partial-wave reactions the kinematic factors
$\tilde D^{(\alpha)\,ff}_{n,\, i_1 i_2}$ for
$\alpha=1,2$ read
\begin{align}
\nonumber
\lefteqn{\tilde D^{(1)\,ff}_{rrrr,\, i_1 i_2}(^1P_1)}&
 \\\nonumber =& \
 \frac{1}{384~(\mmbar)^2}
      \Bigl( \beta^2~(1 + \diffmmbar^2)
             - (3 - 12~\mAhat\mBhat - 3~\Delta_{AB}^2)~(1 - \diffmmbar^2) \Bigr)
 \\\nonumber & \
 + \frac{\beta^2}{48~\mmbar~P_{i_1 AB}}
        \Bigl( 1 + \mAhat + \mBhat
               + \diffmmbar~(\mAhat - \mBhat + \Delta_{AB}) \Bigr)
 \\\nonumber & \
 - \frac{\beta^2}{48~P_{i_1 AB} P_{i_2 BA}}
        \left( 1 + 2~\mAhat + \Delta_{AB} \right)
        \left( 1 + 2~\mBhat - \Delta_{AB} \right)
 \\ & \
 + \sumABij \ ,
\\[2ex]\nonumber
\lefteqn{\tilde D^{(2)\,ff}_{rrrr,\, i_1 i_2}(^1P_1)}&
 \\\nonumber =& \
 \frac{1}{384~(\mmbar)^2}
      \Bigl( \beta^2~(\diffmmbar^2 - 1)
             + (3 - 12~\mAhat\mBhat - 3~\Delta_{AB}^2)~(\diffmmbar^2 + 1)
 \\\nonumber & \ \phantom{  \frac{1}{384~(\mmbar)^2} \Bigl( } \hspace{1em}
             - 12~\diffmmbar~(\mAhat-\mBhat - (\mAhat + \mBhat)~\Delta_{AB})
      \Bigr)
 \\\nonumber & \
 - \frac{\beta^2}{48~\mmbar~P_{i_1 AB}}
        \Bigl( 1+\mAhat+\mBhat -\diffmmbar~(\mAhat - \mBhat +\Delta_{AB})\Bigr)
 \\ & \
 + \frac{\beta^2}{48~P_{i_1 AB} P_{i_2 AB}}
        \left( 1 + 2~\mAhat + \Delta_{AB} \right)
        \left( 1 + 2~\mBhat - \Delta_{AB} \right)
 + \sumij \ ,
\\[2ex]\nonumber
\lefteqn{\tilde D^{(1)\,ff}_{rrqq,\, i_1 i_2}(^1P_1)}&
 \\\nonumber =& \
 \frac{1}{192~(\mmbar)^2}
      \Bigl( \beta^2 - 3 + 3~\Delta_{AB}^2 
             + 12~\diffmmbar \left( \mAhat-\mBhat -(\mAhat + \mBhat)~\Delta_{AB}
                             \right)
 \\\nonumber & \ \phantom{ \frac{1}{384~(\mmbar)^2} \Bigl( }
              - \diffmmbar^2~(\beta^2  + 3 + 24~\mAhat\mBhat - 3~\Delta_{AB}^2) 
       \Bigr)
 \\\nonumber & \
 + \frac{\beta^2}{48~\mmbar~P_{i_1 AB}}
        \Bigl( 1 + \mAhat + \mBhat - \diffmmbar~(\mAhat - \mBhat + \Delta_{AB})
        \Bigr)
 \\\nonumber & \
 + \frac{\beta^2}{48~\mmbar~P_{i_2 BA}}
        \Bigl( 1 - \mAhat - \mBhat - \diffmmbar~(\mAhat - \mBhat - \Delta_{AB})
        \Bigr)
 \\ & \
 - \frac{\beta^2}{24~P_{i_1 AB} P_{i_2 BA}} \left( \beta^2 - 4~\mAhat\mBhat \right)
 \ ,
\\[2ex]\nonumber
\lefteqn{\tilde D^{(2)\,ff}_{rrqq,\, i_1 i_2}(^1P_1)}&
 \\\nonumber =& \
 - \frac{1}{192~(\mmbar)^2}
      \Bigl( \beta^2 - 3 + 3~\Delta_{AB}^2
            + \diffmmbar^2~(\beta^2 + 3+24~\mAhat\mBhat - 3~\Delta_{AB}^2) \Bigr)
 \\\nonumber & \
 - \frac{\beta^2}{48~\mmbar~P_{i_1 AB}}
        \Bigl( 1 + \mAhat + \mBhat + \diffmmbar~(\mAhat - \mBhat + \Delta_{AB})
        \Bigr)
 \\\nonumber & \
 - \frac{\beta^2}{48~\mmbar~P_{i_2 AB}}
        \Bigl( 1 - \mAhat - \mBhat - \diffmmbar~(\mAhat - \mBhat - \Delta_{AB})
        \Bigr)
 \\ & \
 + \frac{\beta^2}{24~P_{i_1 AB} P_{i_2 AB}} \left( \beta^2 - 4~\mAhat\mBhat \right) \ .
\end{align}
The corresponding expressions for combined $^3P_{\cal J}$ partial-wave
reactions are
\begin{align}
\tilde D^{(1)\,ff}_{rrrr,\, i_1 i_2}(^3P_{\cal J}) \ =& \
 -3~\tilde D^{(1)\,ff}_{rrrr,\, i_1 i_2}(^1P_1)
 + \frac{\beta^2}{24~(\mmbar)^2}~( 1 + \diffmmbar^2) \ ,
\\
\tilde D^{(2)\,ff}_{rrrr,\, i_1 i_2}(^3P_{\cal J}) \ =& \
 \phantom{-} 3~\tilde D^{(2)\,ff}_{rrrr,\, i_1 i_2}(^1P_1)
 + \frac{\beta^2}{24~(\mmbar)^2}~( 1 - \diffmmbar^2) \ ,
\\
\tilde D^{(1)\,ff}_{rrqq,\, i_1 i_2}(^3P_{\cal J}) \ =& \
 \tilde D^{(1)\,ff}_{rrqq,\, i_1 i_2}(^1P_1)
 - \frac{\mAhat\mBhat}{2~(\mmbar)^2}
 - \frac{\beta^2}{3~P_{i_1 AB} P_{i_2 BA}}~2~\mAhat\mBhat \ ,
\\
\tilde D^{(2)\,ff}_{rrqq,\, i_1 i_2}(^3P_{\cal J}) \ =& \
 - \tilde D^{(2)\,ff}_{rrqq,\, i_1 i_2}(^1P_1)
 - \frac{\mAhat\mBhat}{2~(\mmbar)^2}
 - \frac{\beta^2}{3~P_{i_1 AB}P_{i_2 AB}}~2~\mAhat\mBhat \ .
\end{align}
The following relations can be used to obtain the remaining non-vanishing
$\tilde D^{(\alpha)\,ff}_{n,\, i_1 i_2}$ expressions in case of diagram topology
$\alpha = 1$. Note that they hold for any $^{2s+1}L_J$ partial-wave
configuration.
\begin{align}
\nonumber
\tilde D^{(1)\,ff}_{qqqq,\, i_1 i_2}(^{2s+1}L_J) \ =& \
 \tilde D^{(1)\,ff}_{rrrr,\, i_1 i_2}(^{2s+1}L_J) \vert_{m_{A,B} \to - m_{A,B}} \ ,
\\\nonumber
\tilde D^{(1)\,ff}_{qqrr,\, i_1 i_2}(^{2s+1}L_J) \ =& \
 \tilde D^{(1)\,ff}_{rrqq,\, i_1 i_2}(^{2s+1}L_J) \vert_{m_{A,B} \to - m_{A,B}} \ ,
\\\nonumber
\tilde D^{(1)\,ff}_{rqqr,\, i_1 i_2}(^{2s+1}L_J) \ =& \
 \tilde D^{(1)\,ff}_{rrqq,\, i_1 i_2}(^{2s+1}L_J) \vert_{m_A \to - m_A} \ ,
\\\nonumber
\tilde D^{(1)\,ff}_{qrrq,\, i_1 i_2}(^{2s+1}L_J) \ =& \
 \tilde D^{(1)\,ff}_{rrqq,\, i_1 i_2}(^{2s+1}L_J) \vert_{m_B \to - m_B} \ ,
\\\nonumber
\tilde D^{(1)\,ff}_{rqrq,\, i_1 i_2}(^{2s+1}L_J) \ =& \
 \tilde D^{(1)\,ff}_{rrrr,\, i_1 i_2}(^{2s+1}L_J) \vert_{m_A \to - m_A} \ ,
\\
\tilde D^{(1)\,ff}_{qrqr,\, i_1 i_2}(^{2s+1}L_J) \ =& \
 \tilde D^{(1)\,ff}_{rrrr,\, i_1 i_2}(^{2s+1}L_J) \vert_{m_B \to - m_B} \ .
\end{align}
In case of diagram topology $\alpha = 2$ analogous relations exist:
\begin{align}
\nonumber
\tilde D^{(2)\,ff}_{qqqq,\, i_1 i_2}(^{2s+1}L_J) \ =& \
 \tilde D^{(2)\,ff}_{rrrr,\, i_1 i_2}(^{2s+1}L_J) \vert_{m_{A,B} \to - m_{A,B}} \ ,
\\\nonumber
\tilde D^{(2)\,ff}_{qqrr,\, i_1 i_2}(^{2s+1}L_J) \ =& \
 \tilde D^{(2)\,ff}_{rrqq,\, i_1 i_2}(^{2s+1}L_J) \vert_{m_{A,B} \to - m_{A,B}} \ ,
\\\nonumber
\tilde D^{(2)\,ff}_{rqrq,\, i_1 i_2}(^{2s+1}L_J) \ =& \
 \tilde D^{(2)\,ff}_{rrqq,\, i_1 i_2}(^{2s+1}L_J) \vert_{m_A \to - m_A} \ ,
\\\nonumber
\tilde D^{(2)\,ff}_{qrqr,\, i_1 i_2}(^{2s+1}L_J) \ =& \
 \tilde D^{(2)\,ff}_{rrqq,\, i_1 i_2}(^{2s+1}L_J) \vert_{m_B \to - m_B} \ ,
\\\nonumber
\tilde D^{(2)\,ff}_{rqqr,\, i_1 i_2}(^{2s+1}L_J) \ =& \
 \tilde D^{(2)\,ff}_{rrrr,\, i_1 i_2}(^{2s+1}L_J) \vert_{m_A \to - m_A} \ ,
\\
\tilde D^{(2)\,ff}_{qrrq,\, i_1 i_2}(^{2s+1}L_J) \ =& \
 \tilde D^{(2)\,ff}_{rrrr,\, i_1 i_2}(^{2s+1}L_J) \vert_{m_B \to - m_B} \ .
\end{align}

\subsubsection{$P$-wave kinematic factors for $X_A X_B = \eta\overline\eta$}

The use of Feynman gauge for our computation of the absorptive parts of the
Wilson coefficients requires to consider unphysical particles in the final states, such as 
pseudo-Goldstone Higgs bosons and ghosts. While the results for final states with pseudo-Goldstone Higgses 
can be obtained from the $VS$ and $SS$ kinematic factors and corresponding coupling structures,
the ghosts constitute a different class ($\eta\bar{\eta}$). 
In order to properly construct the coupling factors that go along the
kinematic factors $\tilde B^{\eta\overline\eta}_{n,\,i_1 i_2}$ 
presented below, we refer
the reader to the rules set up in section A.3.5 of paper I.

For $^1P_1$ partial-wave processes, there is only one 
non-vanishing kinematic factor with ghosts in the final state:
\begin{align}
\tilde B^{\eta\overline\eta}_{qq,VV}(^1P_1) \ =& \
 -\frac{~\beta^2}{48}~\diffmmbar^2 \ .
\end{align}
The corresponding kinematic factors in combined $^3P_{\cal J}$ partial-wave processes
read
\begin{align}
\tilde B^{\eta\overline\eta}_{rr,VV}(^3P_{\cal J}) \ =& \
 \frac{\diffmmbar^2}{16}~(1 - \Delta_{AB}^2) \ ,
\\
\tilde B^{\eta\overline\eta}_{qq,VV}(^3P_{\cal J}) \ =& \ -\frac{\beta^2}{24} \ ,
\\
\tilde B^{\eta\overline\eta}_{rr,VS}(^3P_{\cal J}) \ =& \
 \frac{\widehat{m}_W}{8}~\diffmmbar~(1+\Delta_{AB}) \ ,
\\
\tilde B^{\eta\overline\eta}_{rr,SV}(^3P_{\cal J}) \ =& \
 - \frac{\widehat{m}_W}{8}~\diffmmbar~(1-\Delta_{AB}) \ ,
\\
\tilde B^{\eta\overline\eta}_{rr,SS}(^3P_{\cal J}) \ =& \ -\frac{\widehat{m}_W^2}{4} \ .
\end{align}

\section{Notation for the kinematic factors in the electronic supplement}
\label{sec:appendixsuppl}
The analytic expressions for the kinematic factors needed to construct the
absorptive part of the  Wilson coefficients up to next-to-next-to-leading order 
have been stored in the {\tt Mathematica} package attached to this paper.
They can be loaded into a {\tt Mathematica} session using the command
\vspace{0.2cm}\\
\fbox{\parbox{0.98\textwidth}{
\includegraphics[scale=1.1]{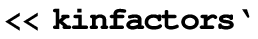}
}
}\vspace{0.5cm}\\
The introductory text in the file explains in detail the notation used for the
kinematic factors, which we summarise in Tab.~\ref{tab:efilenotation}.
\begin{table}[h!]
\renewcommand{\arraystretch}{2.}
\centering
\footnotesize
\begin{tabular}{| l | l | l |}
\hline
   Kinematic factor
 & Name in electronic supplement & Coupling string {\tt n}
 \\[1ex]
\hline
\multirow{2}{*} {$\tilde B^{\, X_A  X_B}_{n, \, i_1 i_2}(^{2s+1}L_J)$}
 &  {\tt Btilde["(i1i2)XAXB", n, \{2s+1\}\^{}L\_J]}  & {\scriptsize {\tt "rr","qq"} $(X_AX_B=VV,VS,SS,GG)$ } \\
&  {\tt i1,i2~=~V,S}  & \!\!\!\! {\scriptsize $\left. 
					  \begin{array}{l}
					  \text{{\tt "rrrr","rqqr",}} \\
					  \text{{\tt "qrrq","qqqq"}}
					  \end{array}
				    \right\}  (X_AX_B=ff)$ }
  \\[3ex]
\hline
\multirow{2}{*} {  $\tilde C^{(\alpha)\, X_A  X_B}_{n, \, i_1 X}(^{2s+1}L_J)$}
 & {\tt Ctilde[alpha,"(X)XAXB", n, \{2s+1\}\^{}L\_J]} & \!\!\!\! {\scriptsize $\left. 
					  \begin{array}{l}
					  \text{{\tt "rrr","qqr",}} \\
					  \text{{\tt "rqq","qrq"}}
					  \end{array}
				    \right\}  (X_AX_B=VV,VS,SS)$ } \\
&  {\tt X~=~V,S}  & \!\!\!\! {\scriptsize $\left. 
					  \begin{array}{l}
					  \text{{\tt "rrrr","qqqq","rrqq",}} \\
					  \text{{\tt "qqrr","rqqr","qrrq",}} \\
					  \text{{\tt "rqrq","qrqr"}}
					  \end{array}
				    \right\}  (X_AX_B=ff)$ }
  \\[2ex]
\hline
  $\tilde D^{(\alpha)\, X_A  X_B}_{n, \, i_1 i_2}(^{2s+1}L_J)$
 & {\tt Dtilde[alpha,"XAXB", n, \{2s+1\}\^{}L\_J]} &  \!\!\!\! {\scriptsize $
					  \begin{array}{l}
					  \text{{\tt "rrrr","qqqq","rrqq","qqrr",}} \\
					  \text{{\tt "rqqr","qrrq","rqrq","qrqr"}}
					  \end{array} $ }
  \\[3ex]
\hline
\end{tabular}
\caption{Notation for the kinematic factors used in the {\tt Mathematica}
         package}
\label{tab:efilenotation}
\end{table}

\noindent
The argument {\tt XAXB} inside the kinematic factors in
Tab.~\ref{tab:efilenotation} can be given the values
\begin{center}
{\tt XAXB\,=\,VV,VS,SS,ff,GG}
\end{center}
depending on the type of particles in the final state. The partial-wave configuration
{\tt \{2s+1\}\^{}L\_J} is specified by one of the following strings:
$$ 
{\small
\text{{\tt \{2s+1\}\^{}L\_J}} \ = \ \left\{
\begin{array}{l}
 \text{{\tt "1S0","3S1",} for the leading-order $S$-wave coefficients} \\
 \text{{\tt "1P1","3P0","3P1","3P2","3PJ",} for the $P$-wave coefficients} \\
 \text{{\tt "1S0,p2","3S1,p2",} for the $g\left( ^{2s+1}S_s \right)$ coefficients }  \\
 \text{{\tt "1S0,dm","1S0,dmbar","3S1,dm","3S1,dmbar",}  for $h_{i}\left( ^{2s+1}S_s \right)$ coefficients.}
\end{array}
\right.
}
$$
\noindent 
Note that for $S$-wave partial-wave configurations, the label {\tt \{2s+1\}\^{}L\_J} 
also contains the information about the type of Wilson coefficient ($f$, $g$ or $h_i$,
with $g$ and $h_i$ describing NNLO $S$-wave coefficients, see Eq.~(\ref{eq:basisSPwave})).
The argument {\tt alpha} inside the kinematic factors {\tt Ctilde} and
{\tt Dtilde} in Tab.~\ref{tab:efilenotation} can get as input
\begin{center}
{\tt alpha\,=\,1,2,3,4}
\end{center}
referring to our enumeration scheme for the respective four triangle and
box topologies.
Finally, the equivalence between the mass variables and propagator structures
introduced in Appendix~\ref{sec:appendix} that enter the expressions for the
kinematic factors and the corresponding names in the  {\tt Mathematica} package
are collected in Tab.~\ref{tab:massesefile}.
\begin{table}[t!]
\renewcommand{\arraystretch}{1.5}
\centering
\normalsize
\begin{tabular}{| c | c | }
\hline
   Quantity
 & Name in electronic supplement
  \\[1ex]
\hline
$\rescm m_{i_1},\, \rescm m_{i_2}$
 &  {\tt mi1}, {\tt mi2}  \\[1ex]
 \hline
$\rescm m_{A},\, \rescm m_{B}$
 &  {\tt mA}, {\tt mB}  \\[1ex]
 \hline
 $\rescm m_{W}$
 &  {\tt mWr}  \\[1ex]
 \hline
$ \Delta_m$
 &  {\tt Dm}  \\[1ex]
 \hline
$\Delta_{AB}$
 &  {\tt DAB}  \\[1ex]
 \hline
 $P_{i_1}^s,\,P_{i_2}^s$
 &  {\tt Psi1}, {\tt Psi2}
 \\[1ex]
 \hline
 $P_X^s$
 &  {\tt PsX} 
 \\[1ex]
 \hline
$P_{i_1 AB},\,P_{i_2 AB}$
 &  {\tt Pti1[A,B]}, {\tt Pti2[A,B]}  \\[1ex]
\hline
\end{tabular}
\caption{Equivalence between the variables in the kinematic factors introduced
         in Appendix~\ref{sec:appendix} and the corresponding names in the
         {\tt Mathematica} package.}
\label{tab:massesefile}
\end{table}

\section{Annihilation rates in the pure-wino NRMSSM at \boldmath ${\cal O}(v_{\rm rel}^2)$}
\label{sec:appendixexample}
In this appendix we illustrate the usage of the kinematic and coupling
factor results given in paper I and in this work by presenting a detailed
end-to-end calculation of the non-relativistic annihilation
cross section for the $\chi^+_1 \chi^-_1 \to W^+ W^-$ reaction including up
to  ${\cal O}(v_{\rm rel}^2)$ effects. The calculation
is performed in the idealised case of the pure-wino NRMSSM, which allows to
present compact analytic results. For completeness, we also provide 
in Sec.~\ref{sec:app_gammaspurewino} the results for the Wilson 
coefficients needed to determine all exclusive (off-)diagonal (co-)annihilation
rates $\chi_{e_1}\chi_{e_2} \to X_A X_B \to \chi_{e_4} \chi_{e_3}$ 
in the  decoupling limit of the pure-wino scenario. 
To the best of our knowledge the analytic results for the $P$- and
$\mathcal O(v_\text{rel}^2)$ $S$-wave (off-)diagonal annihilation rates in the
pure-wino NRMSSM have not been given before in the literature and could be of
interest in the study of next-to-next-to-leading order effects in
Sommerfeld-enhanced pure-wino dark matter annihilations in the Early Universe.

The pure-wino (toy-)NRMSSM scenario is characterised by the mass-degenerate
$SU(2)_L$ fermion triplet states $\chi^0_1, \chi^\pm_1$ (winos) with mass scale
$M_2 > 0$, where the latter denotes the soft SUSY-breaking wino mass. All other 
SUSY mass-parameters including the Bino soft mass $M_1$ and the Higgsino mass
parameter $\mu$ as well as all sfermion mass parameters are assumed to be much
larger than $M_2$, namely $M_1,\vert\mu\vert \gg M_2$. Consequently all
heavier states $\chi^0_i$, $i=2,3,4$ and $\chi^\pm_2$ as well as all sfermion
states are treated as completely decoupled. According to the $SU(2)_L$ symmetric
limit the $SU(2)_L$ gauge bosons as well as all Standard Model fermions
are treated as massless, in agreement with the complete mass-degeneracy between
the non-relativistic states $\chi^0_1$ and $\chi^\pm_1$.
The neutralino and chargino mixing matrix entries relevant to the calculation in
the pure-wino NRMSSM read
\begin{align}
 \tilde Z_{N\, i1} \ = \ \delta_{i2} \, ,
& &
 \tilde Z_{\pm\, i1} \ = \ \delta_{i1} \, ,
\end{align}
where the $\tilde Z_N, \tilde Z_\pm$ derive from the conventionally defined
neutralino and chargino mixing matrices $Z_N, Z_\pm$ by accounting for a
potentially necessary rotation to positive mass-parameters in the NRMSSM,
as defined through Eqs.~(38--41) in paper I. Such a rotation
does however not affect the above mixing-matrix entries relevant in the
pure-wino NRMSSM with $M_2 > 0$. Finally, let us introduce the notation
$ m_\chi = M_2$ for the only mass parameter present in the pure-wino NRMSSM
scenario.

Our goal is the determination of the Wilson coefficients $\hat f(^{2s+1}L_J)$
that enter the coefficients $a$ and $b$ in the non-relativistic expansion of
the $\chi^+_1 \chi^-_1 \to W^+ W^-$ annihilation cross section, see
(\ref{eq:res_sigmavrel}). The $\hat f(^{2s+1}L_J)$ are determined from coupling
and kinematic factors using the master formula
(\ref{eq:genericstructureWilson}). We discuss the construction of the relevant
coupling factors in Sec.~\ref{sec:app_couplingspurewino}. The corresponding
kinematic factors, the resulting absorptive part of the Wilson coefficients as
well as the final result for the $\chi^+_1 \chi^-_1 \to W^+ W^-$ annihilation
cross section in the non-relativistic regime are given in
Sec.~\ref{sec:app_kinfactorspurewino}.
The co-annihilation rates into all other exclusive final states in this scenario
including $P$- and next-to-next-to-leading order
$S$-wave corrections are obtained from the contributions to the Wilson coefficients
from these exclusive final states
that we present in Sec.~\ref{sec:app_gammaspurewino}.

\subsection{Coupling factors}
\label{sec:app_couplingspurewino}
Let us recall from paper I that  each of the coupling factors
$b_n, c^{(\alpha)}_n, d^{(\alpha)}_n$ in (\ref{eq:genericstructureWilson}) related to
a specific $\chi_{e_1} \chi_{e_2} \to X_A X_B \to \chi_{e_4} \chi_{e_3}$ reaction is 
given by a product of two coupling factors associated with the two vertices
occurring in the tree-level annihilation amplitude
$\mathcal A^{(0)}_{\chi_{e_1} \chi_{e_2}\to X_A X_B}$ and the complex conjugate of
another such two-coupling factor product related to the tree-level amplitude
$\mathcal A^{(0)}_{\chi_{e_4} \chi_{e_3} \to X_A X_B}$.
Hence, the building blocks of the $b_n, c^{(\alpha)}_n, d^{(\alpha)}_n$ relevant
in $\chi^+_1 \chi^-_1 \to X_A X_B$ annihilation rates are 
given by the (axial-) vector or (pseudo-) scalar vertex factors in the
$\chi^+_1 \chi^-_1 \to X_A X_B$ tree-level annihilation amplitudes.
Since our results for the kinematic factors refer to Feynman
gauge, in order to determine the annihilation rates into a physical $W^+ W^-$
final state we have to consider $\chi^+_1 \chi^-_1$ 
annihilations into the exclusive final states
$X_A X_B = W^+ W^-, W^+ G^-, W^- G^+$,
$G^+ G^-, \eta^+ \overline\eta^+,\eta^- \overline\eta^-$, with 
$G^\pm$ the charged pseudo-Goldstone Higgs and $\eta^\pm$ the charged ghost particles.
In the pure-wino NRMSSM, the only non-vanishing amplitudes are given by
the diagrams depicted in Fig.~\ref{fig:treelevelannihilation_WWfinalstate},
which we should compare with the generic $\chi \chi \to X_A X_B$
diagrams drawn in Fig.~9 of paper I in order to extract the coupling factors
in accordance to the conventions established therein. 
Note the fermion flow in these diagrams, which coincides with the convention 
used in the generic $\chi_{e_1}\chi_{e_2} \to X_A X_B$ diagrams in Fig.~9 of
paper I. In the case of diagram $t_2$
in Fig.~\ref{fig:treelevelannihilation_WWfinalstate}, which contributes both
to the box and triangle coupling factors, $d^{(\alpha)}_{n,\,i_1 i_2}$ and
$c^{(\alpha)}_{n,\,i_1 i_2}$, the vertex factors $V^{\rho (t_2)}_{ei}$ read
\begin{align}
 V^{\mu (t_2)}_{e_1 i_1}  \ =& \
       \gamma^\mu \left( v^{W *}_{e_1 i_1} + a^{W *}_{e_1 i_1} \gamma_5 \right) \ ,
&
 V^{\nu (t_2)}_{e_2 i_1}  \ =& \
       \gamma^\nu \left( v^{W}_{e_2 i_1} + a^{W}_{e_2 i_1} \gamma_5 \right) \ ,
\end{align}
where $e_1, e_2 = 1$ as these indices refer to the external states
$\chi_{e_1} = \chi^+_1$ and $\chi_{e_2} = \chi^-_1$. In the pure-wino NRMSSM, the
only possible $t$-channel exchanged particle in diagram $t_2$ is the $\chi^0_1$,
therefore $i_1 = 1$. Comparing to the generic form of the vertex factor
$V^{\rho (d)}_{ei} = \gamma^\rho (r^{(d)}_{ei} + q^{(d)}_{ei} \gamma_5)$, we 
identify the expressions that substitute the respective place-holder couplings
$r^{(d)}_{ei}$ and $q^{(d)}_{ei}$:
\begin{figure}[t!]
\begin{center}
\includegraphics[width=0.99\textwidth]{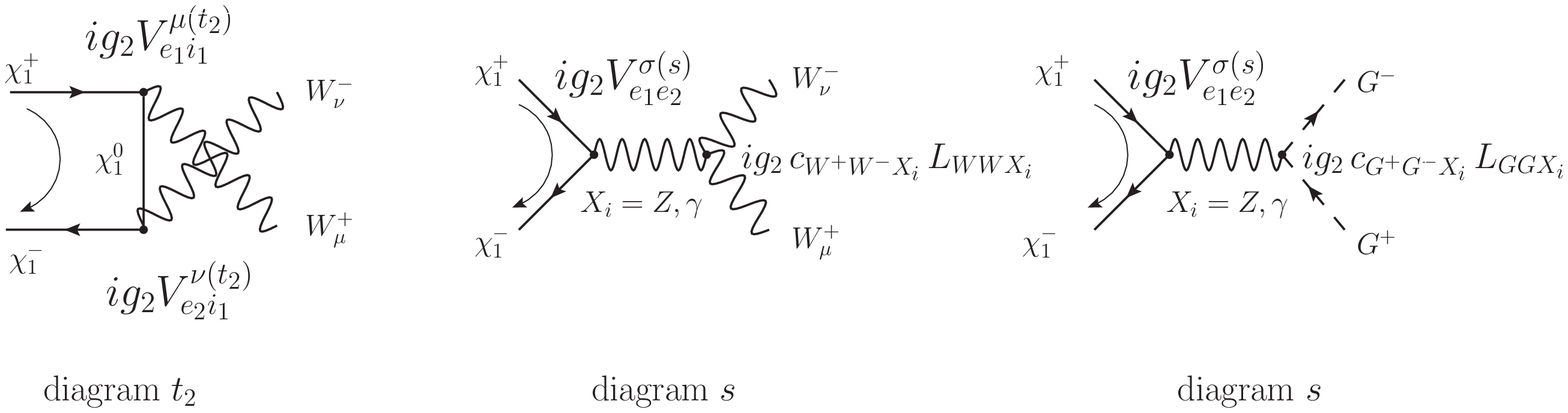}\\[2em]
\includegraphics[width=0.38\textwidth]{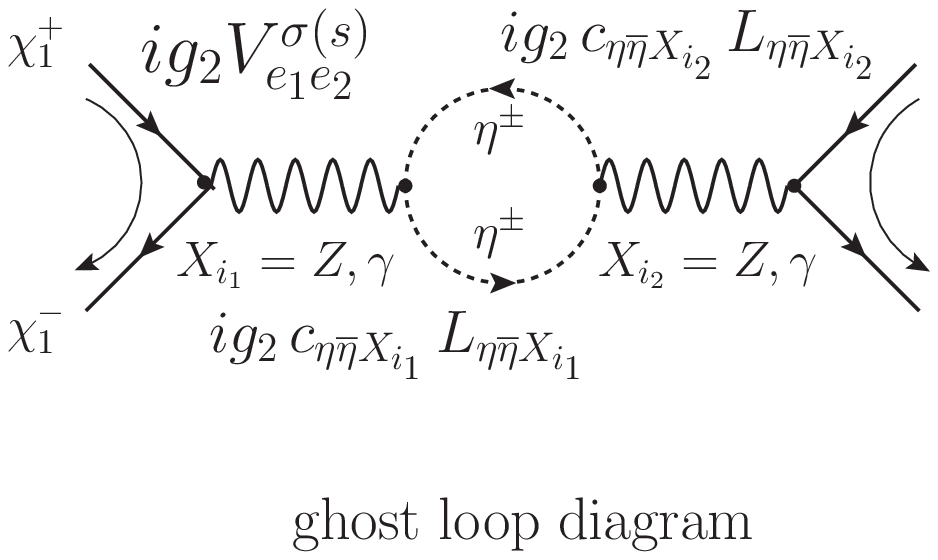}
\caption{ Amplitudes contributing to the physical $\chi^+_1\chi^-_1 \to W^+ W^-$
          annihilation reaction in Feynman gauge.
          Note the fermion flow, that has been fixed to match with the conventions
          established in paper I.}
\label{fig:treelevelannihilation_WWfinalstate}
\end{center}
\end{figure}
\begin{align}
\Bigl( 
 \lbrace r^{(t_2)}_{e_1 i_1} ,\, q^{(t_2)}_{e_1 i_1} \rbrace
 ,\,
 \lbrace r^{(t_2)}_{e_2 i_1} ,\, q^{(t_2)}_{e_2 i_1} \rbrace
\Bigr)
\ \rightarrow \
\Bigl(
 \lbrace v^{W *}_{11} ,\, a^{W *}_{11} \rbrace
 ,\,
 \lbrace v^{W}_{11} ,\, \ a^{W}_{11} \rbrace
\Bigr)\ .
\end{align}
Let us obtain first the coupling factors $d^{(\alpha)}_{n,\,i_1 i_2}$ related 
to the four box amplitudes shown in Fig.~\ref{fig:genericamplitudes}.
As there is no $t$-channel exchange diagram $t_1$, the only non-vanishing
coupling factors $d^{(\alpha)}_{n,\,i_1 i_2}$ are those with label $\alpha = 4$:
$d^{(4)}_{n,\,i_1 i_2}$ expressions arise from the product of coupling factors in
$\chi_{e_1}\chi_{e_2} \to X_A X_B$ annihilation diagrams of type $t_2$ with the
complex conjugate of the coupling factors associated with
$\chi_{e_4}\chi_{e_3} \to X_A X_B$ annihilation via diagram type
$t_2$.\footnote{For further conventions on the enumeration label $\alpha$
see Fig.~\ref{fig:genericamplitudes}.}
The constituent coupling factors for the $d^{(4)}_{n,\, i_1 i_2}$ in
$\chi^+_1 \chi^-_1 \to W^+ W^- \to \chi^+_1 \chi^-_1$ scattering are collected in
the following table:
\begin{align}
\nonumber
 \alpha = 4 \ : \hspace{1cm} \ \
 &\Bigl(
  \lbrace r^{(t_2)}_{e_1 i_1}, q^{(t_2)}_{e_1 i_1} \rbrace, \
  \lbrace r^{(t_2)}_{e_2 i_1}, q^{(t_2)}_{e_2 i_1} \rbrace, \
  \lbrace r^{(t_2) *}_{e_3 i_2}, q^{(t_2) *}_{e_3 i_2} \rbrace, \
  \lbrace r^{(t_2) *}_{e_4 i_2}, q^{(t_2) *}_{e_4 i_2} \rbrace
 \Bigr)
 \\
 \rightarrow \ \
 &\Bigl(
  \lbrace v^{W *}_{11}, a^{W *}_{11} \rbrace, \
  \lbrace v^{W}_{11}, a^{W}_{11} \rbrace, \
  \lbrace v^{W *}_{11}, a^{W *}_{11} \rbrace, \
  \lbrace v^{W}_{11}, a^{W}_{11} \rbrace
 \Bigr) \ .
\label{eq:app_d4couplings}
\end{align}
Selecting one element from each of the four subsets and multiplying these
selected elements with each other gives rise to the $d^{(4)}_{n,\,i_1 i_2}$. The
label $n$ denotes a string of four characters, that indicates which coupling
(type $r$ or $q$) was selected from the $i$th subset in
(\ref{eq:app_d4couplings}). For instance
\begin{align}
 d^{(4)\, \chi^+_1 \chi^-_1 \to W^+ W^- \to \chi^+_1 \chi^-_1}_{rrrr,\, 11} 
 = \ v^{W *}_{11} v^{W}_{11} v^{W *}_{11} v^{W}_{11} \ .
\end{align}
Turning to the coupling factors in triangle and selfenergy amplitudes,
$c^{(\alpha)}_{n,\, i_1 i_2}$ and $b^{}_{n,\, i_1 i_2}$, they receive contributions 
from the $s$-channel diagrams in Fig.~\ref{fig:treelevelannihilation_WWfinalstate}.
We proceed in a similar way as done for the diagram $t_2$ and identify the
following coupling factors for the case of single $s$-channel $Z$-exchange
(first line) and single $s$-channel $\gamma$-exchange (second line):
\begin{align}\nonumber
 V^{\sigma(s)}_{11} \ &= \
    \gamma^\sigma \left( v^Z_{11} + a^Z_{11} \gamma_5 \right) \, ,
&
 c_{W^+W^-Z} \ &= \ c_W \, ,
\\
 V^{\sigma(s)}_{11} \ &= \
    \gamma^\sigma \left( v^\gamma_{11} + a^\gamma_{11} \gamma_5 \right) \, ,
&
 c_{W^+W^-\gamma} \ &= \ s_W \, .
\end{align}
The building blocks for the $b^{}_{n,\, i_1 i_2},c^{(\alpha)}_{n,\, i_1 i_2}$ and finally
these expressions themselves can now be obtained in a similar manner
as described for the $d^{(\alpha)}_{n,\, i_1 i_2}$ expressions. However, before
proceeding with their explicit construction, significant
simplifications can be performed by noting that the pure-wino NRMSSM exhibits a 
particularly simple coupling structure: the (axial-)vector couplings of the
$\chi^0_1$ and $\chi^\pm_1$ to the Standard Model gauge bosons are given by
\begin{align}\nonumber
 v^W_{11} \ &= \ 1\, , \ \
 a^W_{11} \ = \ 0 \ ,
&
 v^\gamma_{11} \ &= \ -s_W\, , \ \
 a^\gamma_{11} \ = \ 0 \ ,
\\
 v^Z_{11} \ &= \ -c_W\, , \ \
 a^Z_{11} \ = \ 0 \ .
\end{align}
With the vanishing of all axial-vector couplings the only non-vanishing
coupling factor $d^{(\alpha)}_{n,\, i_1 i_2}$ for
$\chi^+_1 \chi^-_1 \to W^+ W^- \to \chi^+_1 \chi^-_1$ in the pure-wino NRMSSM
hence reads
\begin{align}
 d^{(4)\,\chi^+_1 \chi^-_1 \to W^+ W^- \to \chi^+_1 \chi^-_1}_{rrrr,\, 11} \ = \ 1 \, .
\label{d4rrrr}
\end{align}
The absence of a $t$-channel exchange diagram $t_1$ implies, that only
$c^{(\alpha)}_{n,\, i_1 i_2}$ factors with $\alpha = 3, 4$ can be non-vanishing, as
these are built from vertex coupling factors associated with diagram type $t_2$ 
and diagram type $s$, see Fig.~\ref{fig:genericamplitudes}.
In the pure-wino NRMSSM, we find the following expressions
\begin{align}
 c^{(\alpha = 3,4)\, \chi^+_1\chi^-_1 \to W^+ W^- \to \chi^+_1\chi^-_1}_{rrr,\,1Z}\ &=\
 - c_W^2 \, ,
&
 c^{(\alpha = 3,4)\, \chi^+_1\chi^-_1 \to W^+ W^- \to \chi^+_1\chi^-_1}_{rrr,\,1\gamma}\ &=\
 - s_W^2 \, ,
\end{align}
and all other $c^{(\alpha)}_{n,\,1 i_2}$ vanish. Finally, the
non-zero factors $b^{}_{n,\, i_1 i_2}$ read
\begin{align}
 b^{}_{rr,\, ZZ} \ &= \ c_W^4 \, ,
&
  b^{}_{rr,\, Z\gamma} \ = \ b^{}_{rr,\, \gamma Z} \ &= \ c_W^2 s_W^2 \, ,
&
 b^{}_{rr,\, \gamma\gamma} \ &= \ s_W^4 \, ,
\end{align}
where we have suppressed the superscript
$\chi^+_1\chi^-_1 \to W^+ W^- \to \chi^+_1\chi^-_1$ to shorten the notation.
A similar procedure leads to the coupling factors in
$\chi^+_1\chi^-_1\to X_A X_B$ rates with the (unphysical) final states
$X_A X_B = G^+ G^-, \eta^+ \overline\eta^+$ and $\eta^-\overline\eta^-$. We quote
the non-vanishing results for the coupling factors related to
$\chi^+_1 \chi^-_1 \to G^+ G^- \to \chi^+_1 \chi^-_1$ reactions:
\begin{align}\nonumber
 b^{}_{rr,\,ZZ} \ &= \ \frac{1}{4}~\left( c_W^2 - s_W^2 \right)^2 \, ,
&
 b^{}_{rr,\,Z\gamma} \ = \  b^{}_{rr,\,\gamma Z}
\ &= \ \frac{s_W^2}{2}~\left( c_W^2 - s_W^2 \right) \, ,
\\
 b^{}_{rr,\, \gamma\gamma} \ &= \ s_W^4 \, .
\end{align}
In case of $\chi^+_1 \chi^-_1 \to \eta^+ \overline\eta^+ \to \chi^+_1 \chi^-_1$
and $\chi^+_1 \chi^-_1 \to \eta^- \overline\eta^- \to \chi^+_1 \chi^-_1$
reactions we find in both cases the same result
(again suppressing the process-specifying superscripts):
\begin{align}
 b^{}_{rr,\, ZZ} \ &= \ c_W^4 \, ,
&
  b^{}_{rr,\, Z\gamma} \ = \ b^{}_{rr,\, \gamma Z} \ &= \ c_W^2 s_W^2 \, ,
&
 b^{}_{rr,\, \gamma\gamma} \ &= \ s_W^4 \, .
\end{align}

\subsection{Kinematic factors}
\label{sec:app_kinfactorspurewino}
As for the coupling factors, the kinematic factors
$B_{n,\, i_1 i_2}, C^{(\alpha)}_{n,\, i_1 X}, D^{(\alpha)}_{n,\, i_1 i_2}$
reduce to very simple expressions in the pure-wino NRMSSM.
As the pure-wino NRMSSM refers to the limit of vanishing $SU(2)_L$ gauge boson
masses, the relevant (mass-)parameters in any of the 
$\chi_{e_1}\chi_{e_2} \to X_A X_B \to \chi_{e_4} \chi_{e_3}$ scattering reactions
with $\chi_{e_a} = \chi^0_1$, $\chi^\pm_1$ read
\begin{align}\nonumber
 m \ &= \ \overline m \ = \ m_\chi \, ,
&
 M \ &= \ 2~m_\chi \, ,
&
 \Delta_{AB} \ &= \ 0 \, ,
\\
 \beta \ &= \ 1 \, ,
&
 P^s_{Z,\gamma} \ &= \ 1 \, ,
&
 P_{1\, AB} \ &= \ \frac{1}{2} \, .
\label{eq:app_parameterspurewino}
\end{align}
Further, the rescaled quantity $\hat m_{i_{1,2}}$ in the pure-wino limit reads
$\hat m_{1} = 1/2$ if it refers to the $\chi^0_1$ or $\chi^\pm_1$ species and it
vanishes if related to $Z$ and $\gamma$, $\hat m_{Z,\gamma} = 0$.
Taking the relations (\ref{eq:app_parameterspurewino}) into
account, we obtain concise analytic results for the kinematic factors relevant in
$\chi^+_1 \chi^-_1 \to X_A X_B \to \chi^+_1 \chi^-_1$
scattering. These are collected in Tab.~\ref{tab:app_kinfactorspurewino}.
\begin{table}[t]
\centering
\begin{tabular}{|c|c|c|c|c|c|c|}
\hline
 &&&&&&
\\[-2.75ex]
 & $^1S_0$
 & $^3S_1$
 & $^1P_1$
 & $^3P_{\cal J}$
 & $^1S_0^{(p^2)}$
 & $^3S_1^{(p^2)}$
\\
\hline\hline
 $B^{VV}_{rr,VV}({}^{2s+1}L_J)$
 & $0$
 & $-~\frac{19}{6}$
 & $0$
 & $0$
 & $0$
 & $\frac{152}{9}$
\\[0.2cm]
 $C^{(\alpha = 3,4) VV}_{rrr, 1V}({}^{2s+1}L_J)$
 & $0$
 & $-~\frac{4}{3}$
 & $0$
 & $0$
 & $0$
 & $\frac{64}{9}$
\\[0.2cm]
 $D^{(4) VV}_{rrrr, 11}({}^{2s+1}L_J)$
 & $2$
 & $\frac{2}{3}$
 & $\frac{8}{3}$
 & $\frac{56}{3}$
 & $-\frac{32}{3}$
 & $-\frac{32}{9}$
\\[0.2cm]
\hline\hline
 $B^{SS}_{rr,VV}({}^{2s+1}L_J)$
 & $0$
 & $\frac{1}{3}$
 & $0$
 & $0$
 & $0$
 & $-\frac{16}{9}$
\\[0.2cm]
\hline\hline
 $B^{\eta\overline\eta}_{rr,VV}({}^{2s+1}L_J)$
 & $0$
 & $-\frac{1}{12}$
 & $0$
 & $0$
 & $0$
 & $\frac{4}{9}$
\\[0.2cm]
\hline
\end{tabular}
\caption{Kinematic factors for partial wave reactions up to
         $\mathcal O(v_\text{rel}^2)$ in the pure-wino NRMSSM, relevant for the
         determination of the $\chi^+_1 \chi^-_1 \to W^+ W^-$ annihilation rate.
         The subscript label $V$ on the kinematic factors $B$ and $C$ above
         refers to both the cases of $Z$ and $\gamma$ single $s$-channel
         exchange in the (tree-level) annihilation amplitudes.
         The results for the kinematic factor $B$ in the last line apply to 
         $\eta\overline\eta = \eta^+\overline\eta^+, \eta^-\overline\eta^-$.}
\label{tab:app_kinfactorspurewino}
\end{table}
Note that we have
given only those kinematic factors that are associated with non-vanishing
coupling factors in the physical
$\chi^+_1 \chi^-_1 \to W^+ W^- \to \chi^+_1 \chi^-_1$ reaction.
Assembling and inserting the above results into the master formula
(\ref{eq:genericstructureWilson}) we find the results for the absorptive
part of the Wilson coefficients that provide
the $\chi^+_1 \chi^-_1 \to W^+ W^-$ annihilation cross-section 
(\ref{eq:res_sigmavrel}). For ${}^3S_1$ annihilation we have
\begin{eqnarray}
\nonumber
&&
\hspace*{-2cm} 
  \hat{f}^{\, \chi^+_1\chi^-_1 \to W^+ W^- \to \chi^+_1\chi^-_1}_{\lbrace 11 \rbrace \lbrace 11 \rbrace}(^{3}S_1)
\\[0.2cm]\nonumber
 \ & &= \ 
 \frac{\pi \alpha_2^2}{4 m_\chi^2} ~
 \Biggl(
 \sum\limits_{n = rr} \ \ \sum\limits_{i_1, i_2 = Z,\gamma}
       b^{\, \chi^+_1\chi^-_1 \to W^+ W^- \to \chi^+_1\chi^-_1 }_{n, \, i_1 i_2} \
       B^{\, VV }_{n, \, i_1 i_2} (^{3}S_1)
\\\nonumber
& &
\hspace{2cm}+
 \sum\limits_{\alpha = 3,4}^{} \ \ \sum\limits_{n = rrr} \
 \sum\limits_{i_1 = 1, i_2 = Z, \gamma}
       c^{(\alpha) \, \chi^+_1\chi^-_1 \to W^+ W^- \to \chi^+_1\chi^-_1 }_{n, \, i_1 i_2} \
       C^{(\alpha) \,  VV }_{n, \, i_1 i_2} (^{3}S_1)
\\\nonumber
& &
\hspace{2cm}+
 \sum\limits_{\alpha = 4}^{}  \ \sum\limits_{n = rrrr} \sum\limits_{i_1, i_2 = 1}
       d^{(\alpha) \, \chi^+_1\chi^-_1 \to W^+ W^- \to \chi^+_1\chi^-_1}_{n, \, i_1 i_2} \
       D^{(\alpha) \, VV }_{n, \, i_1 i_2} (^{3}S_1)
\\\nonumber
& &
\hspace{2cm}+
 \sum\limits_{n = rr} \ \ \sum\limits_{i_1, i_2 = Z,\gamma}
       b^{\, \chi^+_1\chi^-_1 \to G^+ G^- \to \chi^+_1\chi^-_1 }_{n, \, i_1 i_2} \
       B^{\, SS}_{n, \, i_1 i_2} (^{3}S_1)
\\\nonumber
& &
\hspace{2cm}+
 \sum\limits_{\eta = \eta^\pm}\ \ \sum\limits_{n = rr}\ \ \sum\limits_{i_1, i_2 = Z,\gamma}
       b^{\, \chi^+_1\chi^-_1 \to \eta \overline\eta \to \chi^+_1\chi^-_1 }_{n, \, i_1 i_2} \
       B^{\, \eta\overline\eta}_{n, \, i_1 i_2} (^{3}S_1)
 \Biggr)
\\[0.2cm]\nonumber
& &= \ 
 \frac{\pi \alpha_2^2}{4 m_\chi^2} ~
 \Biggl(
 \left( c_W^4 + c_W^2 s_W^2 + s_W^4 \right) \times \left(-\frac{19}{6}\right)
 \ + \
 2 \left( -c_W^2 - s_W^2 \right) \times \left(-\frac{4}{3}\right)
\\\nonumber
& &\hspace{2cm}+ 
 \ 1 \times \frac{2}{3} \
 \ + \
 \frac{1}{4} \times \frac{1}{3}
 \ - \
 2 \times \frac{1}{12}
 \Biggr)
\\[0.2cm]
& &= \ 
 \frac{1}{48}~\frac{\pi \alpha_2^2}{m_\chi^2} \ ,
\end{eqnarray}
where we have summed over all (unphysical) final states in  Feynman gauge,
$X_A X_B = W^+ W^-, G^+ G^-, \eta^+\overline\eta^+, \eta^-\overline\eta^-$, that
contribute to the physical $\chi^+_1 \chi^-_1 \to W^+ W^-$ rate in the pure-wino
NRMSSM scenario.
In case of the $^1S_0$ annihilation reaction only the pieces related to 
the $\alpha = 4$ box-amplitude contribute, and the only non-vanishing coupling
factor $d^{(4)}_{n, \, i_1 i_2}$
is $d^{(4)}_{rrrr \,, 11}$ given in~(\ref{d4rrrr}), therefore
\begin{align}\nonumber
  \hat{f}^{\, \chi^+_1\chi^-_1 \to W^+ W^- \to \chi^+_1\chi^-_1}_{\lbrace 11 \rbrace \lbrace 11 \rbrace}(^{1}S_0) \ &=
 \frac{\pi \alpha_2^2}{4 m_\chi^2} \ \
       d^{(4) \, \chi^+_1\chi^-_1 \to W^+ W^- \to \chi^+_1\chi^-_1}_{rrrr, \, 11} \
       D^{(4) \, VV }_{rrrr, \, 11} (^{1}S_0)
\\
&= \
  \frac{\pi \alpha_2^2}{2 m_\chi^2} \, .
\end{align}
Finally, the absorptive parts of the $\mathcal O(v_\text{rel}^2)$ partial-wave Wilson coefficients  read
\begin{align}\nonumber
  \hat{f}^{\, \chi^+_1\chi^-_1 \to W^+ W^- \to \chi^+_1\chi^-_1}_{\lbrace 11 \rbrace \lbrace 11 \rbrace}(^{1}P_1) \ &= \
 \frac{2\pi \alpha_2^2}{3 m_\chi^2} ,
&
  \hat{f}^{\, \chi^+_1\chi^-_1 \to W^+ W^- \to \chi^+_1\chi^-_1}_{\lbrace 11 \rbrace \lbrace 11 \rbrace}(^{3}P_{\cal J}) \ &= \
 \frac{14 \pi \alpha_2^2}{3 m_\chi^2} ,
\\
  \hat{g}^{\, \chi^+_1\chi^-_1 \to W^+ W^- \to \chi^+_1\chi^-_1}_{\lbrace 11 \rbrace \lbrace 11 \rbrace}(^{1}S_0) \ &= \
 -~\frac{8\pi \alpha_2^2}{3 m_\chi^2} ,
&
  \hat{g}^{\, \chi^+_1\chi^-_1 \to W^+ W^- \to \chi^+_1\chi^-_1}_{\lbrace 11 \rbrace \lbrace 11 \rbrace}(^{3}S_1) \ &= \
 -~\frac{\pi \alpha_2^2}{9 m_\chi^2} .
\end{align}
Hence, following (\ref{eq:res_sigmavrel}),
the non-relativistic expansion of the $\chi^+_1\chi^-_1\to W^+W^-$
annihilation cross section in the pure-wino NRMSSM is given by
\begin{align}\nonumber
 \sigma^{\chi^+_1 \chi^-_1 \to W^+ W^-} v_\text{rel} \ &= \
 a \ + \ \left( b_P + b_S  \right) v_\text{rel}^2 \ + \ \mathcal O(v_\text{rel}^4)
\\\nonumber
&= \
 \frac{9}{16} ~ \frac{\pi \alpha_2^2}{m_\chi^2}
 \ + \
 \left( \frac{1}{3} -\frac{3}{16} \right) \frac{\pi \alpha_2^2}{m_\chi^2} ~
 v_\text{rel}^2
 \ + \
 \mathcal O(v_\text{rel}^4) \ ,
\\
&= \
 \frac{9}{16} ~ \frac{\pi \alpha_2^2}{m_\chi^2}
 \ + \
 \frac{7}{48} ~ \frac{\pi \alpha_2^2}{m_\chi^2} ~ v_\text{rel}^2
 \ + \
 \mathcal O(v_\text{rel}^4) \ .
\end{align}
The values for the parameters $a, b_P$ and $b_S$, that one obtains for a
pure-wino NRMSSM mass scale $m_\chi = 2748.92\,$GeV read
$a = 3.06\cdot 10^{-27}\,$ cm$^3$ s$^{-1}$,
$b_P\,c^2 = 1.81\cdot 10^{-27}$ cm$^3$ s$^{-1}$
and $b_S\,c^2 = -1.02\cdot10^{-27}$ cm$^3$ s$^{-1}$.
The mass scale $m_\chi$ agrees with the neutralino LSP mass of
the MSSM scenario introduced in Sec.~\ref{sec:results}.
The latter MSSM scenario features a small but non-vanishing Higgsino
admixture to the wino-like $\chi^0_1$ and $\chi^\pm_1$: the Higgsino-like
neutralino and chargino states are not at all decoupled but reside at the
scale of $\sim 2.9 - 3\,$TeV.
Thus we should not expect the results for the wino-like scenario of
Sec.~\ref{sec:results} to be approximated by the pure-wino NRMSSM. This is in
fact what the comparison of the parameters $a, b_P$ and $b_S$ for the
$\chi^+_1\chi^-_1 \to W^+ W^-$ annihilation cross section shows:
the corresponding parameters in the MSSM scenario
investigated in Sec.~\ref{sec:results} were given by
$a = 2.65 \cdot10^{-27}\,$cm$^3$ s$^{-1}$,
$b_P\,c^2 = 1.86\cdot10^{-27}\,$cm$^3$ s$^{-1}$,
$b_S\,c^2 = -0.88\cdot10^{-27}\,$cm$^3$ s$^{-1}$.
The results for the $S$-wave parameters $a$ and $b_S$ in the pure-wino
$\chi^+_1\chi^-_1 \to W^+ W^-$ reaction are a bit larger, which is a consequence
of the larger couplings of the pure-wino neutralino and chargino states to the
$SU(2)_L$ gauge bosons and the absence of $t$-channel annihilation into the
(unphysical) final state $G^+ G^-$.
Due to the non-decoupled higgsino-like neutralino states in the scenario of
Sec.~\ref{sec:results} the latter contribution is present and interferes
destructively with the corresponding $s$-channel exchange contribution also
present in the pure-wino NRMSSM limit. This leads to a suppression of the $a$
and $b_S$ cross section parameters in the wino-like scenario of
Sec.~\ref{sec:results} with respect to the pure-wino NRMSSM.
On the contrary the parameter $b_P$ turns out to be somewhat larger in the
Sec.~\ref{sec:results} scenario which traces back to the non-vanishing
$P$-wave $t$-channel annihilations into $G^+G^-$ final states that are absent
in the pure-wino NRMSSM.
Note that the $\chi^+\chi^-\to W^+ W^-$ annihilation cross section for the
Sec.~\ref{sec:results} scenario in addition exhibits non-vanishing contributions
from the (unphysical) $VS = W^\pm G^\mp$ final states not present in the
pure-wino NRMSSM. These are however suppressed with respect to the
$X_A X_B = W^+W^-, G^+G^-$ contributions.

\subsection{Exclusive (co-)annihilation rates in the pure-wino NRMSSM}
\label{sec:app_gammaspurewino}
\begin{table}[t!]
\centering
\begin{tabular}{|c|c|c|c|c|c|c|}
\hline
\multicolumn{7}{|c|}{$\chi^+_1 \chi^-_1 \to \chi^+_1 \chi^-_1$ reactions}
 \\[0.1cm]
\hline
 &&&&&&\\[-0.55cm]
physical final state $X_A X_B$
 & $c(^1S_0)$
 & $c(^3S_1)$
 & $c(^1P_1)$
 & $c(^3P_{\cal J})$
 & $c(^1S_0^{(p^2)})$
 & $c(^3S_1^{(p^2)})$
\\
\hline\hline
 $W^+ W^-$
 & $\frac{1}{2}$
 & $\frac{1}{48}$
 & $\frac{2}{3}$
 & $\frac{14}{3}$
 & $-~\frac{8}{3}$
 & $-\frac{1}{9}$
\\[0.2cm]
 $Z Z$
 & $c_W^4$
 & $0$
 & $0$
 & $\frac{28}{3}~c_W^4$
 & $-~\frac{16}{3}~c_W^4$
 & $0$
\\[0.2cm]
 $Z \gamma$
 & $2\,c_W^2 s_W^2$
 & $0$
 & $0$
 & $\frac{56}{3}\,c_W^2 s_W^2$
 & $-\frac{32}{3}\,c_W^2 s_W^2$
 & $0$
\\[0.2cm]
 $\gamma\gamma$
 & $\,s_W^4$
 & $0$
 & $0$
 & $\frac{28}{3}\,s_W^4$
 & $-\frac{16}{3}\,s_W^4$
 & $0$
\\[0.2cm]
 $Z h^0$
 & $0$
 & $\frac{1}{48}$
 & $0$
 & $0$
 & $0$
 & $-~\frac{1}{9}$
\\[0.2cm]
 $q \overline q$
 & $0$
 & $\frac{1}{8}$
 & $0$
 & $0$
 & $0$
 & $-~\frac{2}{3}$
\\[0.2cm]
 $l^+ l^-, \ \nu \overline\nu$
 & $0$
 & $\frac{1}{24}$
 & $0$
 & $0$
 & $0$
 & $-~\frac{2}{9}$
\\[0.2cm]
 $\sum X_A X_B$
 & $\frac{3}{2}$
 & $\frac{25}{24}$
 & $\frac{2}{3}$
 & $14$
 & $-8$
 & $-~\frac{50}{9}$
\\[0.2cm]
\hline\hline
\multicolumn{7}{|c|}{}
\\[-0.4cm]
\multicolumn{7}{|c|}{$\chi^0_1 \chi^0_1 \to \chi^0_1 \chi^0_1$ reactions}
 \\[0.2cm]
\hline\hline
 $W^+ W^-$
 & $2$
 & $0$
 & $0$
 & $\frac{56}{3}$
 & $-~\frac{32}{3}$
 & $0$
\\[0.2cm]
\hline\hline
\multicolumn{7}{|c|}{}
\\[-0.4cm]
\multicolumn{7}{|c|}{$\chi^0_1 \chi^0_1 \to \chi^+_1 \chi^-_1$ and
                     $\chi^+_1 \chi^-_1 \to \chi^0_1 \chi^0_1$ reactions}
 \\[0.2cm]
\hline\hline
 $W^+ W^-$
 & $1$
 & $0$
 & $0$
 & $\frac{28}{3}$
 & $-~\frac{16}{3}$
 & $0$
\\
\hline
\end{tabular}
\caption{$c(^{2s+1}L_J)$ factors that enter the contributions to the
         pure-wino NRMSSM Wilson coefficients in neutral
         $\chi_{e_1}\chi_{e_2} \to X_A X_B \to \chi_{e_4}\chi_{e_3}$ processes
         with exclusive (physical) final states $X_A X_B$.
         In case of $\chi^+_1 \chi^-_1 \to X_A X_B \to \chi^+_1 \chi^-_1$ rates
         where several two-particle final states $X_A X_B$ are accessible the
         inclusive result is also given.}
\label{tab:app_x1x1_0_purewino}
\end{table}

This section collects the results for the exclusive (physical) $X_A X_B$
final state contributions to the Wilson coefficients $\hat f, \hat g$
that determine the (off-)diagonal (co-)annihilation rates
$\chi_{e_1}\chi_{e_2} \to X_A X_B \to \chi_{e_4} \chi_{e_3}$ in the pure-wino NRMSSM. 
The non-relativistic expansion of the respective exclusive rates can then be
obtained from (\ref{eq:res_Gammarate}).  For convenience we
write the pure-wino NRMSSM Wilson coefficients as
\begin{align}
 \hat f^{\,\chi_{e_1}\chi_{e_2} \to X_A X_B \to \chi_{e_4} \chi_{e_3}}
    \left( {}^{2s+1}L_J \right)
 \ = \
 \frac{\pi \alpha_2^2}{m_\chi^2} ~
 c^{\,\chi_{e_1}\chi_{e_2} \to X_A X_B \to \chi_{e_4} \chi_{e_3}} \left( {}^{2s+1}L_J \right) \ .
 \label{eq:app_genericf_structure}
\end{align}
In case of the next-to-next-to-leading order $S$-wave coefficients we establish
a similar notation with $\hat f$ replaced by $\hat g$ on the l.h.s. of (\ref{eq:app_genericf_structure})
and the ${}^{2s+1}L_J = {}^1S_0, {}^3S_1$ label of the factor
$c$ on the r.h.s. substituted by $^{1}S_0^{(p^2)}, {}^{3}S_1^{(p^2)}$.
Note that the Wilson coefficients
$\hat h_i$ always vanish in the pure-wino NRMSSM due to the complete
mass-degeneracy of the $\chi^0_1$ and $\chi^\pm_1$ states.

We have already noted at the beginning of Sec.~\ref{sec:appendixexample} that
the pure-wino NRMSSM toy-scenario features massless SM gauge bosons and SM
fermions. These can hence appear as possible $X_A X_B$ final state particles in
the $\chi_{e_1}\chi_{e_2} \to X_A X_B \to \chi_{e_4} \chi_{e_3}$ reactions.
As far as the Higgs-sector is concerned we present in this section results that
refer to the decoupling limit \cite{Gunion:2002zf} in the underlying MSSM
scenario: we assume a SM-like $CP$-even Higgs boson $h^0$ in the
low-energy spectrum of the theory while the heavier Higgs states
$A^0, H^0, H^\pm$ are  entirely decoupled
($m_{A^0} \sim m_{H^0} \sim m_{H^+} \gg m_\chi \gg 0$).
As generically $m_{h^0} < m_Z$ at tree-level in the MSSM, the $h^0$ is
consequently treated as massless in the pure-wino NRMSSM.
According to their overall charge the (co-)annihilation processes can be
arranged into three charge-sectors: neutral, positive and double positive
charged.
The results for the corresponding (double) negative charged reactions are
identical to the results for (double) positive charged processes.
We collect our results for the factors $c(^{2s+1}L_J)$ in
Tables~\ref{tab:app_x1x1_0_purewino}--\ref{tab:app_x1x1_p2_purewino}.

\begin{table}[t]
\centering
\begin{tabular}{|c|c|c|c|c|c|c|}
\hline
\multicolumn{7}{|c|}{$\chi^0_1 \chi^+_1 \to \chi^0_1 \chi^+_1$ reactions}
 \\[0.1cm]
\hline
 &&&&&&\\[-0.55cm]
physical final state $X_A X_B$
 & $c(^1S_0)$
 & $c(^3S_1)$
 & $c(^1P_1)$
 & $c(^3P_{\cal J})$
 & $c(^1S_0^{(p^2)})$
 & $c(^3S_1^{(p^2)})$
\\
\hline\hline
 $W^+ Z$
 & $\frac{1}{2}~c_W^2$
 & $\frac{1}{48}$
 & $\frac{2}{3}~c_W^2$
 & $\frac{14}{3}~c_W^2$
 & $-~\frac{8}{3}\,c_W^2$
 & $-\frac{1}{9}$
\\[0.2cm]
 $W^+ \gamma$
 & $\frac{1}{2}~s_W^2$
 & $0$
 & $\frac{2}{3}~s_W^2$
 & $\frac{14}{3}~s_W^2$
 & $-~\frac{8}{3}~s_W^2$
 & $0$
\\[0.2cm]
 $W^+ h^0$
 & $0$
 & $\frac{1}{48}$
 & $0$
 & $0$
 & $0$
 & $-\frac{1}{9}$
\\[0.2cm]
 $u \overline d$
 & $0$
 & $\frac{1}{4}$
 & $0$
 & $0$
 & $0$
 & $-~\frac{4}{3}$
\\[0.2cm]
 $\nu l^+$
 & $0$
 & $\frac{1}{12}$
 & $0$
 & $0$
 & $0$
 & $-~\frac{4}{9}$
\\[0.2cm]
 $\sum X_A X_B$
 & $\frac{1}{2}$
 & $\frac{25}{24}$
 & $\frac{2}{3}$
 & $\frac{14}{3}$
 & $-~\frac{8}{3}$
 & $-~\frac{50}{9}$
\\[0.2cm]
\hline
\end{tabular}
\caption{$c(^{2s+1}L_J)$ expressions associated with the
         pure-wino NRMSSM Wilson coefficients in exclusive
         single charged
         $\chi^0_1\chi^+_1 \to X_A X_B \to \chi^0_1\chi^+_1$
         reactions.
         The last line is the inclusive result.}
\label{tab:app_x1n1_p1_purewino}
\end{table}
\begin{table}[t]
\centering
\begin{tabular}{|c|c|c|c|c|c|c|}
\hline
\multicolumn{7}{|c|}{$\chi^+_1 \chi^+_1 \to \chi^+_1 \chi^+_1$ reactions}
 \\[0.1cm]
\hline
 &&&&&&\\[-0.55cm]
physical final state $X_A X_B$
 & $c(^1S_0)$
 & $c(^3S_1)$
 & $c(^1P_1)$
 & $c(^3P_{\cal J})$
 & $c(^1S_0^{(p^2)})$
 & $c(^3S_1^{(p^2)})$
\\
\hline\hline
 $W^+ W^+$
 & $1$
 & $0$
 & $0$
 & $\frac{28}{3}$
 & $-~\frac{16}{3}$
 & $0$
\\[0.2cm]
\hline
\end{tabular}
\caption{$c(^{2s+1}L_J)$ factors related to the pure-wino
         NRMSSM Wilson coefficients in double charged
         $\chi^+_1\chi^+_1 \to X_A X_B \to \chi^+_1\chi^+_1$
         processes.}
\label{tab:app_x1x1_p2_purewino}
\end{table}

In case of inclusive leading-order ${}^1S_0$ and ${}^3S_1$ (co-)
annihilations we find agreement between the results of
Tables~\ref{tab:app_x1x1_0_purewino}--\ref{tab:app_x1x1_p2_purewino} and the
corresponding expressions given in \cite{Hisano:2006nn} for the same 
scenario.
In addition, we reproduce the leading-order ${}^1S_0$ wave annihilation rates
into the exclusive final states $W^+W^-,\, ZZ,\, Z\gamma$ and $\gamma\gamma$
given by the same authors in~\cite{Hisano:2004ds}, apart from the
$W^+W^-$ off-diagonal rates, where our findings are a factor of $2$ larger.
The results for the $P$- and
$\mathcal O(v_\text{rel}^2)$ $S$-wave Wilson coefficients are new.


\begin{thebibliography}{10}




\bibitem{Gondolo:2004sc}
P.~Gondolo, J.~Edsjo, P.~Ullio, L.~Bergstrom, M.~Schelke, {\em et al.},
  \href{http://dx.doi.org/10.1088/1475-7516/2004/07/008}{{\em JCAP} {\bf 0407}
  (2004)  008},
\href{http://arxiv.org/abs/astro-ph/0406204}{{\tt arXiv:astro-ph/0406204
  [astro-ph]}}.

\bibitem{Belanger:2010gh}
G.~Belanger, F.~Boudjema, P.~Brun, A.~Pukhov, S.~Rosier-Lees, {\em et al.},
  \href{http://dx.doi.org/10.1016/j.cpc.2010.11.033}{{\em Comput.Phys.Commun.}
  {\bf 182} (2011)  842--856},
\href{http://arxiv.org/abs/1004.1092}{{\tt arXiv:1004.1092 [hep-ph]}}.

\bibitem{Herrmann:2007ku}
B.~Herrmann and M.~Klasen,
  \href{http://dx.doi.org/10.1103/PhysRevD.76.117704}{{\em Phys.Rev.} {\bf D76}
  (2007)  117704},
\href{http://arxiv.org/abs/0709.0043}{{\tt arXiv:0709.0043 [hep-ph]}}.

\bibitem{Herrmann:2009wk}
B.~Herrmann, M.~Klasen, and K.~Kovarik,
  \href{http://dx.doi.org/10.1103/PhysRevD.79.061701}{{\em Phys.Rev.} {\bf D79}
  (2009)  061701},
\href{http://arxiv.org/abs/0901.0481}{{\tt arXiv:0901.0481 [hep-ph]}}.

\bibitem{Herrmann:2009mp}
B.~Herrmann, M.~Klasen, and K.~Kovarik,
  \href{http://dx.doi.org/10.1103/PhysRevD.80.085025}{{\em Phys.Rev.} {\bf D80}
  (2009)  085025},
\href{http://arxiv.org/abs/0907.0030}{{\tt arXiv:0907.0030 [hep-ph]}}.

\bibitem{Baro:2007em}
N.~Baro, F.~Boudjema, and A.~Semenov,
  \href{http://dx.doi.org/10.1016/j.physletb.2008.01.031}{{\em Phys.Lett.} {\bf
  B660} (2008)  550--560},
\href{http://arxiv.org/abs/0710.1821}{{\tt arXiv:0710.1821 [hep-ph]}}.

\bibitem{Baro:2009na}
N.~Baro, F.~Boudjema, G.~Chalons, and S.~Hao,
  \href{http://dx.doi.org/10.1103/PhysRevD.81.015005}{{\em Phys.Rev.} {\bf D81}
  (2010)  015005},
\href{http://arxiv.org/abs/0910.3293}{{\tt arXiv:0910.3293 [hep-ph]}}.

\bibitem{Boudjema:2011ig}
F.~Boudjema, G.~Drieu La~Rochelle, and S.~Kulkarni,
  \href{http://dx.doi.org/10.1103/PhysRevD.84.116001}{{\em Phys.Rev.} {\bf D84}
  (2011)  116001},
\href{http://arxiv.org/abs/1108.4291}{{\tt arXiv:1108.4291 [hep-ph]}}.

\bibitem{Chatterjee:2012db}
A.~Chatterjee, M.~Drees, and S.~Kulkarni,
\href{http://arxiv.org/abs/1209.2328}{{\tt arXiv:1209.2328 [hep-ph]}}.

\bibitem{Drees:2013er}
  M.~Drees and J.~Gu,
  arXiv:1301.1350 [hep-ph].

\bibitem{Hisano:2004ds}
J.~Hisano, S.~Matsumoto, M.~M. Nojiri, and O.~Saito,
  \href{http://dx.doi.org/10.1103/PhysRevD.71.063528}{{\em Phys.Rev.} {\bf D71}
  (2005)  063528},
\href{http://arxiv.org/abs/hep-ph/0412403}{{\tt arXiv:hep-ph/0412403
  [hep-ph]}}.

\bibitem{Hisano:2006nn}
J.~Hisano, S.~Matsumoto, M.~Nagai, O.~Saito, and M.~Senami,
  \href{http://dx.doi.org/10.1016/j.physletb.2007.01.012}{{\em Phys.Lett.} {\bf
  B646} (2007)  34--38},
\href{http://arxiv.org/abs/hep-ph/0610249}{{\tt arXiv:hep-ph/0610249
  [hep-ph]}}.

\bibitem{Cirelli:2007xd}
M.~Cirelli, A.~Strumia, and M.~Tamburini,
  \href{http://dx.doi.org/10.1016/j.nuclphysb.2007.07.023}{{\em Nucl.Phys.}
  {\bf B787} (2007)  152--175},
\href{http://arxiv.org/abs/0706.4071}{{\tt arXiv:0706.4071 [hep-ph]}}.


\bibitem{Hryczuk:2010zi}
A.~Hryczuk, R.~Iengo, and P.~Ullio,
  \href{http://dx.doi.org/10.1007/JHEP03(2011)069}{{\em JHEP} {\bf 1103} (2011)
   069},
\href{http://arxiv.org/abs/1010.2172}{{\tt arXiv:1010.2172 [hep-ph]}}.

\bibitem{Hryczuk:2011vi}
A.~Hryczuk and R.~Iengo, \href{http://dx.doi.org/10.1007/JHEP01(2012)163}{{\em
  JHEP} {\bf 1201} (2012)  163},
\href{http://arxiv.org/abs/1111.2916}{{\tt arXiv:1111.2916 [hep-ph]}}.

\bibitem{Beneke:2012tg}
  M.~Beneke, C.~Hellmann and P.~Ruiz-Femenia,
  JHEP {\bf 1303} (2013) 148
  [arXiv:1210.7928 [hep-ph]].

\bibitem{Chen:2013bi}
  J.~Chen and Y.~-F.~Zhou,
  arXiv:1301.5778 [hep-ph].

\bibitem{dotmfile} See the \texttt{Mathematica} package ``kinfactors.m'' available with the arXiv source of this paper.


\bibitem{Bodwin:1994jh}
G.~T. Bodwin, E.~Braaten, and G.~P. Lepage,
  \href{http://dx.doi.org/10.1103/PhysRevD.51.1125, 10.1103/PhysRevD.55.5853,
  10.1103/PhysRevD.51.1125, 10.1103/PhysRevD.55.5853}{{\em Phys.Rev.} {\bf D51}
  (1995)  1125--1171},
\href{http://arxiv.org/abs/hep-ph/9407339}{{\tt arXiv:hep-ph/9407339
  [hep-ph]}}.


\bibitem{Alwall:2011uj}
J.~Alwall, M.~Herquet, F.~Maltoni, O.~Mattelaer, and T.~Stelzer,
  \href{http://dx.doi.org/10.1007/JHEP06(2011)128}{{\em JHEP} {\bf 1106} (2011)
   128},
\href{http://arxiv.org/abs/1106.0522}{{\tt arXiv:1106.0522 [hep-ph]}}.

\bibitem{paperIII}
M.~Beneke, C.~Hellmann, and P.~Ruiz-Femen\'ia. In preparation.

\bibitem{Gunion:2002zf}
  J.~F.~Gunion and H.~E.~Haber,
  Phys.\ Rev.\ D {\bf 67} (2003) 075019
  [hep-ph/0207010].


\end{thebibliography}

\providecommand{\href}[2]{#2}\begingroup\raggedright\endgroup

\end{document}